\documentclass[10pt,a4paper]{book}

\PassOptionsToPackage{table,svgnames,dvipsnames}{xcolor}

\usepackage[a4paper, portrait, margin=1.1811in]{geometry}
\usepackage[english]{babel}
\usepackage[utf8]{inputenc}
\usepackage[T1]{fontenc}
\usepackage{lmodern}
\usepackage{etoolbox}
\usepackage{graphicx}
\usepackage{titlesec}
\usepackage{caption}
\usepackage{booktabs}
\usepackage{xcolor} 
\usepackage[colorlinks, citecolor=cyan]{hyperref}
\usepackage{caption}
\graphicspath{ {./images/} }
\usepackage{scrextend}
\usepackage{fancyhdr}
\usepackage{graphicx}
\newcounter{lemma}

\newcounter{theorem}

\hypersetup{
    linkcolor=blue
}

\makeatletter
\patchcmd{\@maketitle}{\LARGE \@title}{\fontsize{16}{25}\selectfont\@title}{}{}
\makeatother

\usepackage{authblk}

\titlespacing\section{0pt}{12pt plus 4pt minus 2pt}{6pt}
\titlespacing\subsection{0pt}{12pt plus 4pt minus 2pt}{4pt}
\titlespacing\subsubsection{12pt}{12pt plus 4pt minus 2pt}{4pt}

\usepackage{enumitem}
\setlist[itemize]{topsep=0pt} 

\titleformat{\section}{\normalfont\fontsize{14}{17}\bfseries}{\thesection.}{1em}{}
\titleformat{\subsection}{\normalfont\fontsize{12}{15}\bfseries}{\thesubsection.}{1em}{}
\titleformat{\subsubsection}{\normalfont\fontsize{10}{15}\bfseries}{\thesubsubsection.}{1em}{}

\titleformat{\author}{\normalfont\fontsize{10}{15}\bfseries}{\thesection}{1em}{}

\usepackage{parskip}
\newlength\paragraphmargin
\usepackage{caption}  
\usepackage{subcaption} 
\usepackage{floatrow} 
\usepackage{csquotes} 

\usepackage{amsmath,bm}
\usepackage{amsfonts}
\usepackage{fancyhdr}
\usepackage{amssymb}

\usepackage{upgreek} 
\usepackage[ nocomma ]{optidef} 
\usepackage[mathscr]{euscript} 
\usepackage{wrapfig} 

\usepackage{blindtext}

\usepackage{enumitem} 
\usepackage{wrapfig}

\usepackage{longtable}
\usepackage{array} 

\usepackage{titlesec}

\titleformat{\paragraph}
{\normalfont\normalsize\bfseries}{\theparagraph}{1em}{}
\titlespacing*{\paragraph}
{0pt}{3.25ex plus 1ex minus .2ex}{1.5ex plus .2ex}

\usepackage{acro} 

\newlist{inparaenum}{enumerate}{2}
\setlist[inparaenum]{nosep}
\setlist[inparaenum,1]{label=\bfseries\arabic*.}
\setlist[inparaenum,2]{label=\emph{\alph*})}

\definecolor{Prune}{RGB}{99,0,60} 
\definecolor{B1}{RGB}{49,62,72} 
\definecolor{C1}{RGB}{124,135,143}
\definecolor{D1}{RGB}{213,218,223}
\definecolor{A2}{RGB}{198,11,70}
\definecolor{B2}{RGB}{237,20,91}
\definecolor{C2}{RGB}{238,52,35}
\definecolor{D2}{RGB}{243,115,32}
\definecolor{A3}{RGB}{124,42,144}
\definecolor{B3}{RGB}{125,106,175}
\definecolor{C3}{RGB}{198,103,29}
\definecolor{D3}{RGB}{254,188,24}
\definecolor{A4}{RGB}{0,78,125}
\definecolor{B4}{RGB}{14,135,201}
\definecolor{C4}{RGB}{0,148,181}
\definecolor{D4}{RGB}{70,195,210}
\definecolor{A5}{RGB}{0,128,122}
\definecolor{B5}{RGB}{64,183,105}
\definecolor{C5}{RGB}{140,198,62}
\definecolor{D5}{RGB}{213,223,61}
\definecolor{SkyBlue}{RGB}{135,206,250}
\definecolor{MedAqua}{RGB}{102,205,170}
\definecolor{SteelBlue}{RGB}{70,130,180}
\definecolor{MyDarkGreen}{RGB}{0,153,0}
\definecolor{MyGreen}{RGB}{187,255,153}
\definecolor{Grey}{RGB}{192,192,192}
\definecolor{LightGrey}{RGB}{200,200,200}

\usepackage{graphicx}
\usepackage{pgfplots}
\usepackage{pgfplotstable}
\usepackage[final]{microtype}

\usepackage{eurosym}

\usepackage{bm}

\numberwithin{equation}{section}

\usepackage{algorithm}
\usepackage{algpseudocode}

\usepackage{eurosym} 

\usepackage{cases}

\usepackage{setspace}

\pgfplotsset{compat=1.18}
	
\newsavebox\affbox
\author[]{\textbf{Florent Cogen}}
\author[]{\textbf{Emily Little}}
\author[]{\textbf{Virginie Dussartre}}
\author[]{\textbf{Quentin Bustarret}}
\affil[]{\textbf{RTE} \\
7C Place du Dôme \\
La Défense \\France
}

\title{\textbf{\huge ATLAS: A Model of Short-term European Electricity Market Processes under Uncertainty - Balancing Modules}}
\date{\today}

\begin{document}

\pagestyle{headings}	
\newpage
\setcounter{page}{1}
\renewcommand{\thepage}{\arabic{page}}

\newgeometry{left=2cm,bottom=2cm, top=2cm, right=2cm}

\maketitle
	
%

\tableofcontents

\chapter{Introduction and Context}

This paper is part of the documentation of the ATLAS agent-based electricity market model. It follows and complements \cite{little_atlas_nodate} by focusing on modules specifically developed for the balancing section of the electricity market process. Certain steps of the balancing section, such as the Clearing, are done using modules also applied for day-ahead or intraday market simulations that are already explained in the aforementioned article, and therefore not detailed here.\\

\section{Overview of balancing markets characteristics}
\label{sec:Introduction - markets characteristics}

The European power system is currently experiencing a significant change in its balancing stage. Historically managed locally by each European Transmission System Operator (TSO) using diverse processes, it is now switching to common balancing markets occurring within the last hour before real-time, and that are structured around 4 different types of reserves:

\begin{itemize}
    \item Frequency Containment Reserves (FCR) are the fastest amongst them, being automatically activated within a few seconds when an imbalance arises to stop the induced frequency deviation.
    \item automatic Frequency Restoration Reserves (aFRR) are automatically activated in 30 seconds in order to bring the frequency back to 50 Hz.
    \item manual Frequency Restoration Reserves (mFRR) take over the role of aFRR in 12.5 minutes, with a manual activation process.
    \item Finally, Replacement Reserves (RR) are manually activated in 30 minutes to compensate for the activation of all previously described reserves.
\end{itemize}

Amongst these reserve types, mFRR and RR stand out for being activated manually. They are consequently part of energy markets, respectively traded on the cross-border platforms MARI and TERRE. Both markets include two kinds of actors, Balancing Services Providers (BSPs) that offer reserve energy and TSOs that formulate balancing needs.\\

\subsection{The RR market}
\label{sec:Introduction - RR market}
\begin{figure}[ht!]
    \centering
    \includegraphics[width = \textwidth]{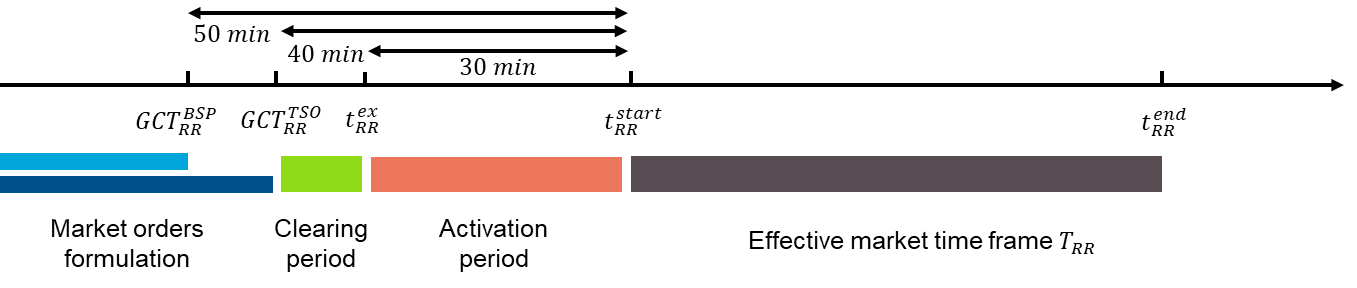}
    \caption{RR market schematic time frame}
    \label{fig:rr_market_time_frame}
\end{figure}

A sequential representation of the RR market operational process is given in Figure \ref{fig:rr_market_time_frame}, based on public descriptions available in \cite{entsoe2023terre}. It is executed at an hourly frequency, and spans over an effective time frame $T_{RR}$ (meaning the time frame over which the traded reserves are activated) that lasts for 1 hour between $t_{RR}^{start}$ and $t_{RR}^{end}$. The whole market process is comprised of the following stages:

\begin{enumerate}
    \item BSPs send their reserve orders to their respective TSOs before their Gate Closure Time ($GCT_{RR}^{BSP}$), whose value is decided by each TSO for its own area. Consequently, different BSP GCTs exist in Europe, usually comprised between 55 and 50 minutes before $t_{RR}^{start}$.
    \item TSOs can filter out certain BSP marker orders if a strong justification is given, mainly related to internal congestions that may be created by these orders and that cannot be addressed by the RR market process. In addition, they need to compute their balancing needs, convert them into market orders and send them to the TERRE platform before $GCT_{RR}^{TSO}$, universally set at 40 minutes before $t_{RR}^{start}$, alongside all BSP orders that were not filtered.
    \item The TERRE platform conducts the Clearing stage between $GCT_{RR}^{TSO}$ and $t_{RR}^{ex}$. In this stage, market orders are activated using a common merit order list and the market clearing price is determined using the marginal pricing method.
    \item Clearing results are transmitted back to all actors 30 minutes before $t_{RR}^{start}$.
\end{enumerate}

\subsection{The mFRR market}
\label{sec:Introduction - mFRR market}
Compared to RR markets, mFRR markets add another layer of complexity as they combine two different activation processes, as described in \cite{entsoe2023mari}: "direct" and "scheduled" activations. The direct activation mode is a continuous market, where a pool of BSP orders is always available. A TSO can submit a balancing need to the platform at any time, and it is instantly cleared with a merit order list constituted of BSP orders of the relevant direction (for instance, given an upward balancing need--i.e. a need for additional generation--, only upward BSP orders are considered in the merit order list). In contrast, the scheduled activation mode is basically identical to the RR market process, with different time frames. This activation mode is the only one implemented in ATLAS, as the direct activation mode was deemed too complex to model and it is currently not integrated into it.

\begin{figure}[ht!]
    \centering
    \includegraphics[width = \textwidth]{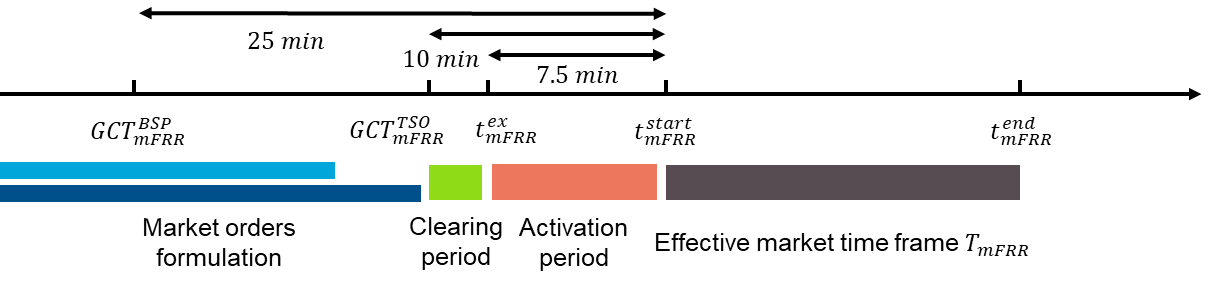}
    \caption{mFRR market schematic time frame}
    \label{fig:mfrr_market_time_frame}
\end{figure}

The sequential process of the scheduled mFRR market is described in Figure \ref{fig:mfrr_market_time_frame}. It is executed at a quarter-hourly frequency, and spans over an effective time frame $T_{mFRR}$ that lasts for 15 minutes between $t_{mFRR}^{start}$ and $t_{mFRR}^{end}$. The whole process is divided into the same 4 steps as the ones presented in Section \ref{sec:Introduction - RR market}, with the following specific values:

\begin{itemize}
    \item The BSP Gate Closure Time ($GCT_{mFRR}^{BSP}$)
    is set at 25 minutes before $t_{mFRR}^{start}$.
    \item The TSO $GCT_{mFRR}^{TSO}$ is set at 10 minutes before $t_{mFRR}^{start}$.
    \item The Clearing is conducted on the MARI platform between $GCT_{mFRR}^{TSO}$ and $t_{mFRR}^{ex}$, and results are transmitted back to all actors at $t_{mFRR}^{ex}$ (7.5 minutes before $t_{mFRR}^{start}$).
\end{itemize}

\section{Balancing stage within ATLAS}
\label{sec:Introduction - balancing in ATLAS}

\begin{figure}[H]
    \centering
    \includegraphics[width=\textwidth]{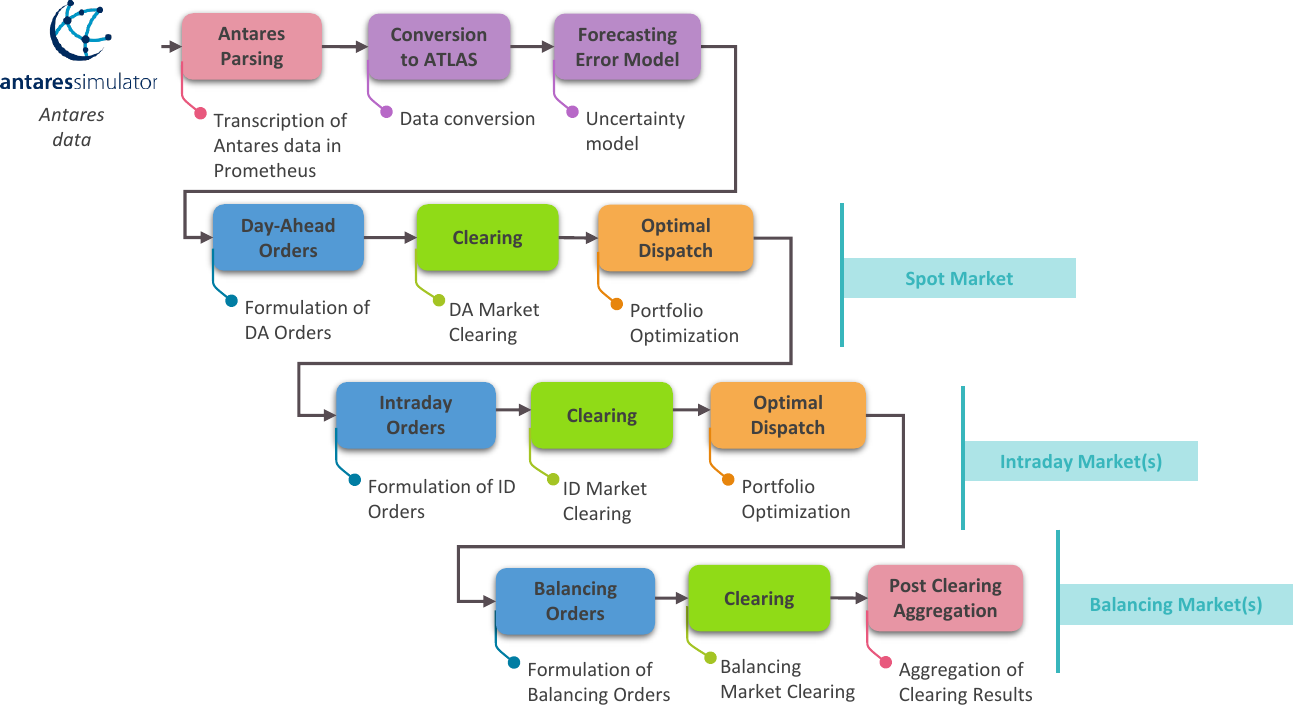}
    \caption{ATLAS Modules}
    \label{fig:atlasModules}
\end{figure}

The ATLAS model consists of several modules (coded in Python) that can be used in a variety of combinations in order to model the chain of electricity markets from day ahead to close-to-real time, and the basic structure for the most common market simulation pattern is shown in Figure \ref{fig:atlasModules}. All modules of the data preparation, Spot market and Intraday market stages are described in \cite{little_atlas_nodate}.\\ 

Figure \ref{fig:atlasModules} only offers an overview of the balancing stage. Several balancing configurations are modeled in ATLAS: 
\begin{enumerate}
    \item Relying on balancing markets only (depicted in \ref{fig:balancing_markets_only}). This figuration is comprised of  4 successive steps. The formulation of BSP orders is described in Chapter \ref{ch:Balancing_Market_BSP_Orders}, while the formulation of TSO orders is covered by Chapter \ref{ch:Balancing_Market_TSO_Orders}. The Clearing process is described in Chapter of \cite{little_atlas_nodate}, and finally Chapter. \ref{ch:Post Clearing Aggregation} of this document explains the Post Clearing Aggregation.
    \begin{figure}[H]
        \centering
        \includegraphics[width = 0.82\textwidth]{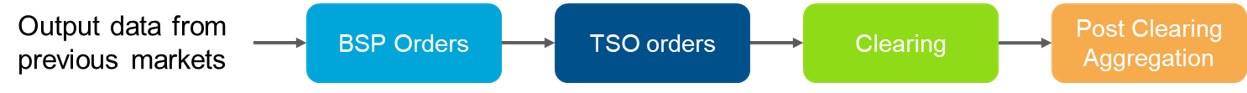}
        \caption{ATLAS - Detailed balancing market}
        \label{fig:balancing_markets_only}
    \end{figure}
    \item Using a local balancing process. In ATLAS, this is modeled by the Balancing Mechanism (described in Chapter \ref{ch:4. - BalancingMechanism}), which is implemented based on the historical French balancing process (Figure \ref{fig:ma_alone}). The Balancing Mechanism will serve as the reference for all local balancing processes in ATLAS. 
    \begin{figure}[H]
        \centering
        \includegraphics[width = 0.35\textwidth]{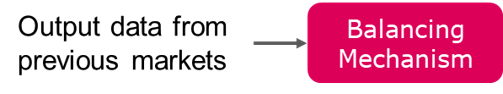}
        \caption{ATLAS - Balancing Mechanism used alone}
        \label{fig:ma_alone}
    \end{figure}
    
    \item Using balancing markets followed by a local balancing process, a configuration that integrates all modules previously described (Figure \ref{fig:balancing_markets_plus_ma}).
    \begin{figure}[H]
        \centering
        \includegraphics[width = \textwidth]{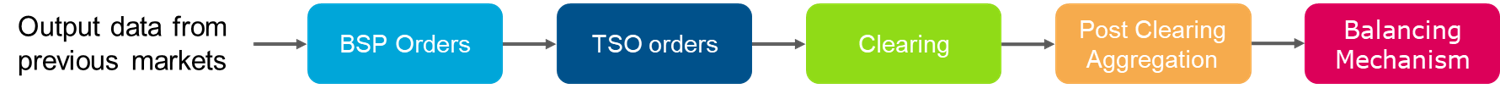}
        \caption{ATLAS - Detailed balancing market followed by a Balancing Mechanism}
        \label{fig:balancing_markets_plus_ma}
    \end{figure}
\end{enumerate}

\section{Nomenclature} 
This section describes notations used across all chapters of this document. Some elements correspond to a specific type, which is indicated in bold and italics:
\begin{itemize}
    \item \textbf{\textit{Parameter}} refers to a parameter, which is indicated by the user before the execution of the module.
    \item \textbf{\textit{Input data}} refers to an element that is extracted from the input dataset.
    \item \textbf{\textit{Variable}} refers to an optimization variable.
\end{itemize}

Remark: For sets, the notation $A_{b}$ refers to the subset of $A$ linked with variable $b$. For instance, $Z_{ca}$ indicates the subset of market areas belonging to the control area $ca$.

\renewcommand{\arraystretch}{1.3} 
\begin{longtable}[H]{!{\color{Grey}\vrule}>{\centering\arraybackslash}p{3cm} !{\color{Grey}\vrule} p{13cm}!{\color{Grey}\vrule} }
    \arrayrulecolor{Grey} \hline
    \rowcolor{Grey} \multicolumn{2}{|c|}{\large{\textbf{Sets}}} \label{tab:nomenclature_sets}\\ 
    \hline
    \rowcolor{Grey} 
    \textbf{Notation} & \textbf{Description}\\
    \hline
    $CA$ & Set of all control areas (or control blocks)\\
    \hline
    $Z$ & Set of all market areas.\\
    \hline
    $PF$ & Set of all portfolios.\\
    \hline 
    $U$ & Set of all units.\\
    \hline 
    $U^{unit\_type}$ & Set of all units of type $unit\_type \in [g, l, th, h, st, w, pv]$. ($g$ = generation, $l$ = flexible load, $th$ = thermal, $h$ = hydraulic, $st$ = storage, $w$ = wind, $pv$ = photovoltaic)\\
    \hline 
    $ID_{u}$ & Set of all combinatorial indexes for unit $u \in U$\\
    \hline
    $O$ & Set of all market orders. $O^{up}$ refers to the subset of upward orders and $O^{dn}$ to the subset of downward orders\\
    \hline
    $C^{excl}$ & Set of all coupling instances of type $Exclusion$\\
    \hline 
    $C^{pc}$ & Set of all coupling instances of type $Parent Children$\\
    \hline 
    $C^{idr}$ & Set of all coupling instances of type $Identical Ratio$\\
    \hline
\end{longtable}

\begin{longtable}[H]{!{\color{Grey}\vrule}>{\centering\arraybackslash}p{3cm} !{\color{Grey}\vrule} p{10.5cm}!{\color{Grey}\vrule} >{\centering\arraybackslash} p{2cm}!{\color{Grey}\vrule}}
    \arrayrulecolor{Grey} \hline
    \rowcolor{Grey} \multicolumn{3}{|c|}{\large{\textbf{Temporal variables}}} \label{tab:nomenclature_temporal}\\
    \hline
    \rowcolor{Grey} 
    \textbf{Notation} & \textbf{Description} & \textbf{Units}\\
    \hline
    $t_{m}^{ex}$ & Execution date of the market $m \in \{RR, mFRR\}$ & -\\
    \hline
    $t_{m}^{start}$ & Start date of the effective period of the market $m$ & -\\
    \hline
    $t_{m}^{end}$ & End date of the effective period of the market $m$ & -\\
    \hline
    $\Delta t_{m}$ & Time step of the market $m$ & min\\
    \hline
    $T_{m}$ & Effective time frame of the market $m$, i.e. $T_{m} = [t^{start}_m, t^{start}_m + \Delta t_{m},\, \dots \,, t^{end}_m - \Delta t_{m}]$ & -\\
    \hline
    $t^{start}_{id}$ & Start date of the combinatorial index $id$ & -\\
    \hline
    $t^{end}_{id}$ & End date of the combinatorial index $id$ & -\\
    \hline
    $T_{id}$ & Time frame of the combinatorial index $id$, i.e. $T_{id} = [t^{start}_{id}, t^{start}_{id} + \Delta t_{m},\, \dots \,, t^{end}_{id} - \Delta t_{m}]$ & -\\
    \hline
    $\Delta T_{id}$ & Length of the combinatorial index $id$, i.e. $\Delta T_{id} = t^{end}_{id} - t^{start}_{id}$ & min\\
    \hline
\end{longtable}

\begin{longtable}[H]{!{\color{Grey}\vrule}>{\centering\arraybackslash}p{3cm} !{\color{Grey}\vrule} p{10.5cm}!{\color{Grey}\vrule} >{\centering\arraybackslash} p{2cm}!{\color{Grey}\vrule}}
    \arrayrulecolor{Grey} \hline
    \rowcolor{Grey} \multicolumn{3}{|c|}{\large{\textbf{TSO-specific characteristics}}} \label{tab:nomenclature_zonal_charac}\\
    \hline
    \rowcolor{Grey} 
    \textbf{Notation} & \textbf{Description} & \textbf{Units}\\
    \hline
    $bn_{m,ca,t,t_{m}^{ex}}$ \newline (short: $\mkern3mu bn_{t}$) & Balancing needs for market $m$ in control area $ca \in CA$ for time $t \in T_{m}$, seen from $t_{m}^{ex}$. For readability, it is shortened as $bn_{t}$ & MW\\
    \hline
    $\sigma_{ca,t,t_{m}^{ex}}^{bn}$ \newline (short: $\mkern3mu \sigma^{bn}_{t}$) & Binary parameter indicating the direction of TSO (area $ca$) balancing needs at time $t$ seen from $t_{m}^{ex}$. $\sigma_{ca,t,t_{m}^{ex}}^{bn} = 1 \mkern6mu if \mkern6mu bn_{m,ca,t,t_{m}^{ex}} > 0$, and $0$ otherwise. For readability, it is shortened as $\sigma^{bn}_{t}$& - \\
    \hline
    $\delta_{ca}^{for}$ & Binary parameter indicating if TSO associated with control area $ca \in CA$ takes into account forecasts errors in its imbalance needs computation. \textbf{\textit{Parameter}} & - \\
    \hline
    $\delta_{ca}^{elas}$ & Binary parameter indicating if the demand of TSO associated with control area $ca \in CA$ is formulated as elastic ($\delta_{ca}^{elas} = 1$) or not ($\delta_{ca}^{elas} = 0$). \textbf{\textit{Parameter}} & - \\
    \hline
    $alt_{ca}$ & Alternative chosen to compute prices and volumes of the demand of TSO associated with control area $ca \in CA$. Possible values are $\{"mFRRalt", "FrBMalt"\}$. \textbf{\textit{Parameter}}. & - \\
    \hline
    $V^{s}$ & Maximum quantity of each slice of TSO demand formulated on markets. Only relevant if $\delta_{ca}^{elas} = 1$. \textbf{\textit{Parameter}} & MW \\
    \hline
    $\delta_{ca}^{risk}$ & Binary parameter indicating if the demand of TSO associated with control area $ca \in CA$ is taking into account volume-based risk aversion ($\delta_{ca}^{risk} = 1$) or not ($\delta_{ca}^{risk} = 0$). Only relevant if $\delta_{ca}^{elas} = 1$. \textbf{\textit{Parameter}} & - \\
    \hline
    $\epsilon_{ca}^{alt_{ca}}$ & List of quantiles of the distribution of the TSO forecast error function of area $ca$ associated with alternative $alt_{ca}$. Only relevant if $\delta_{ca}^{risk} = 1$. \textbf{\textit{Parameter}} & - \\
    \hline
\end{longtable}

\begin{longtable}[H]{!{\color{Grey}\vrule}>{\centering\arraybackslash}p{3cm} !{\color{Grey}\vrule} p{10.5cm}!{\color{Grey}\vrule} >{\centering\arraybackslash} p{2cm}!{\color{Grey}\vrule}}
    \arrayrulecolor{Grey} \hline
    \rowcolor{Grey} \multicolumn{3}{|c|}{\large{\textbf{Zonal-specific characteristics}}} \label{tab:A.3.2_nomenclature_zonal_charac}\\
    \hline
    \rowcolor{Grey} 
    \textbf{Notation} & \textbf{Description} & \textbf{Units}\\
    \hline
    $q_{m,ca,t}^{\sigma^{bn}_{t},max}$ & Overall quantity for orders in direction opposite to $\sigma^{bn}_{t}$ formulated by BSPs included in control area $ca$ on market $m$, at time $t$ & MW\\
    \hline
    $(\Delta q)^{bal}_{z,t}$ & Commercial (power) balance of area $z \in Z$ at time $t$, equal to the sum of all power exports minus the sum of all power imports & MW \\
    \hline
    $\rho_{m, ca}^{FrBMalt}$ & Percentage of BSP available capacity in area $ca$ that is not submitted on the market $m$, and is consequently directly sent to the FrBM market. Only relevant if $\delta_{ca}^{elas} = 1$, with $alt_{ca} = "FrBMalt"$. \textbf{\textit{Parameter}} & - \\
    \hline
\end{longtable}

\begin{longtable}[H]{!{\color{Grey}\vrule}>{\centering\arraybackslash}p{3cm} !{\color{Grey}\vrule} p{10.5cm}!{\color{Grey}\vrule} >{\centering\arraybackslash} p{2cm}!{\color{Grey}\vrule}}
    \arrayrulecolor{Grey} \hline
    \rowcolor{Grey} \multicolumn{3}{|c|}{\large{\textbf{Global unit characteristics}}} \label{tab:nomenclature_unit_charac}\\
    \hline
    \rowcolor{Grey} 
    \textbf{Notation} & \textbf{Description} & \textbf{Units}\\
    \hline
    $P_{u,t,t_{m}^{ex}}^{plan}$ & Power output of unit $u \in U$ at time $t \in T_{m}$, seen from time $t_{m}^{ex}$. \textbf{\textit{Input data}} & MW\\
    \hline
    $P_{u,t}^{max}$ & Maximum power output of unit $u \in U$ at time $t \in T_{m}$. \textbf{\textit{Input data}} & MW\\
    \hline
    $P_{u,t}^{min}$ & Minimum power output of unit $u \in U$ at time $t \in T_{m}$. \textbf{\textit{Input data}} & MW\\
    \hline
    $\Delta P_{u,t}^{max}$ & Maximum ramping limit of unit $u \in U$ at time $t \in T_{m}$. \textbf{\textit{Input data}} & MW/min\\
    \hline
    $c_{u}^{VAR}$ & Variable cost of unit $u \in U$. \textbf{\textit{Input data}} & €/MWh\\
    \hline
    $c_{u}^{SU}$ & Startup cost of unit $u \in U$. \textbf{\textit{Input data}} & €\\
    \hline
    $\delta_{u,t}^{turned\_on}$ & Binary variable indicating if unit $u \in U^{th}$ is starting at $t \in T_{m}$ & -\\
    \hline
    $\delta_{u,t}^{turned\_off}$ & Binary variable indicating if unit $u \in U^{th}$ is shutting down at $t \in T_{m}$. \textbf{\textit{Variable}} & -\\
    \hline
    $d_{u}^{notice}$ & Notice delay of unit $u \in U$. \textbf{\textit{Input data}} & min\\
    \hline
    $Res_{ut,t_{m}^{ex}}^{ResType, ResDir}$ & Procured reserves of type $ResType \in [FCR, aFRR, mFRR, RR]$ in direction $ResDir \in [up, down]$  on unit $u$ at time $t$, seen from $t_{m}^{ex}$. \textbf{\textit{Input data}} & -\\
    \hline
    $Q_{u,id}^{ResDir,max}$ & Maximum quantity in direction $ResDir \in [up, down]$ that can be offered on unit $u$, for the combinatorial index $id$ & MW \\
    \hline
    $Q_{u,id}^{ResDir,min}$ & Minimum quantity in direction $ResDir \in [up, down]$ that can be offered on unit $u$, for the combinatorial index $id$ & MW \\
    \hline
\end{longtable}

\begin{longtable}[H]{!{\color{Grey}\vrule}>{\centering\arraybackslash}p{3cm} !{\color{Grey}\vrule} p{10.5cm}!{\color{Grey}\vrule} >{\centering\arraybackslash} p{2cm}!{\color{Grey}\vrule}}
    \arrayrulecolor{Grey} \hline
    \rowcolor{Grey} \multicolumn{3}{|c|}{\large{\textbf{Thermal-specific unit characteristics}}} \label{tab:nomenclature_thermal}\\
    \hline
    \rowcolor{Grey} 
    \textbf{Notation} & \textbf{Description} & \textbf{Units}\\
    \hline
    $d_{u}^{SU}$ & Startup duration of unit $u \in U^{th}$. \textbf{\textit{Input data}} & min\\
    \hline
    $d_{u}^{SD}$ & Shutdown duration of unit $u \in U^{th}$. \textbf{\textit{Input data}} & min\\
    \hline
    $d_{u}^{minOn}$ & Minimum time on duration of unit $u \in U^{th}$. \textbf{\textit{Input data}} & min\\
    \hline
    $d_{u}^{minOff}$ & Minimum time off duration of unit $u \in U^{th}$. \textbf{\textit{Input data}} & min\\
    \hline
    $d_{u}^{minStable}$ & Minimum stable power duration of unit $u \in U^{th}$. \textbf{\textit{Input data}} & min\\
    \hline
    $\delta_{u,T_{id},t_{m}^{ex}}^{SU}$ & Binary variable indicating if orders formulated on the unit $u$, for the combinatorial index $id$ seen from $t_{m}^{ex}$ should be treated as startup orders ($\delta_{u,T_{id},t_{m}^{ex}}^{SU} = 1$) or not ($ = 0$)& -\\
    \hline
    $\delta_{u,T_{id},t_{m}^{ex}}^{SD}$ & Binary variable indicating if the unit $u$ is able to be shutdown at time $t$ seen from $t_{m}^{ex}$. If written $\delta_{u,T_{id},t_{m}^{ex}}^{SD}$, it signifies that the variable concerns the entire set of times in $T_{id}$. & -\\
    \hline
    $\sigma_{u,T_{id}}^{SU}$ & Binary variable indicating if orders formulated on the combinatorial index $id$ induced ($\sigma_{u,T_{id}}^{SU} = 1$) or canceled ($\sigma_{u,T_{id}}^{SU} = -1$) a startup, or had no impact on startups ($\sigma_{u,T_{id}}^{SU} = 0$) of unit $u$ & -\\
    \hline
\end{longtable}

\begin{longtable}[H]{!{\color{Grey}\vrule}>{\centering\arraybackslash}p{3cm} !{\color{Grey}\vrule} p{10.5cm}!{\color{Grey}\vrule} >{\centering\arraybackslash} p{2cm}!{\color{Grey}\vrule}}
    \arrayrulecolor{Grey} \hline
    \rowcolor{Grey} \multicolumn{3}{|c|}{\large{\textbf{Hydraulic- and Storage-specific unit characteristics}}} \label{tab:nomenclature_hydro_storage}\\
    \hline
    \rowcolor{Grey} 
    \textbf{Notation} & \textbf{Description} & \textbf{Units}\\
    \hline
    $E_{u,t,t_{m}^{ex}}^{stored}$ & Stored energy in the reservoir of unit $u \in U^{h} \, \cup \, U^{st}$ at time $t \in T_{m}$, seen from time $t_{m}^{ex}$ & MWh\\
    \hline
    $E_{u,t}^{max}$ & Maximum storage level in the reservoir of unit $u \in U^{h} \, \cup \, U^{st}$ at time $t \in T_{m}$. \textbf{\textit{Input data}} & MWh\\
    \hline
    $E_{u,t}^{min}$ & Minimum storage level in the reservoir of unit $u \in U^{h} \, \cup \, U^{st}$ at time $t \in T_{m}$. \textbf{\textit{Input data}} & MWh\\
    \hline
    $d_{u}^{tran}$ & Transition duration between pumping and turbining of unit $u$ of type Pumped Hydraulic Storage (PHS), that are modeled as storage units in ATLAS (i.e. $u \in U^{st}$). \textbf{\textit{Input data}} & min\\
    \hline
    $WV_{u,t,E_{u,t,t_{m}^{ex}}^{stored}}$ & Marginal storage value of unit $u  \in U^{h}$ at time $t$, for stored energy level $E_{u,t,t_{m}^{ex}}^{stored}$. \textbf{\textit{Input data}} & €/MWh\\
    \hline
    $h_{DA}^{ex}$ & Execution hour of the day-ahead market, used to apply energy constraints to Hydraulic and Storage units in the BSP Order Formulation module. \textbf{\textit{Parameter}} & -\\
    \hline
\end{longtable}

\begin{longtable}[H]{!{\color{Grey}\vrule}>{\centering\arraybackslash}p{3cm} !{\color{Grey}\vrule} p{10.5cm}!{\color{Grey}\vrule} >{\centering\arraybackslash} p{2cm}!{\color{Grey}\vrule}}
    \arrayrulecolor{Grey} \hline
    \rowcolor{Grey} \multicolumn{3}{|c|}{\large{\textbf{Load-, Photovoltaic- and Wind-specific unit characteristics}}} \label{tab:nomenclature_pv_wind}\\
    \hline
    \rowcolor{Grey} 
    \textbf{Notation} & \textbf{Description} & \textbf{Units}\\
    \hline
    $P_{u,t,t_{m}^{ex}}^{for}$ & Forecast of the maximum power output of unit $u \in \{U^{w}, U^{pv}, U^{l}\}$ at time $t \in T_{m}$. \textbf{\textit{Input data}} & MW\\
    \hline
    $Curt_{u}$ & Percentage of curtailed power allowed on unit $u \in \{U^{w}, U^{pv}\}$. \textbf{\textit{Input data}} & - \\
    \hline
\end{longtable}

\begin{longtable}[!ht]{!{\color{Grey}\vrule}>{\centering\arraybackslash}p{3cm} !{\color{Grey}\vrule} p{13cm}!{\color{Grey}\vrule} }
    \arrayrulecolor{Grey} \hline
    \rowcolor{Grey} \multicolumn{2}{|c|}{\large{\textbf{Market order characteristics}}} \label{tab:nomenclature_orders}\\ 
    \hline
    \rowcolor{Grey} 
    \textbf{Notation} & \textbf{Meaning}\\ \hline
    $p_o$ & Price of order $o$ \\
    \hline
    $q_o^{min}$ & Minimum quantity of power offered for order $o$ \\
    \hline
    $q_o^{max}$ & Maximum quantity of power offered for order $o$ \\
    \hline
    $t_o^{start}$ & Start date of order \\
    \hline
    $t_o^{end}$ & End date of order \\
    \hline
    $t_o^{ex}$ & Execution date of order \\
    \hline
    $d_{o}$ & Duration of order \\
    \hline
    $\sigma_{o}$ & Sale/Purchase indicator, $\sigma = 1$ for purchase, -1 for sale \\ 
    \hline
    $\delta_o^{TSO}$ & Binary variable indicating if order $o$ is from a TSO ($\delta_o^{TSO} = 1$) or not. \\
    \hline
    $q_o^{acc}$ & Amount of power accepted by the Clearing stage for order $o$ \\
    \hline
    $\delta_{o}^{acc}$ & Binary variable indicating if order $o$ is activated by the Clearing stage ($\delta_{o}^{acc} = 1$ if $q_o^{acc} > 0$)\\
    \hline
\end{longtable}

\begin{longtable}[H]{!{\color{Grey}\vrule}>{\centering\arraybackslash}p{3cm} !{\color{Grey}\vrule} p{10.5cm}!{\color{Grey}\vrule} >{\centering\arraybackslash} p{2cm}!{\color{Grey}\vrule}}
    \arrayrulecolor{Grey} \hline
    \rowcolor{Grey} \multicolumn{3}{|c|}{\large{\textbf{Other notations}}} \label{tab:nomenclature_others}\\
    \hline
    \rowcolor{Grey} 
    \textbf{Notation} & \textbf{Description} & \textbf{Units}\\
    \hline
    $VoLL$ & Value of loss load. \textbf{\textit{Parameter}} & €/MWh\\
    \hline
    $p^{spill}$ & Spillage penalty. \textbf{\textit{Parameter}} & \euro/MWh\\
    \hline
    $p^{redispatch}$ & Redispatch penalty. \textbf{\textit{Parameter}} & \euro/MWh\\
    \hline
    $E^{VoLL}_{ca,t}$ & Loss load energy in control area $ca$ at time $t$. \textbf{\textit{Variable}} & MWh\\
    \hline
    $E_{ca,t}^{spill}$ & Spillage power in control area $ca$ at time $t$.\textbf{\textit{Variable}} & MW\\
    \hline
\end{longtable}

\renewcommand{\arraystretch}{1.1} 

\chapter{Balancing Market - BSP Orders} \label{ch:Balancing_Market_BSP_Orders}

\section{Overview of the module}
\label{sec:bsp_orders_overview}
The objective of this module is to formulate BSP balancing energy market orders, for RR or mFRR markets. Submitted market orders are required to display all information indicated in Table \ref{tab:tso_orders_nomenclature_orders}. In ATLAS, the formulation of balancing market orders is done with a unit-based hypothesis. This is justified by the extremely short full activation time of balancing markets (at most 30 minutes for RR orders), combined with their high frequency. Under these constraints, running a portfolio optimization process to determine the unit dispatch after each market would be unfeasible for BSPs.

In addition, this module does not include any optimization problem, even for the modeling of operating constraints. This is a deliberate choice: we observed that using an optimization problem for each unit of our input dataset, and for each balancing market simulated, led to significant computation time, mainly because of the high frequency of said markets. Consequently, the computation of the available capacity for each unit is performed through a non-optimization algorithm.

This chapter first introduces the concept of combinatorial orders that will be used to integrate detailed operating constraints in the formulation algorithm of thermic units (Section \ref{sec:bsp_orders_combi_indexes}). The complete order formulation is then presented in Section \ref{sec:bsp_orders_formulation_main}, which includes the calculation of upward and downward available power, and the definition of both the quantity and price of market orders. Finally, Section \ref{sec:bsp_orders_coupling_links} presents the coupling links used in the module and the different situations they cover. 

\section{Nomenclature and inputs}
\label{sec:bsp_orders_nomenclature_inputs}
This section describes notations used across this chapter. Some elements correspond to a specific type, which is indicated in bold and italics:
\begin{itemize}
    \item \textbf{\textit{Parameter}} refers to a parameter, which is indicated by the user before the execution of the module.
    \item \textbf{\textit{Input data}} refers to an element that is extracted from the input dataset.
    \item \textbf{\textit{Variable}} refers to an optimization variable.
\end{itemize}

Remark: For sets, the notation $A_{b}$ refers to the subset of $A$ linked with variable $b$. For instance, $Z_{ca}$ indicates the subset of market areas belonging to the control area $ca$.

\renewcommand{\arraystretch}{1.3} 
\begin{longtable}[H]{!{\color{Grey}\vrule}>{\centering\arraybackslash}p{3cm} !{\color{Grey}\vrule} p{13cm}!{\color{Grey}\vrule} }
    \arrayrulecolor{Grey} \hline
    \rowcolor{Grey} \multicolumn{2}{|c|}{\large{\textbf{Sets}}} \label{tab:bsp_orders_nomenclature_sets}\\ 
    \hline
    \rowcolor{Grey} 
    \textbf{Notation} & \textbf{Description}\\
    \hline
    $CA$ & Set of all control areas (or control blocks)\\
    \hline
    $Z$ & Set of all market areas.\\
    \hline
    $PF$ & Set of all portfolios.\\
    \hline 
    $U$ & Set of all units.\\
    \hline 
    $U^{unit\_type}$ & Set of all units of type $unit\_type \in [g, l, th, h, st, w, pv]$. ($g$ = generation, $l$ = flexible load, $th$ = thermal, $h$ = hydraulic, $st$ = storage, $w$ = wind, $pv$ = photovoltaic)\\
    \hline 
    $ID_{u}$ & Set of all combinatorial indexes for unit $u \in U$\\
    \hline
    $O$ & Set of all market orders. $O^{up}$ refers to the subset of upward orders and $O^{dn}$ to the subset of downward orders\\
    \hline
    $C^{excl}$ & Set of all coupling instances of type $Exclusion$\\
    \hline 
    $C^{pc}$ & Set of all coupling instances of type $Parent Children$\\
    \hline 
    $C^{idr}$ & Set of all coupling instances of type $Identical Ratio$\\
    \hline
\end{longtable}

\begin{longtable}[H]{!{\color{Grey}\vrule}>{\centering\arraybackslash}p{3cm} !{\color{Grey}\vrule} p{10.5cm}!{\color{Grey}\vrule} >{\centering\arraybackslash} p{2cm}!{\color{Grey}\vrule}}
    \arrayrulecolor{Grey} \hline
    \rowcolor{Grey} \multicolumn{3}{|c|}{\large{\textbf{Temporal variables}}} \label{tab:bsp_orders_nomenclature_temporal}\\
    \hline
    \rowcolor{Grey} 
    \textbf{Notation} & \textbf{Description} & \textbf{Units}\\
    \hline
    $t_{m}^{ex}$ & Execution date of the market $m \in \{RR, mFRR\}$ & -\\
    \hline
    $t_{m}^{start}$ & Start date of the effective period of the market $m$ & -\\
    \hline
    $t_{m}^{end}$ & End date of the effective period of the market $m$ & -\\
    \hline
    $\Delta t_{m}$ & Time step of the market $m$ & min\\
    \hline
    $T_{m}$ & Effective time frame of the market $m$, i.e. $T_{m} = [t^{start}_m, t^{start}_m + \Delta t_{m},\, \dots \,, t^{end}_m - \Delta t_{m}]$ & -\\
    \hline
    $t^{start}_{id}$ & Start date of the combinatorial index $id$ & -\\
    \hline
    $t^{end}_{id}$ & End date of the combinatorial index $id$ & -\\
    \hline
    $T_{id}$ & Time frame of the combinatorial index $id$, i.e. $T_{id} = [t^{start}_{id}, t^{start}_{id} + \Delta t_{m},\, \dots \,, t^{end}_{id} - \Delta t_{m}]$ & -\\
    \hline
    $\Delta T_{id}$ & Length of the combinatorial index $id$, i.e. $\Delta T_{id} = t^{end}_{id} - t^{start}_{id}$ & min\\
    \hline
\end{longtable}

\begin{longtable}[H]{!{\color{Grey}\vrule}>{\centering\arraybackslash}p{3cm} !{\color{Grey}\vrule} p{10.5cm}!{\color{Grey}\vrule} >{\centering\arraybackslash} p{2cm}!{\color{Grey}\vrule}}
    \arrayrulecolor{Grey} \hline
    \rowcolor{Grey} \multicolumn{3}{|c|}{\large{\textbf{Zonal- and TSO-specific characteristics}}} \label{tab:bsp_orders_nomenclature_zonal_tso}\\
    \hline
    \rowcolor{Grey} 
    \textbf{Notation} & \textbf{Description} & \textbf{Units}\\
    \hline
    $bn_{m,ca,t,t_{m}^{ex}}$ \newline (short: $\mkern3mu bn_{t}$) & Balancing needs for market $m$ in control area $ca \in CA$ for time $t \in T_{m}$, seen from $t_{m}^{ex}$. For readability, it is shortened as $bn_{t}$ & MW\\
    \hline
    $\sigma_{ca,t,t_{m}^{ex}}^{bn}$ \newline (short: $\mkern3mu \sigma^{bn}_{t}$) & Binary parameter indicating the direction of TSO (area $ca$) balancing needs at time $t$ seen from $t_{m}^{ex}$. $\sigma_{ca,t,t_{m}^{ex}}^{bn} = 1 \mkern6mu if \mkern6mu bn_{m,ca,t,t_{m}^{ex}} > 0$, and $0$ otherwise. For readability, it is shortened as $\sigma^{bn}_{t}$& - \\
    \hline
    $\delta_{ca}^{for}$ & Binary parameter indicating if TSO associated with control area $ca \in CA$ takes into account forecasts errors in its imbalance needs computation. \textbf{\textit{Parameter}} & - \\
    \hline
    $\delta_{ca}^{elas}$ & Binary parameter indicating if the demand of TSO associated with control area $ca \in CA$ is formulated as elastic ($\delta_{ca}^{elas} = 1$) or not ($\delta_{ca}^{elas} = 0$). \textbf{\textit{Parameter}} & - \\
    \hline
    $alt_{ca}$ & Alternative chosen to compute prices and volumes of the demand of TSO associated with control area $ca \in CA$. Possible values are $\{"mFRRalt", "FrBMalt"\}$. \textbf{\textit{Parameter}}. & - \\
    \hline
    $V^{s}$ & Maximum quantity of each slice of TSO demand formulated on markets. Only relevant if $\delta_{ca}^{elas} = 1$. \textbf{\textit{Parameter}} & MW \\
    \hline
    $\delta_{ca}^{risk}$ & Binary parameter indicating if the demand of TSO associated with control area $ca \in CA$ is taking into account volume-based risk aversion ($\delta_{ca}^{risk} = 1$) or not ($\delta_{ca}^{risk} = 0$). Only relevant if $\delta_{ca}^{elas} = 1$. \textbf{\textit{Parameter}} & - \\
    \hline
    $\epsilon_{ca}^{alt_{ca}}$ & List of quantiles of the distribution of the TSO forecast error function of area $ca$ associated with alternative $alt_{ca}$. Only relevant if $\delta_{ca}^{risk} = 1$. \textbf{\textit{Parameter}} & - \\
    \hline
\end{longtable}

\begin{longtable}[H]{!{\color{Grey}\vrule}>{\centering\arraybackslash}p{3cm} !{\color{Grey}\vrule} p{10.5cm}!{\color{Grey}\vrule} >{\centering\arraybackslash} p{2cm}!{\color{Grey}\vrule}}
    \arrayrulecolor{Grey} \hline
    \rowcolor{Grey} \multicolumn{3}{|c|}{\large{\textbf{Zonal-specific characteristics}}} \label{tab:bsp_orders_nomenclature_zonal_charac}\\
    \hline
    \rowcolor{Grey} 
    \textbf{Notation} & \textbf{Description} & \textbf{Units}\\
    \hline
    $q_{m,ca,t}^{\sigma^{bn}_{t},max}$ & Overall quantity for orders in direction opposite to $\sigma^{bn}_{t}$ formulated by BSPs included in control area $ca$ on market $m$, at time $t$ & MW\\
    \hline
    $(\Delta q)^{bal}_{z,t}$ & Commercial (power) balance of area $z \in Z$ at time $t$, equal to the sum of all power exports minus the sum of all power imports & MW \\
    \hline
    $\rho_{m, ca}^{FrBMalt}$ & Percentage of BSP available capacity in area $ca$ that is not submitted on the market $m$, and is consequently directly sent to the FrBM market. Only relevant if $\delta_{ca}^{elas} = 1$, with $alt_{ca} = "FrBMalt"$. \textbf{\textit{Parameter}} & - \\
    \hline
\end{longtable}

\begin{longtable}[H]{!{\color{Grey}\vrule}>{\centering\arraybackslash}p{3cm} !{\color{Grey}\vrule} p{10.5cm}!{\color{Grey}\vrule} >{\centering\arraybackslash} p{2cm}!{\color{Grey}\vrule}}
    \arrayrulecolor{Grey} \hline
    \rowcolor{Grey} \multicolumn{3}{|c|}{\large{\textbf{Global unit characteristics}}} \label{tab:bsp_orders_nomenclature_unit_charac}\\
    \hline
    \rowcolor{Grey} 
    \textbf{Notation} & \textbf{Description} & \textbf{Units}\\
    \hline
    $P_{u,t,t_{m}^{ex}}^{plan}$ & Power output of unit $u \in U$ at time $t \in T_{m}$, seen from time $t_{m}^{ex}$. \textbf{\textit{Input data}} & MW\\
    \hline
    $P_{u,t}^{max}$ & Maximum power output of unit $u \in U$ at time $t \in T_{m}$. \textbf{\textit{Input data}} & MW\\
    \hline
    $P_{u,t}^{min}$ & Minimum power output of unit $u \in U$ at time $t \in T_{m}$. \textbf{\textit{Input data}} & MW\\
    \hline
    $\Delta P_{u,t}^{max}$ & Maximum ramping limit of unit $u \in U$ at time $t \in T_{m}$. \textbf{\textit{Input data}} & MW/min\\
    \hline
    $c_{u}^{var}$ & Variable cost of unit $u \in U$. \textbf{\textit{Input data}} & €/MWh\\
    \hline
    $c_{u}^{SU}$ & Startup cost of unit $u \in U$. \textbf{\textit{Input data}} & €\\
    \hline
    $\delta_{u,t}^{turned\_on}$ & Binary variable indicating if unit $u \in U^{th}$ is starting at $t \in T_{m}$ & -\\
    \hline
    $\delta_{u,t}^{turned\_off}$ & Binary variable indicating if unit $u \in U^{th}$ is shutting down at $t \in T_{m}$. \textbf{\textit{Variable}} & -\\
    \hline
    $d_{u}^{notice}$ & Notice delay of unit $u \in U$. \textbf{\textit{Input data}} & min\\
    \hline
    $R_{ut,t_{m}^{ex}}^{m^{R}, \sigma^{R}}$ & Procured reserves of type $m^{R} \in [FCR, aFRR, mFRR, RR]$ in direction $\sigma^{R} \in [up, dn]$  on unit $u$ at time $t$, seen from $t_{m}^{ex}$. \textbf{\textit{Input data}} & -\\
    \hline
    $Q_{u,id}^{\sigma^{R},max}$ & Maximum quantity in direction $\sigma^{R} \in [up, dn]$ that can be offered on unit $u$, for the combinatorial index $id$\\
    \hline
    $Q_{u,id}^{\sigma^{R},min}$ & Minimum quantity in direction $\sigma^{R} \in [up, dn]$ that can be offered on unit $u$, for the combinatorial index $id$\\
    \hline
\end{longtable}

\begin{longtable}[H]{!{\color{Grey}\vrule}>{\centering\arraybackslash}p{3cm} !{\color{Grey}\vrule} p{10.5cm}!{\color{Grey}\vrule} >{\centering\arraybackslash} p{2cm}!{\color{Grey}\vrule}}
    \arrayrulecolor{Grey} \hline
    \rowcolor{Grey} \multicolumn{3}{|c|}{\large{\textbf{Thermal-specific unit characteristics}}} \label{tab:bsp_orders_nomenclature_thermal}\\
    \hline
    \rowcolor{Grey} 
    \textbf{Notation} & \textbf{Description} & \textbf{Units}\\
    \hline
    $d_{u}^{SU}$ & Startup duration of unit $u \in U^{th}$. \textbf{\textit{Input data}} & min\\
    \hline
    $d_{u}^{SD}$ & Shutdown duration of unit $u \in U^{th}$. \textbf{\textit{Input data}} & min\\
    \hline
    $d_{u}^{minOn}$ & Minimum time on duration of unit $u \in U^{th}$. \textbf{\textit{Input data}} & min\\
    \hline
    $d_{u}^{minOff}$ & Minimum time off duration of unit $u \in U^{th}$. \textbf{\textit{Input data}} & min\\
    \hline
    $d_{u}^{minStable}$ & Minimum stable power duration of unit $u \in U^{th}$. \textbf{\textit{Input data}} & min\\
    \hline
    $\delta_{u,T_{id},t_{m}^{ex}}^{SU}$ & Binary variable indicating if orders formulated on the unit $u$, for the combinatorial index $id$ seen from $t_{m}^{ex}$ should be treated as startup orders ($\delta_{u,T_{id},t_{m}^{ex}}^{SU} = 1$) or not ($ = 0$)& -\\
    \hline
    $\delta_{u,T_{id},t_{m}^{ex}}^{SD}$ & Binary variable indicating if the unit $u$ is able to be shutdown at time $t$ seen from $t_{m}^{ex}$. If written $\delta_{u,T_{id},t_{m}^{ex}}^{SD}$, it signifies that the variable concerns the entire set of times in $T_{id}$. & -\\
    \hline
    $\sigma_{u,T_{id}}^{SU}$ & Binary variable indicating if orders formulated on the combinatorial index $id$ induced ($\sigma_{u,T_{id}}^{SU} = 1$) or canceled ($\sigma_{u,T_{id}}^{SU} = -1$) a startup, or had no impact on startups ($\sigma_{u,T_{id}}^{SU} = 0$) of unit $u$ & -\\
    \hline
\end{longtable}

\begin{longtable}[H]{!{\color{Grey}\vrule}>{\centering\arraybackslash}p{3cm} !{\color{Grey}\vrule} p{10.5cm}!{\color{Grey}\vrule} >{\centering\arraybackslash} p{2cm}!{\color{Grey}\vrule}}
    \arrayrulecolor{Grey} \hline
    \rowcolor{Grey} \multicolumn{3}{|c|}{\large{\textbf{Hydraulic- and Storage-specific unit characteristics}}} \label{tab:bsp_orders_nomenclature_hydro_storage}\\
    \hline
    \rowcolor{Grey} 
    \textbf{Notation} & \textbf{Description} & \textbf{Units}\\
    \hline
    $E_{u,t,t_{m}^{ex}}^{stored}$ & Stored energy in the reservoir of unit $u \in U^{h} \, \cup \, U^{st}$ at time $t \in T_{m}$, seen from time $t_{m}^{ex}$ & MWh\\
    \hline
    $E_{u,t}^{max}$ & Maximum storage level in the reservoir of unit $u \in U^{h} \, \cup \, U^{st}$ at time $t \in T_{m}$. \textbf{\textit{Input data}} & MWh\\
    \hline
    $E_{u,t}^{min}$ & Minimum storage level in the reservoir of unit $u \in U^{h} \, \cup \, U^{st}$ at time $t \in T_{m}$. \textbf{\textit{Input data}} & MWh\\
    \hline
    $d_{u}^{tran}$ & Transition duration between pumping and turbining of unit $u$ of type Pumped Hydraulic Storage (PHS), that are modeled as storage units in ATLAS (i.e. $u \in U^{st}$). \textbf{\textit{Input data}} & min\\
    \hline
    $WV_{u,t,E_{u,t,t_{m}^{ex}}^{stored}}$ & Marginal storage value of unit $u  \in U^{h}$ at time $t$, for stored energy level $E_{u,t,t_{m}^{ex}}^{stored}$. \textbf{\textit{Input data}} & €/MWh\\
    \hline
    $h_{DA}^{ex}$ & Execution hour of the day-ahead market, used to apply energy constraints to Hydraulic and Storage units. \textbf{\textit{Parameter}} & -\\
    \hline
\end{longtable}

\begin{longtable}[H]{!{\color{Grey}\vrule}>{\centering\arraybackslash}p{3cm} !{\color{Grey}\vrule} p{10.5cm}!{\color{Grey}\vrule} >{\centering\arraybackslash} p{2cm}!{\color{Grey}\vrule}}
    \arrayrulecolor{Grey} \hline
    \rowcolor{Grey} \multicolumn{3}{|c|}{\large{\textbf{Load-, Photovoltaic- and Wind-specific unit characteristics}}} \label{tab:bsp_orders_nomenclature_pv_wind}\\
    \hline
    \rowcolor{Grey} 
    \textbf{Notation} & \textbf{Description} & \textbf{Units}\\
    \hline
    $P_{u,t,t_{m}^{ex}}^{for}$ & Forecast of the maximum power output of unit $u \in \{U^{w}, U^{pv}, U^{l}\}$ at time $t \in T_{m}$. \textbf{\textit{Input data}} & MW\\
    \hline
    $Curt_{u}$ & Percentage of curtailed power allowed on unit $u \in \{U^{w}, U^{pv}\}$. \textbf{\textit{Input data}} & - \\
    \hline
\end{longtable}

\begin{longtable}[!ht]{!{\color{Grey}\vrule}>{\centering\arraybackslash}p{3cm} !{\color{Grey}\vrule} p{13cm}!{\color{Grey}\vrule} }
    \arrayrulecolor{Grey} \hline
    \rowcolor{Grey} \multicolumn{2}{|c|}{\large{\textbf{Market order characteristics}}} \label{tab:bsp_orders_nomenclature_orders}\\ 
    \hline
    \rowcolor{Grey} 
    \textbf{Notation} & \textbf{Meaning}\\ \hline
    $p_o$ & Price of order $o$ \\
    \hline
    $q_o^{min}$ & Minimum quantity of power offered for order $o$ \\
    \hline
    $q_o^{max}$ & Maximum quantity of power offered for order $o$ \\
    \hline
    $t_o^{start}$ & Start date of order \\
    \hline
    $t_o^{end}$ & End date of order \\
    \hline
    $t_o^{ex}$ & Execution date of order \\
    \hline
    $\sigma_{o}$ & Sale/Purchase indicator, $\sigma = 1$ for purchase, -1 for sale \\ 
    \hline
    $\delta_o^{TSO}$ & Binary variable indicating if order $o$ is from a TSO ($\delta_o^{TSO} = 1$) or not. \\
    \hline
    $q_o^{acc}$ & Amount of power accepted by the Clearing stage for order $o$ \\
    \hline
    $\delta_{o}^{acc}$ & Binary variable indicating if order $o$ is activated by the Clearing stage ($\delta_{o}^{acc} = 1$ if $q_o^{acc} > 0$)\\
    \hline
\end{longtable}

\begin{longtable}[H]{!{\color{Grey}\vrule}>{\centering\arraybackslash}p{3cm} !{\color{Grey}\vrule} p{10.5cm}!{\color{Grey}\vrule} >{\centering\arraybackslash} p{2cm}!{\color{Grey}\vrule}}
    \arrayrulecolor{Grey} \hline
    \rowcolor{Grey} \multicolumn{3}{|c|}{\large{\textbf{Other notations}}} \label{tab:bsp_orders_nomenclature_others}\\
    \hline
    \rowcolor{Grey} 
    \textbf{Notation} & \textbf{Description} & \textbf{Units}\\
    \hline
    $p^{spill}$ & Spillage penalty. \textbf{\textit{Parameter}} & \euro/MWh\\
    \hline
    $p^{redispatch}$ & Redispatch penalty. \textbf{\textit{Parameter}} & \euro/MWh\\
    \hline
    $E_{ca,t}^{spill}$ & Spillage power in control area $ca$ at time $t$.\textbf{\textit{Variable}} & MW\\
    \hline
\end{longtable}

\renewcommand{\arraystretch}{1.1} 

\clearpage
\section{Concepts of combinatorial orders and indexes} \label{sec:bsp_orders_combi_indexes}
\subsection{Context and objective}
\label{sec:bsp_orders_combi_indexes_obj}
\begin{figure}[H]
    \centering
    \includegraphics[width=\textwidth]{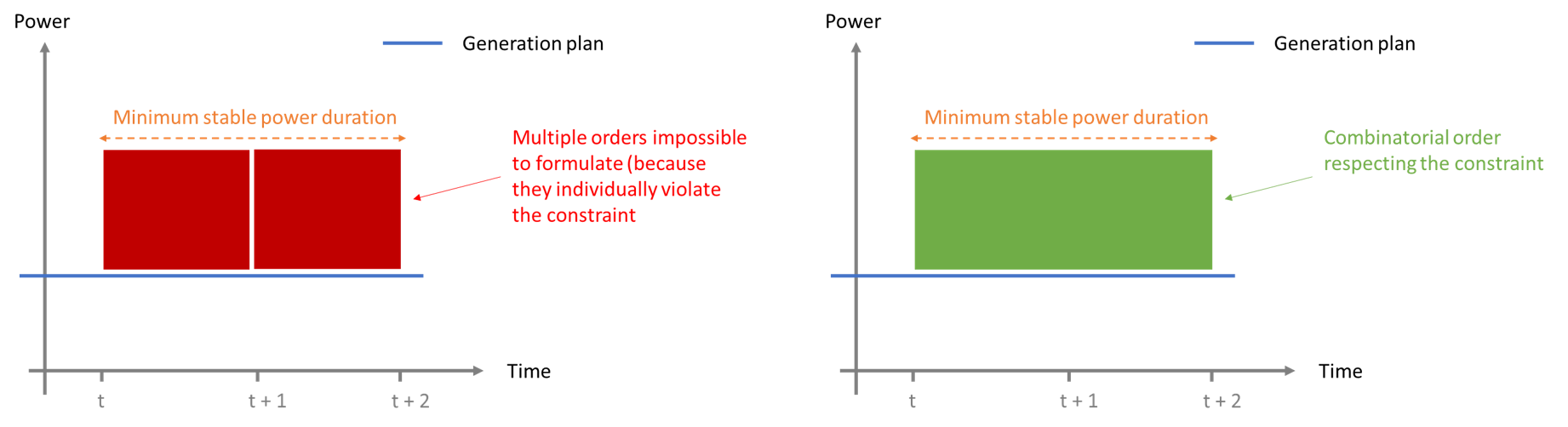}
    \caption{Example of operating constraint inducing an order over multiple time steps}
    \label{fig:combinatorial_orders_example}
\end{figure}

On balancing markets, BSPs may have to formulate orders over several time steps in order to respect operating constraints. This principle is illustrated in Figure \ref{fig:combinatorial_orders_example}, in the case where a generation group is constrained by a stable power duration constraint (see details about this constraint in Section \ref{sec:bsp_orders_thermic_stable_power}).\\

The objective of combinatorial orders is to represent this behavior and the resulting orders. To do so, the module considers subsets of successive time steps and tries to formulate feasible orders on them, linked with various coupling instances (see Section \ref{sec:bsp_orders_coupling_links} for more information on coupling instances). These subsets of time steps are called combinatorial indexes. Except in very specific cases, all orders from the same combinatorial index share identical price and volume, to eventually represent a single order on the entire index.

The behavior of the BSP Order Formulation module regarding combinatorial orders is defined by a boolean parameter. If it is set to True, the module will formulate combinatorial orders. Otherwise, it will only compute the available capacity on individual time steps, which usually leads to lower submitted volumes as it has less leeway to account for operating constraints.\\

\subsection{Method used to identify combinatorial indexes}
\label{sec:bsp_orders_combi_indexes_method}
In order to formulate every combinatorial order possible on a given balancing time frame, the module determines all possible combinations of indexes $id \in ID$ composed of successive time steps over which the studied unit $u$ has available power (either upward or downward). A first approximation of the available power is done by considering the maximum or the minimum power (see Section \ref{sec:bsp_orders_max_min_power}) and the previously procured reserves (see Section \ref{sec:bsp_orders_proc_reserves}). Upward ($ID_{up}$) and downward ($ID_{dn}$) combinatorial indexes are then computed based on this first estimation of available power, using the following method:
\begin{itemize}
    \item Identification of all time steps within the balancing time frame, for which there is available power.
    \item Determination of all combinations of successive time steps within those previously identified.
\end{itemize}

\begin{figure}[H]
    \centering
    \includegraphics[width=\textwidth]{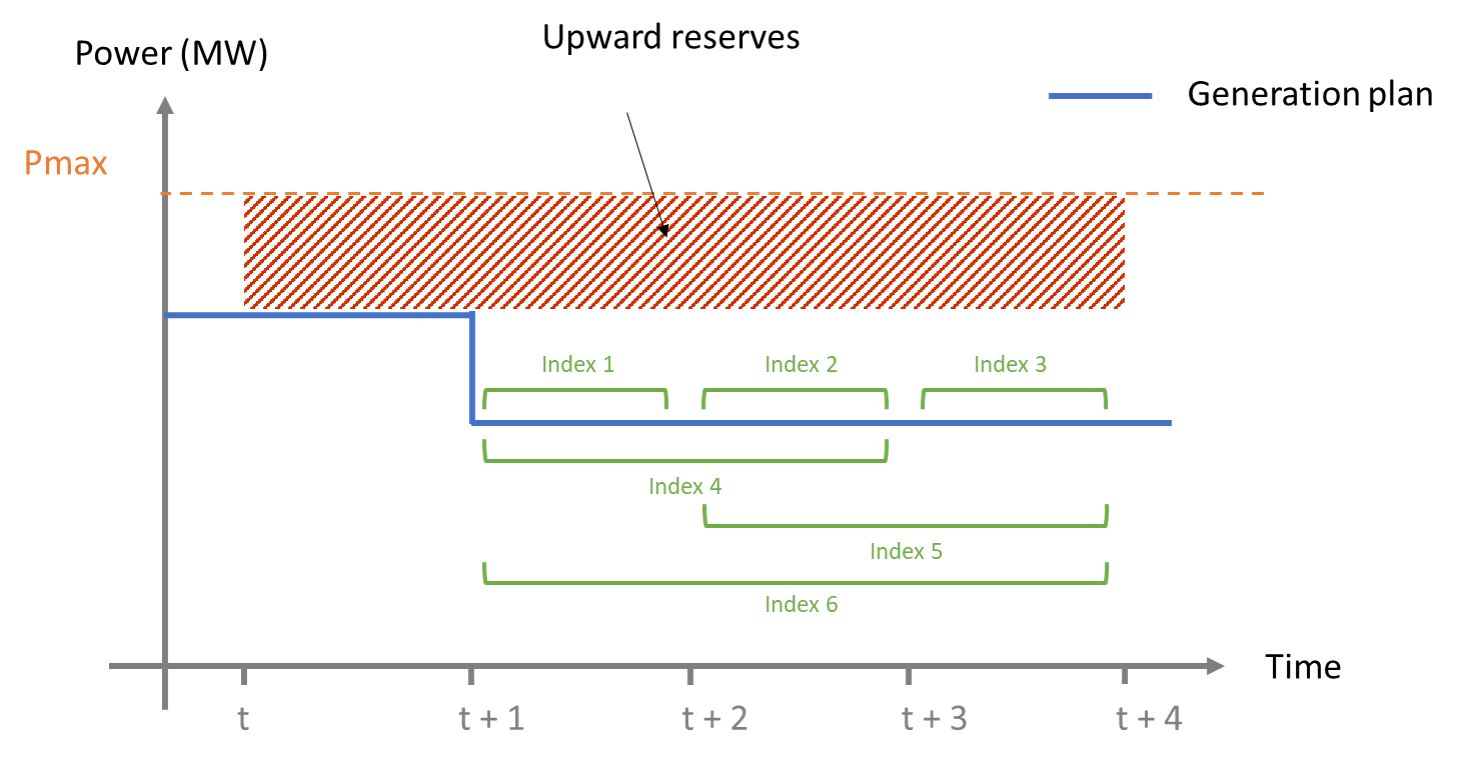}
    \caption{Combinatorial indexes determination for upward orders}
    \label{fig:combinatorial_indexes_determination}
\end{figure}

Figure \ref{fig:combinatorial_indexes_determination} illustrates this method on an example of a balancing time frame of 4 time steps (by only looking at upward orders).\\

In this example, the maximum power value and previously procured upward reserves constraints imply that there is no power available in the first time period. For the following 3 periods, 6 combinatorial indexes exist and are represented in green.\\

A third type of combinatorial index can be created by the module. It is specific to thermal units and aims at identifying available indexes for shutdown orders, which are a specific kind of downward orders that force the equipment to stop\footnote{For thermic units, shutdown orders are separated from usual downward orders. For a given time step, a classic downward order and a shutdown order can be formulated simultaneously, and later be linked by an \textit{Exclusion} coupling as detailed in Section \ref{sec:bsp_orders_coupling_excl_diff_indexes}. Other types of units are not concerned by shutdown orders as they are assumed to be able to turn their power output to $0$ at will}. The identification of time steps over which a shutdown order is feasible differs from that of downward orders. Indeed, the minimum power constraint doesn't have the same effect on shutdown orders. The requirements for a period to be part of a shutdown combinatorial index are the following:

\begin{itemize}
    \setlength\itemsep{-0.2em}
    \item The unit must be thermal, hence $u \in U^{th}$. Other types of units are not concerned by shutdown orders and simply formulate downward orders.
    \item The unit must produce at least its minimum power over the entire period. In some situations, the power output of the unit may be between $0$ and its minimum power, which means that it is in a starting or shutdown state. It is then considered unable to formulate shutdown orders.
    \item The unit must not have any procured reserve (upward or downward) during this period.
\end{itemize}

Operating constraints are then successively applied on each combinatorial index, following the process described in Section \ref{sec:bsp_orders_formulation_main}, to check the feasibility of orders and to determine their characteristics.

\section{Orders formulation} \label{sec:bsp_orders_formulation_main}
The BSP order formulation process uses a unit-based method: each unit in the input dataset is studied individually to identify available upward and downward capacities. The outcome of this process is the creation of market orders, that should indicate all the characteristics displayed in Table \ref{tab:bsp_orders_nomenclature_orders}, except for $\delta_{o}^{acc}$ and $q_{o}^{acc}$ which are filled by the Clearing stage.

If we take any BSP order formulated on market $m$ for time $t$, some properties can already be filled (Equation \ref{eq:bsp_orders_initial_charac}): its start date is equal to $t$, its end date is equal to $t + \Delta t_{m}$, its execution date corresponds to the execution date of the market $t_{m}^{ex}$, its binary variable tagging a TSO order is set to 0.

\begin{equation}
    \label{eq:bsp_orders_initial_charac}
    \forall o \in O^{BSP}, \begin{cases}
        t_o^{start} = t\\
        t_o^{end} = t + \Delta t_{m}\\
        t_o^{ex} = t_{m}^{ex}\\
        \delta_o^{TSO} = 0
    \end{cases}
\end{equation}

To compute any remaining property, the formulation of the market order takes place in two steps. First, the module determines $q_o^{max}$ and $q_o^{min}$ values based on the available power at time $t_o^{start}$ ($t_o^{end}$ is always assumed to be equal to $t_o^{start} + \Delta t_{m}$), computed according to generation plans and operating constraints applied to units. This module doesn't currently integrate a notion of strategic behavior for the volume of orders, which means that every unit is supposed to offer its entire available power on balancing markets. Then, the order price is computed based on variable and fixed costs. Other properties are easier to fill and do not require specific calculations.\\

\subsection{Determination of the available power for upward and downward orders}
\label{sec:bsp_orders_available_power_determination}
For each time $t \in T$, the module follows a series of steps to precise the available power range for both upward reserves (this range is comprised between upper bound $Q_{u,t,t_{m}^{ex}}^{up,min}$ and lower bound $Q_{u,t,t_{m}^{ex}}^{up,max}$) and downward reserves (ranging between $Q_{u,t,t_{m}^{ex}}^{dn,min}$ and $Q_{u,t,t_{m}^{ex}}^{dn,max}$), based on operating constraints and previously planned power output of the studied equipment. A first set of general constraints are applied regardless of the unit type. They are detailed in Section \ref{sec:bsp_orders_general_constraints}. Specific constraints are then applied to certain types of unit (Section \ref{sec:bsp_orders_thermic_specific_constraints} for thermic units, Section \ref{sec:bsp_orders_hydraulic_storage_constraints} for hydraulic and storage units). For wind, photovoltaic and flexible units, no specific constraints are required\footnote{In particular, it is assumed that the maximum ramping limit of these types of units is infinite.}.\\

In principle, the upper bounds of both upward $Q_{u,t,t_{m}^{ex}}^{up,max}$ and downward $Q_{u,t,t_{m}^{ex}}^{dn,max}$ available power are updated at each following step. In contrast, the lower bounds  $Q_{u,t,t_{m}^{ex}}^{up,min}$ and $Q_{u,t,t_{m}^{ex}}^{dn,min}$ are updated only when necessary, as only specific cases may impose restrictions on the lower bound. In the equations of this document, the update is denoted by the symbol $\gets$, as the calculation of the updated values usually includes the previous value.\\

\subsubsection{General constraints}
\label{sec:bsp_orders_general_constraints}

\paragraph{Maximum power and minimum power}
\label{sec:bsp_orders_max_min_power}
A first estimation of $Q_{u,t,t_{m}^{ex}}^{up,max}$ (resp. $Q_{u,t,t_{m}^{ex}}^{dn,max}$) is computed based on the maximum power (resp. minimum power) of the unit and the last generation plan available. This is straightforward for units whose power output is not associated with forecasts (Equations \ref{eq:bsp_orders_up_pmax} and \ref{eq:bsp_orders_up_pmin}):
\begin{subequations} 
    \begin{align}
        \centering
        \forall u \in U^{th} & \cup U^{h} \cup U^{st}, \mkern9mu \forall t \in T_{m},\notag\\ 
        Q_{u,t,t_{m}^{ex}}^{up,max} & \gets P_{u,t}^{max} - P_{u,t,t_{m}^{ex}}^{plan} \label{eq:bsp_orders_up_pmax} \\
        Q_{u,t,t_{m}^{ex}}^{dn,max} & \gets P_{u,t,t_{m}^{ex}}^{plan} - P_{u,t}^{min} \label{eq:bsp_orders_up_pmin}
    \end{align}
\end{subequations} 

For wind, photovoltaic and flexible load units, the maximum power output evolves according to forecasts. Consequently, the latest maximum power forecast available at $t^{ex}_{m}$ is taken as a reference. The minimum power of flexible load is assumed to always be $0$, which leads to Equations \ref{eq:bsp_orders_up_pmax_load} and \ref{eq:bsp_orders_up_pmin_load}. For wind and photovoltaic units, the minimum power depends on the curtailment ratio $Curt_{u}$ authorized on each unit (Equations \ref{eq:bsp_orders_up_pmax_wind_pv} and \ref{eq:bsp_orders_up_pmin_wind_pv}):
\begin{subequations} 
    \begin{align}
        \centering
        \forall u \in U^{l}, & \mkern9mu \forall t \in T_{m}, \notag\\ 
        Q_{u,t,t_{m}^{ex}}^{up,max} & \gets P_{u,t,t^{ex}_{m}}^{for} - P_{u,t,t_{m}^{ex}}^{plan} \label{eq:bsp_orders_up_pmax_load} \\
        Q_{u,t,t_{m}^{ex}}^{dn,max} & \gets P_{u,t,t_{m}^{ex}}^{plan} \label{eq:bsp_orders_up_pmin_load}
    \end{align}
\end{subequations}
\begin{subequations} 
    \begin{align}
        \centering
        \forall u \in U^{w} \cup & \mkern3mu U^{pv}, \mkern9mu \forall t \in T_{m}, \notag\\ 
        Q_{u,t,t_{m}^{ex}}^{up,max} & \gets P_{u,t,t^{ex}_{m}}^{for} - P_{u,t,t_{m}^{ex}}^{plan} \label{eq:bsp_orders_up_pmax_wind_pv} \\
        Q_{u,t,t_{m}^{ex}}^{dn,max} & \gets P_{u,t,t_{m}^{ex}}^{plan} - P_{u,t,t^{ex}_{m}}^{for} * Curt_{u}
        \label{eq:bsp_orders_up_pmin_wind_pv}
    \end{align}
\end{subequations}

\paragraph{Previously procured reserves}
\label{sec:bsp_orders_proc_reserves}
Previously procured upward (resp. downward) reserves are subtracted from $Q_{u,t,t_{m}^{ex}}^{up,max}$ (resp. $Q_{u,t,t_{m}^{ex}}^{dn,max}$). Procured reserves of the same type of the studied market are not taken into account in this step, as they obviously can be part of the activation market.
\begin{subequations}
    \label{eq:res_up_down}
    \begin{align}
        \centering
        \forall u \in U, \mkern9mu & \forall t \in T_{m}, \notag\\
        Q_{u,t,t_{m}^{ex}}^{up,max} \gets Q_{u,t,t_{m}^{ex}}^{up,max} & - \sum \limits_{m^{R} \neq m} R^{m^{R},up}_{u,t,t_{m}^{ex}}  \label{eq:res_upward} \\  
        Q_{u,t,t_{m}^{ex}}^{dn,max} \gets Q_{u,t,t_{m}^{ex}}^{dn,max} & - \sum \limits_{m^{R} \neq m} R^{m^{R},dn}_{u,t,t_{m}^{ex}} \label{eq:res_downward}
    \end{align}
\end{subequations}

\paragraph{Notice delay}
\label{sec:Orders formulation - notice delay}
The last step before determining combinatorial indexes is to check if certain time steps of the balancing time frame are \textit{de facto} unavailable for the unit because of the notice delay constraint $d_u^{notice}$. This constraint imposes a set time before the unit can modify its power output, and it may conflict with the duration between the end of the Clearing stage and the real-time:
\begin{align}
    \centering
    \forall u \in U, \mkern9mu \forall t \in T_{m}, \mkern9mu & \forall \sigma^{R} \in \{up, dn\}, \notag\\ 
    \textbf{if} \mkern9mu ( t_{id}^{start} - t_{m}^{ex} < d_u^{notice}) \mkern9mu & \textbf{then} \quad \begin{cases}
        Q_{u,t,t_{m}^{ex}}^{\sigma^{R},max} \gets 0\\
        Q_{u,t,t_{m}^{ex}}^{\sigma^{R}, min} =\gets 0
    \end{cases}
\label{eq:notice_delay}
\end{align}

\subsubsection{Thermic-specific constraints}
\label{sec:bsp_orders_thermic_specific_constraints}
This section details all specific constraints applied to thermal units. In particular, if the user chooses the option to formulate combinatorial orders, they are defined at this stage for thermic units (Section \ref{sec:bsp_orders_thermic_combi_indexes_definition}). Constraints applied to thermic units include:
\begin{itemize}
    \setlength\itemsep{-0.2em}
    \item Maximum ramping.
    \item Startup and shutdown duration.
    \item Minimum time ON and minimum time OFF.
    \item Minimum stable power duration.
\end{itemize}

\paragraph{Determination of combinatorial indexes}
\label{sec:bsp_orders_thermic_combi_indexes_definition}
Combinatorial indexes are then determined for every other type of unit, by following the method described in section \ref{sec:bsp_orders_combi_indexes_method}. If the user chooses not to generate combinatorial orders, using the associated parameter, then these indexes are each made of exactly one time period (for instance, such a set of indexes for a RR market could be [$t_1$], [$t_2$], [$t_3$] and [$t_4$]). Once indexes are defined, the module studies each one individually. From now on, all following steps are done a certain combinatorial index $id \in ID_{u}$ whose time frame is $T_{id}$. The notation $T_{id}$ will be used in place of $t$ for some variable to indicate that it applies for any time $t \in T_{id}$.\\

\paragraph{Maximum ramping constraint}
\label{sec:bsp_orders_thermic_max_ramp}
The maximum ramping constraint is verified for both upward and downward orders:
\begin{itemize}
    \item Between $t_{id}^{start} - \Delta t_{m}$ and $t_{id}^{start}$.
    \item Between $t_{id}^{end}$ and $t_{id}^{end} + \Delta t_{m}$
\end{itemize}
Note: to simplify notations, $t \pm n * \Delta t_{m}$ will be noted $t \pm n$.\\

This constraint is written as in Equation \ref{eq:bsp_orders_th_max_ramping}:
\begin{subequations}
    \label{eq:bsp_orders_th_max_ramping}
    \begin{align}
        \centering
        & \forall u \in U^{th}, \mkern9mu \forall T_{id}, \notag\\  
        Q_{u,T_{id},t_{m}^{ex}}^{up,max} \gets min(& Q_{u,T_{id},t_{m}^{ex}}^{up,max}, \notag\\
        & \Delta P_{u,t_{id}^{start}}^{max} * \Delta t_{m} - (P_{u,t_{id}^{start},t_{m}^{ex}}^{plan} - P_{u,t_{id}^{start}-1,t_{m}^{ex}}^{plan}), \notag \\ 
        & \Delta P_{u,t_{id}^{end}}^{max} * \Delta t_{m} - (P_{u,t_{id}^{end}+1,t_{m}^{ex}}^{plan} - P_{u,t_{id}^{end},t_{m}^{ex}}^{plan})) \label{eq:bsp_orders_th_upward_max_ramping} \\ 
        Q_{u,T_{id},t_{m}^{ex}}^{dn,max} \gets min(& Q_{u,T_{id},t_{m}^{ex}}^{dn,max}, \notag\\
        & \Delta P_{u,t_{id}^{start}}^{max} * \Delta t_{m} - (P_{u,t_{id}^{start}-1,t_{m}^{ex}}^{plan} - P_{u,t_{id}^{start},t_{m}^{ex}}^{plan}), \notag \\ 
        & \Delta P_{u,t_{id}^{end}}^{max} * \Delta t_{m} - (P_{u,t_{id}^{end},t_{m}^{ex}}^{plan} - P_{u,t_{id}^{end}+1,t_{m}^{ex}}^{plan})) \label{eq:bsp_orders_th_downward_max_ramping}
    \end{align}
\end{subequations}

In this module, we make the hypothesis that ramping constraints are not taken into account when starting or stopping a thermal unit. Thus, it can be shutdown from any power between its minimum and maximum power, and in a similar way start to any power in this interval.\\

Since combinatorial orders from the same index have similar $Q^{\sigma^{R},max}$ and $Q^{\sigma^{R},min}$, and generation plans in the input market are assumed to always be feasible, it is not necessary to check ramping constraints between time steps within combinatorial indexes.\\

\paragraph{Startup of a thermal unit}
\label{sec:bsp_orders_thermic_startup}
Seen from a time $t_{m}^{ex}$, a thermal unit $u \in U^{th}$ whose power output is 0 for every time step of a combinatorial index needs to start to provide upward reserves. This is expressed by $\forall t \in T_{id}, \quad P_{u,t,t_{m}^{ex}}^{plan} = 0$, and is a prerequisite check before applying any equation of this section. Other types of units are considered to be able to start at will.

Several cases can emerge, depending on the power output that is planned before and after the combinatorial index. In some cases, startup orders can conflict with minimum time OFF, minimum time ON and startup duration constraints and these will be checked if relevant. On top of this, depending on the case, formulating an upward order can induce or cancel a startup, and in these situations the upward order is divided into two parts to properly take into account any additional or reduced costs generated. This is tracked by the variables $\delta_{u,T_{id},t_{m}^{ex}}^{SU}$ and  $\sigma_{u,T_{id}}^{SU}$, initialized as follows when the startup of unit $u$ is checked:
\begin{align}
    \centering
    \forall u \in U^{th}, & \mkern9mu \forall T_{id}, \notag\\  \textbf{if} \mkern9mu (P_{u,t,t_{m}^{ex}}^{plan} = 0) \mkern9mu \textbf{then} \quad &
    \begin{cases}
        \delta_{u,T_{id},t_{m}^{ex}}^{SU} \gets 1\\
        \sigma_{u,T_{id}}^{SU} \gets 0
    \end{cases}
\end{align}

\begin{itemize}
    \item Case 1): The unit is ON at times $t_{id}^{start}-1$ and $t_{id}^{end}+1$. Here, an upward order isn't a startup, it is rather canceling a shutdown. This is a case where two orders need to be created, to properly distribute these cost savings. These orders can be formulated if the maximum ramping constraint is not violated between times $t_{id}^{start}-1$ and $t_{id}^{end}+1$. This principle is illustrated in a simple example, in Figure \ref{fig:start_order_infeasible_ramping}, with an index of 1 time period, for which the maximum ramping constraint is violated and the order cannot be formulated.\\

    \begin{figure}[H]
        \centering
        \includegraphics[width=0.8\textwidth]{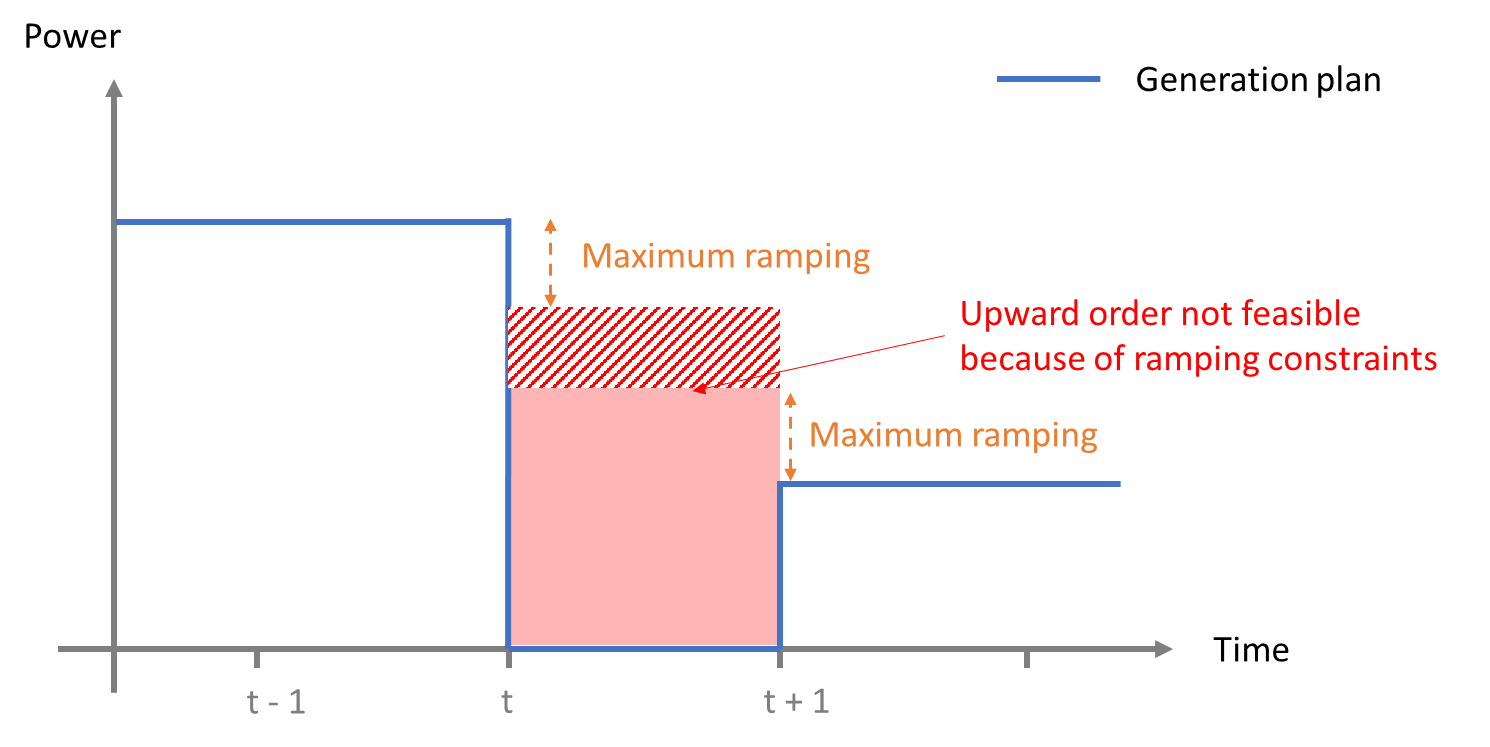}
        \caption{Example of a startup order infeasible because of maximum ramping constraints}
        \label{fig:start_order_infeasible_ramping}
    \end{figure}

    Respecting maximum ramping constraints both before and after the combinatorial index imposes bounds to the order, which means updating values of both $Q^{max}$ and $Q^{min}$. If formulating orders would always violate said constraints (the striped red area in Figure \ref{fig:start_order_infeasible_ramping}, then they cannot be formulated. This is verified by Equation \ref{eq:startup_case_1_infeasible}. There is no need to check for minimum time ON / OFF or startup duration constraints, since there will be no actual startup.

    $\forall u \in U^{th}, \mkern9mu \forall t \in T_{id}, \quad \textbf{if} \mkern9mu (P_{u,t_{id}^{start}-1,t_{m}^{ex}}^{plan} > 0 \mkern9mu \& \mkern9mu P_{u,t_{id}^{end}+1,t_{m}^{ex}}^{plan} > 0) \mkern9mu \textbf{then}$
    
    \begin{subequations}
    \label{eq:startup_case_1}
        \centering
        \begin{numcases}
        \notag
        \textbf{if} \mkern9mu (P_{u,t_{id}^{end}+1,t_{m}^{ex}}^{plan} \geq P_{u,t_{id}^{start}-1,t_{m}^{ex}}^{plan}),\mkern9mu \textbf{then} \quad \begin{cases}
                Q_{u,T_{id},t_{m}^{ex}}^{up,min} \gets max(P_{u,t_{id}^{start}}^{min}, P_{u,t_{id}^{end}+1,t_{m}^{ex}}^{plan} - \Delta P_{u,t_{id}^{start}}^{max} * \Delta t_{m}) \\
                Q_{u,T_{id},t_{m}^{ex}}^{up,max} \gets min(P_{u,t_{id}^{start}}^{max}, P_{u,t_{id}^{start}-1,t_{m}^{ex}}^{plan} - \Delta P_{u,t_{id}^{start}}^{max} * \Delta t_{m})\\
                \sigma_{u,T_{id}}^{SU} \gets -1
            \end{cases}\label{eq:startup_case_1_up}\\
            \textbf{if} \mkern9mu (P_{u,t_{id}^{end}+1,t_{m}^{ex}}^{plan} < P_{u,t_{id}^{start}-1,t_{m}^{ex}}^{plan}) \mkern9mu \textbf{then} \quad \begin{cases}
                Q_{u,T_{id},t_{m}^{ex}}^{up,min} \gets max(P_{u,t_{id}^{start}}^{min}, P_{u,t_{id}^{start}-1,t_{m}^{ex}}^{plan} - \Delta P_{u,t_{id}^{start}}^{max} * \Delta t_{m}) \\
                Q_{u,T_{id},t_{m}^{ex}}^{up,max} \gets min(P_{u,t_{id}^{start}}^{max}, P_{u,t_{id}^{end}+1,t_{m}^{ex}}^{plan} - \Delta P_{u,t_{id}^{start}}^{max} * \Delta t_{m})\\
                \sigma_{u,T_{id}}^{SU} \gets -1
            \end{cases} \label{eq:startup_case_1_down}\\
            \textbf{if} \mkern9mu (| P_{u,t_{id}^{end}+1,t_{m}^{ex}}^{plan} - P_{u,t_{id}^{start}-1,t_{m}^{ex}}^{plan}| > 2 * \Delta P_{u,t_{id}^{start}}^{max} * 60) \mkern9mu \textbf{then} \quad \begin{cases}
                Q_{u,T_{id},t_{m}^{ex}}^{up,min} \gets 0\\
                Q_{u,T_{id},t_{m}^{ex}}^{up,max} \gets 0\\
                \delta_{u,T_{id},t_{m}^{ex}}^{SU} \gets 0
            \end{cases}  \label{eq:startup_case_1_infeasible}
        \end{numcases}
    \end{subequations}

    \item Case 2): The unit is ON at time $t_{id}^{start} -1$, and OFF at $t_{id}^{end} +1$. No startup costs are induced in this case, as the unit was already running during the previous time step. The upward order is feasible if the minimum time OFF constraint is respected after $t_{id}^{end} +1$, and its $Q^{max}$ and $Q^{min}$ values are bounded because of maximum ramping constraints related with $P_{u,t_{id}^{start}-1,t_{m}^{ex}}^{plan}$.

    $\forall u \in U^{th}, \mkern9mu \forall t \in T_{id}, \quad \textbf{if} \mkern9mu (P_{u,t_{id}^{start}-1,t_{m}^{ex}}^{plan} > 0 \mkern9mu \& \mkern9mu P_{u,t_{id}^{end}+1,t_{m}^{ex}}^{plan} = 0) \mkern9mu \textbf{then}$
    
    \begin{equation}
        \centering
        \label{eq:startup_case_2}
        \begin{cases}
            \textbf{if} \mkern9mu (\exists \mkern9mu t' \in \{t_{id}^{end}, \dots, t_{id}^{end} + d_u^{minOff}\} \mkern9mu | \mkern9mu P_{u,t',t_{m}^{ex}}^{plan} > 0) \mkern9mu \textbf{then} \quad  
            \begin{cases}
                Q_{u,t,t_{m}^{ex}}^{up,min} \gets 0\\
                Q_{u,t,t_{m}^{ex}}^{up,max} \gets 0\\
                \delta_{u,T_{id},t_{m}^{ex}}^{SU} \gets 0
            \end{cases}\\
            \textbf{else}, \quad \begin{cases}
                Q_{u,T_{id},t_{m}^{ex}}^{up,min} \gets max(P_{u,t_{id}^{start}}^{min}, P_{u,t_{id}^{start}-1,t_{m}^{ex}}^{plan}) \\
                Q_{u,T_{id},t_{m}^{ex}}^{up,max} \gets min(P_{u,t_{id}^{start}}^{max}, P_{u,t_{id}^{start}-1,t_{m}^{ex}}^{plan} + \Delta P_{u,t_{id}^{start}}^{max} * \Delta t_{m})\\
                \delta_{u,T_{id},t_{m}^{ex}}^{SU} \gets 0
            \end{cases}
        \end{cases}
    \end{equation}

    \item Case 3): The unit is OFF at time $t_{id}^{start} -1$, and ON at $t_{id}^{end} +1$. In that situation, there is indeed a startup but that does not induce startup costs (because it is only advancing the startup planned later). The upward order is feasible if the unit has time to start between $t_{m}^{ex}$ and $t_{id}^{start}$, and if the minimum time OFF constraint is respected before $t_{id}^{start}$.

    $\forall u \in U^{th}, \mkern9mu \forall t\in T_{id}, \quad \textbf{if} \mkern9mu (P_{u,t_{id}^{start}-1,t_{m}^{ex}}^{plan} = 0 \mkern9mu \& \mkern9mu P_{u,t_{id}^{end}+1,t_{m}^{ex}}^{plan} > 0) \mkern9mu \textbf{then}$
    
    \begin{subequations}
        \label{eq:startup_case_3}
        \centering
        \begin{numcases}
            \notag
            \textbf{if} \mkern9mu (t_{id}^{start} - t_{m}^{ex} < d_u^{SU}) \mkern9mu \textbf{then} \quad  
            \begin{cases}
                Q_{u,t,t_{m}^{ex}}^{up,min} \gets 0\\
                Q_{u,t,t_{m}^{ex}}^{up,max} \gets 0\\
                \delta_{u,T_{id},t_{m}^{ex}}^{SU} \gets 0
            \end{cases}  
            \label{eq:startup_case_3_unfeasible_suduration}\\
            \textbf{elif} \mkern9mu (\exists \mkern9mu t' \in \{t_{id}^{start} - d_u^{minOff}, \dots, t_{id}^{start} - 1\} \mkern9mu | \mkern9mu P_{u,t',t_{m}^{ex}}^{plan} > 0) \mkern9mu \textbf{then} \quad \begin{cases}
                Q_{u,t,t_{m}^{ex}}^{up,min} \gets 0\\
                Q_{u,t,t_{m}^{ex}}^{up,max} \gets 0\\
                \delta_{u,T_{id},t_{m}^{ex}}^{SU} \gets 0
            \end{cases}
            \label{eq:startup_case_3_unfeasible_minoff}\\
            \textbf{else}, \quad \begin{cases}
                Q_{u,t,t_{m}^{ex}}^{up,min} \gets max(P_{u,t_{id}^{start}}^{min}, P_{u,t_{id}^{end}+1,t_{m}^{ex}}^{plan}) \\
                Q_{u,t,t_{m}^{ex}}^{up,max} \gets min(P_{u,t_{id}^{start}}^{max}, P_{u,t_{id}^{end}+1,t_{m}^{ex}}^{plan} + \Delta P_{u,t_{id}^{start}}^{max} * \Delta t_{m})\\
                \delta_{u,T_{id},t_{m}^{ex}}^{SU} \gets 0
            \end{cases} \label{eq:startup_case_3_feasible}
        \end{numcases}
            
    \end{subequations}

    \item Case 4): The unit is within a startup already planned, at any time within $T_{id}$. This state is a possible outcome of day-ahead and intraday markets and is defined by a power output between 0 and the minimum power of the unit (meaning that the unit is currently ramping up to its minimum power). No balancing order can be formulated in that case:
    \begin{align}
        \label{eq:startup_case_4}
        \centering
        \forall u \in U^{th}, & \mkern9mu \forall T_{id}, \notag\\ \textbf{if} \mkern9mu (\exists \mkern3mu t \in T_{id} \mkern9mu | \mkern9mu 0 < P_{u,t,t_{m}^{ex}}^{plan} < P_{u,t}^{min}) & \mkern9mu \textbf{then} \quad
        \begin{cases}
            Q_{u,T_{id},t_{m}^{ex}}^{up,min} \gets 0\\ 
            Q_{u,T_{id},t_{m}^{ex}}^{up,max} \gets 0\\
            \delta_{u,T_{id},t_{m}^{ex}}^{SU} \gets 0
        \end{cases} 
    \end{align}

    \item Case 5): The unit is OFF at both $t_{id}^{start} -1$ and $t_{id}^{end} +1$. This is the most basic case of startup, and the only one that actually induces startup costs. An upward order can be formulated if the following conditions are respected:
    \begin{itemize}\setlength\itemsep{-0.2em}
        \item Startup duration between $t_{m}^{ex}$ and $t_{id}^{start}$ (Equation \ref{eq:startup_case_5_checking_durations_1}).
        \item Minimum time OFF before $t_{id}^{start}$ (Equation \ref{eq:startup_case_5_checking_durations_2}) and after $t_{id}^{end}$ (Equation \ref{eq:startup_case_5_checking_durations_3}).
        \item Minimum time ON during $T_{id}$ (Equation \ref{eq:startup_case_5_checking_durations_4}).
    \end{itemize}
    Which can be expressed as:
    
    $\forall u \in U^{th}, \mkern9mu \forall t \in T_{id}, \quad \textbf{if} \mkern9mu (P_{u,t_{id}^{start}-1,t_{m}^{ex}}^{plan} = 0 \mkern9mu \quad \& \mkern9mu P_{u,t_{id}^{end}+1,t_{m}^{ex}}^{plan} = 0) \mkern9mu \textbf{then}$
    
    \begin{subequations}
        \label{eq:startup_case_5_checking_durations}
        \centering
        \begin{numcases}
            \notag
            \textbf{if} \mkern9mu (t_{id}^{start} - t_{m}^{ex} < d_{u}^{SU}) \mkern9mu \textbf{then} \quad \begin{cases}
                Q_{u,T_{id},t_{m}^{ex}}^{up,max} \gets 0\\ Q_{u,T_{id},t_{m}^{ex}}^{up,max} \gets 0\\
                \delta_{u,T_{id},t_{m}^{ex}}^{SU} \gets 0
            \end{cases} \label{eq:startup_case_5_checking_durations_1}\\
            \textbf{elif} \mkern9mu (\exists \mkern9mu t' \in \{t_{id}^{start} - d_u^{minOff}, \dots, t_{id}^{start} - 1\} \mkern9mu | \mkern9mu P_{u,t',t_{m}^{ex}}^{plan} > 0) \mkern9mu \textbf{then} \quad \begin{cases}
                Q_{u,T_{id},t_{m}^{ex}}^{up,max} \gets 0\\ Q_{u,T_{id},t_{m}^{ex}}^{up,max} \gets 0\\
                \delta_{u,T_{id},t_{m}^{ex}}^{SU} \gets 0
            \end{cases} \label{eq:startup_case_5_checking_durations_2}\\
            \textbf{elif} \mkern9mu (\exists \mkern9mu t' \in \{t_{id}^{end}, \dots, t_{id}^{end} + d_u^{minOff}\} \mkern9mu | \mkern9mu P_{u,t',t_{m}^{ex}}^{plan} > 0) \mkern9mu \textbf{then} \quad \begin{cases}
                Q_{u,T_{id},t_{m}^{ex}}^{up,max} \gets 0\\ Q_{u,T_{id},t_{m}^{ex}}^{up,max} \gets 0\\
                \delta_{u,T_{id},t_{m}^{ex}}^{SU} \gets 0
            \end{cases} \label{eq:startup_case_5_checking_durations_3}\\
            \textbf{elif} \mkern9mu (\Delta T_{id} \geq d_{u}^{minOn}) \mkern9mu \textbf{then} \quad \begin{cases}
                Q_{u,T_{id},t_{m}^{ex}}^{up,max} \gets 0\\
                Q_{u,T_{id},t_{m}^{ex}}^{up,max} \gets 0\\
                \delta_{u,T_{id},t_{m}^{ex}}^{SU} \gets 0
            \end{cases} \label{eq:startup_case_5_checking_durations_4}\\
            \textbf{else} \mkern9mu \sigma_{u,T_{id}}^{SU} \gets 1
            \label{eq:startup_case_5_checking_durations_5}
        \end{numcases}
    \end{subequations}
\end{itemize}

\paragraph{Shutdown of a thermal unit}
\label{sec:bsp_orders_thermic_shutdown}
The following constraints are applied to verify if a shutdown order is feasible for thermal units. First, a thermal unit is only concerned by shutdown orders over a shutdown combinatorial index $id \in ID^{shut}$ if it is ON on every time period of $id$, meaning that these checks are only done if $\forall t \in T_{id}, \quad P_{u,t,t_{m}^{ex}}^{plan} > P_{u,t,t_{m}^{ex}}^{min}$.

Similarly to startup orders, variables $\delta_{u,T_{id},t_{m}^{ex}}^{SD}$ and $\sigma_{u,T_{id}}^{SU}$ are respectively used to indicate if the shutdown order is feasible, and if it generates or cancels startup costs, and are initialized as follows:
\begin{align}
    \centering
    \forall u \in U^{th}, & \mkern9mu  \forall t \in T_{id}, \notag\\ 
    \textbf{if} \mkern9mu (P_{u,t,t_{m}^{ex}}^{plan} \geq P_{u,t,t_{m}^{ex}}^{min}) \mkern9mu & \textbf{then} \quad
    \begin{cases}
        \delta_{u,T_{id},t_{m}^{ex}}^{SD} & \gets 1\\
        \sigma_{u,T_{id}}^{SU} & \gets 0
    \end{cases}
\end{align}

First, if the unit has any amount of procured reserves (either upward or downward) during at least one time step $t \in T_{id}$, then it cannot be shutdown on $id$ as it would no longer be able to provide said reserves (Equation \ref{eq:bsp_orders_res_shutdown}). 
\begin{align}
    \centering
    \forall u \in U^{th}, \mkern9mu \forall t \in T_{m}, \mkern9mu & \forall m^{R}, \mkern9mu \forall \mkern6mu \sigma^{R} \in \{up, dn\} \notag\\ 
    \textbf{if} \mkern9mu (R_{u,t,t_{m}^{ex}}^{m^{R}, \sigma^{R}} > 0) \quad &\textbf{then} \quad \delta_{u,t}^{SD} \gets 0
\label{eq:bsp_orders_res_shutdown}
\end{align}

Then, if the notice delay constraint checked in section \ref{sec:Orders formulation - notice delay} was not respected, the shutdown order is not feasible (Equation \ref{eq:bsp_orders_thermic_shut_notice}):
\begin{align}
    \centering
    \label{eq:bsp_orders_thermic_shut_notice}
    \forall u \in U^{th}, & \mkern9mu \forall t \in T_{id}, \notag\\
    \textbf{if} \mkern9mu (t^{start}_{id} - t^{ex}_{id} < d^{notice}_{u}) \mkern9mu & \textbf{then} \quad \delta_{u,T_{id},t_{m}^{ex}}^{SD} \gets 0
\end{align}

Once these first checks are complete, several cases emerge:
\begin{itemize}
    \item Case 1): The unit is OFF at times $t_{id}^{start}-1$ and $t_{id}^{end}+1$. In this case, the shutdown order is always feasible, and in practice it cancels a startup:
    \begin{align}
        \centering
        \forall u  & \in U^{th}, \mkern9mu  \forall t \in T_{id},  \notag\\ 
        \textbf{if} \mkern9mu (P_{u,t_{id}^{start}-1,t_{m}^{ex}}^{plan} = 0  \mkern9mu \& & \mkern9mu P_{u,t_{id}^{end}+1,t_{m}^{ex}}^{plan} = 0) \mkern9mu  \textbf{then} \quad 
        \sigma_{u,T_{id}}^{SU} \gets -1
        \label{eq:shutdown_case_1}
    \end{align}

    \item Case 2): The unit is OFF at time $t_{id}^{start}-1$ and ON at $t_{id}^{end}+1$. The shutdown order can be formulated if the minimum time ON is respected after $t_{id}^{end}$, $Q^{max}$ is not modified and no additional startup is created by the order.
    \begin{align}
        \label{eq:shutdown_case_2}
        \centering
        \forall u \in U^{th}, \mkern9mu  \forall t \in T_{id}, \quad  \textbf{if} \mkern9mu (P_{u,t_{id}^{start}-1,t_{m}^{ex}}^{plan} & = 0 \mkern9mu \& \mkern9mu P_{u,t_{id}^{end}+1,t_{m}^{ex}}^{plan} > 0) \mkern9mu \textbf{then} \notag\\
        \textbf{if} \mkern9mu (\exists \mkern3mu t' \in \{t_{id}^{end}, \dots, t_{id}^{end} + d_{u}^{minOn}\} \mkern9mu |  & \mkern9mu P_{u,t',t_{m}^{ex}}^{plan} = 0) \mkern9mu \textbf{then} \quad \delta_{u,T_{id},t_{m}^{ex}}^{SD} \gets 0
    \end{align}

    \item Case 3): The unit is ON at $t_{id}^{start}-1$ and OFF at $t_{id}^{end}+1$. Here, the shutdown order can be formulated if the minimum time ON is respected before $t_{id}^{start}$. If it is verified, the shutdown order can be formulated and no additional startup is generated.
    \begin{align}
        \label{eq:shutdown_case_3}
        \centering
        \forall u \in U^{th}, \mkern9mu  \forall t \in T_{id}, \quad  \textbf{if} \mkern9mu (P_{u,t_{id}^{start}-1,t_{m}^{ex}}^{plan} & > 0 \mkern9mu \& \mkern9mu P_{u,t_{id}^{end}+1,t_{m}^{ex}}^{plan} = 0) \mkern9mu \textbf{then} \notag\\
        \textbf{if} \mkern9mu (\exists \mkern3mu t' \in \{t_{id}^{start}- d_{u}^{minOn}, t_{id}^{start}\} \mkern9mu | & \mkern9mu P_{u,t',t_{m}^{ex}}^{plan} = 0) \mkern9mu \textbf{then}\notag \\
        \delta_{u,T_{id},t_{m}^{ex}}^{SD} \gets 0 & 
    \end{align}

    \item Case 4): The unit is starting during $T_{id}$. Same situation as for case 4 of startup orders (see Section \ref{sec:bsp_orders_thermic_startup}), no order is permitted on $T_{id}$.
    \begin{align}
        \label{eq:shutdown_case_4}
        \centering
        \forall u \in U^{th}, & \mkern9mu  \forall t \in T_{id}, \notag\\ 
        \textbf{if} \mkern9mu (\exists \mkern3mu t' \in T_{id} \mkern9mu | \mkern9mu 0 < P_{u,t',t_{m}^{ex}}^{plan} & < P_{u,t'}^{min}) \mkern9mu \textbf{then} \quad  \delta_{u,T_{id},t_{m}^{ex}}^{SD} \gets 0 
    \end{align}

    \item Case 5): The unit is ON at $t_{id}^{start}-1$ and at $t_{id}^{end}+1$. A shutdown order can be formulated if the following conditions are satisfied:
    \begin{itemize}\setlength\itemsep{-0.1em}
        \item The startup duration is less than a time step, to make sure that it can restart in time at the end of the order activation.
        \begin{align}
            \label{eq:shutdown_case_5_verif1}
            \centering
            \forall u \in U^{th}, & \mkern9mu  \forall t \in T_{id}, \notag\\  
            \textbf{if} \mkern9mu (P_{u,t_{id}^{start}-1 ,t_{m}^{ex}}^{plan} > 0 \mkern9mu \& \mkern9mu P_{u,t_{id}^{end}+1,t_{m}^{ex}}^{plan} & > 0 \mkern9mu \& \mkern9mu d_u^{SU} > \Delta t_{m}) \mkern9mu \textbf{then} \quad \delta_{u,T_{id},t_{m}^{ex}}^{SD} \gets 0
        \end{align}
        \item The minimum time OFF is not violated during the order:
        \begin{align}
            \label{eq:shutdown_case_5_verif2}
            \centering
            \forall u \in U^{th}, & \mkern9mu  \forall t \in T_{id}, \notag\\  
            \textbf{if} \mkern9mu (P_{u,t_{id}^{start}-1,t_{m}^{ex}}^{plan} > 0 \mkern9mu \& \mkern9mu P_{u,t_{id}^{end}+1,t_{m}^{ex}}^{plan} > & 0 \mkern9mu \& \mkern9mu d_u^{minOff} > \Delta T_{id}) \mkern9mu \textbf{then} \quad            \delta_{u,T_{id},t_{m}^{ex}}^{SD} \gets 0
        \end{align}
        \item The minimum time ON is respected before $t_{id}^{start}$ and after $t_{id}^{end}$:

        {\setstretch{0}
        \begin{equation}
            \centering
            \forall u \in U^{th}, \mkern9mu  \forall t \in T_{id}, \quad  \textbf{if} \mkern9mu (P_{u,t_{id}^{start}-1,t_{m}^{ex}}^{plan} > 0 \mkern9mu \& \mkern9mu  P_{u,t_{id}^{end}+1,t_{m}^{ex}}^{plan} > 0) \mkern9mu \textbf{then}\notag
        \end{equation}
        }
        \begin{subequations}
            \label{eq:shutdown_case_5_verif3}
            \centering
            \begin{numcases}
                \notag
                \textbf{if} \mkern9mu (\exists \mkern3mu t' \in \{t_{id}^{start} - 1 -  d_u^{minOn}, \dots, t_{id}^{start} - 1 \} \mkern9mu | \mkern9mu P_{u,t',t_{m}^{ex}}^{plan} = 0) \mkern9mu \textbf{then} \quad \delta_{u,T_{id},t_{m}^{ex}}^{SD} \gets 0 \label{eq:shutdown_case_5_verif3_1}\\
                \textbf{elif} \mkern9mu (\exists \mkern3mu t' \in \{t_{id}^{end}, \dots, t_{id}^{end} + d_u^{minOn}\} \mkern9mu | \mkern9mu P_{u,t',t_{m}^{ex}}^{plan} = 0) \mkern9mu \textbf{then} \quad \delta_{u,T_{id},t_{m}^{ex}}^{SD} \gets 0 \label{eq:shutdown_case_5_verif3_2}
            \end{numcases}
        \end{subequations}
        
    \end{itemize}
    If all requirements are met, the shutdown order is feasible on index $id$ and it induces a startup cost:
    \begin{equation}
        \centering
        \textbf{if} \mkern9mu (\delta_{u,T_{id},t_{m}^{ex}}^{SD} = 1) \mkern9mu \textbf{then} \quad \sigma_{u,T_{id}}^{SU} \gets 1
    \label{eq:shutdown_case_5_feasible}
    \end{equation}
\end{itemize}

\paragraph{Minimum stable power duration constraint}
\label{sec:bsp_orders_thermic_stable_power}
The minimum stable power duration constraint forces units to keep a constant power output for a certain duration before being able to modify it. In the current version of ATLAS, it is only applied to thermal units. When trying to formulate orders on unit $u \in U_{th}$ on the combinatorial index $id \in ID_{u}$, the module checks if these orders cannot generate violations of this constraint before, during, and after $id$. 

The first step is to check if $d_u^{minStable} > \Delta t_{m}$, to make sure that this constraint is relevant for the studied market. If it is, the module proceeds to a series of checks for available power in upward and downward directions:
\begin{itemize}
    \item Stable power level before the index:
    \begin{align}
        \label{eq:stable_power_before}
        \centering
        \forall u \in U^{th}, & \mkern9mu \forall t \in T_{id}, \notag\\ 
        \textbf{if} \mkern9mu (\exists \mkern3mu t \in  \{t_{id}^{start} - 1 - d_u^{minStable}, \dots, t_{id}^{start} - 1\} \mkern9mu | & \mkern9mu P_{u,t,t_{m}^{ex}}^{plan} \neq P_{u,t_{id}^{start} - 1,t_{m}^{ex}}^{plan}) \mkern9mu \textbf{then} \quad
        \begin{cases}
            Q_{u,T_{id},t_{m}^{ex}}^{up,max} \gets 0\\
            Q_{u,T_{id},t_{m}^{ex}}^{dn,max} \gets 0\\
            \delta_{u,T_{id},t_{m}^{ex}}^{SD} \gets 0
        \end{cases} 
    \end{align}

    \item Stable power level after the index:
    \begin{align}
        \label{eq:stable_power_after}
        \centering
        \forall u \in U^{th}, & \mkern9mu \forall t \in T_{id}, \notag\\ 
        \textbf{if} \mkern9mu (\exists \mkern3mu t \in \{t_{id}^{end} +1, \dots, t_{id}^{end} + 1 + d_u^{minStable}\} \mkern9mu | & \mkern9mu P_{u,t,t_{m}^{ex}}^{plan} \neq P_{u,t_{id}^{end},t_{m}^{ex}}^{plan}) \mkern9mu \textbf{then} \quad \begin{cases}
            Q_{u,T_{id},t_{m}^{ex}}^{up,max} \gets 0\\
            Q_{u,T_{id},t_{m}^{ex}}^{dn,max} \gets 0\\
            \delta_{u,T_{id},t_{m}^{ex}}^{SD} \gets 0
        \end{cases}
    \end{align}
\end{itemize}

\begin{figure}[H]
    \centering
    \includegraphics[width = \textwidth]{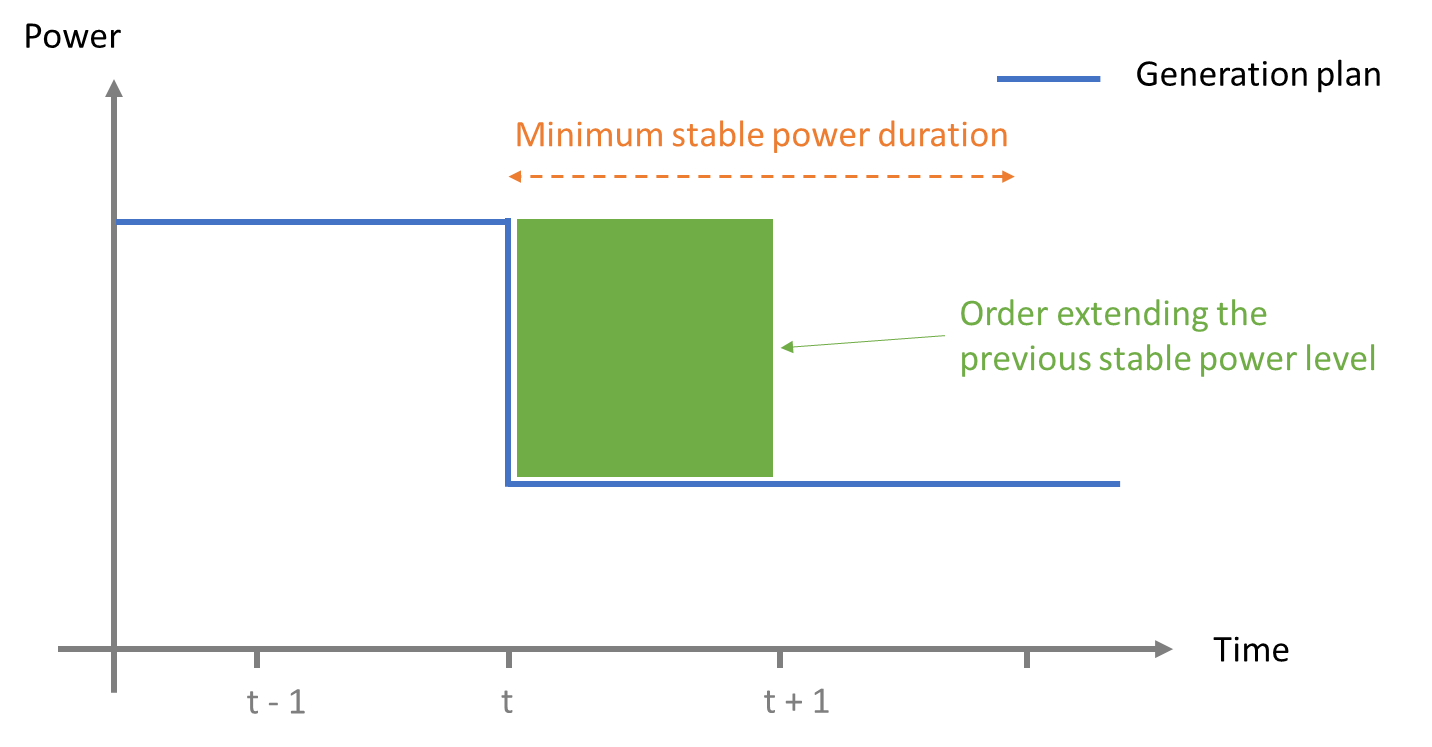}
    \caption{Extension of a previous stable power level}
    \label{fig:stable_power_level_extension}
\end{figure}

If the previous conditions are verified, but the duration of the index is shorter than the minimum stable power duration, the offer can still be formulated as an extension of a previous or later stable power level (similar to cases 1 and 2 of both startup or shutdown orders). This principle is illustrated in Figure \ref{fig:stable_power_level_extension}, in which the possibility of formulating an order between t and t+1 while having a minimum stable power duration of 2 time steps is looked at. The only feasible order is indicated in green. Such an order is necessarily indivisible, and its $Q^{max}$ is also imposed, and this means that if it conflicts with other constraints, the order cannot be formulated. 

Once again, several cases emerge, depending on how the stable power level extension is done (before or after, and upward or downward). In all cases, it leads to indivisible orders:
\begin{itemize}
    \item If the unit is within a ramp, no order can be formulated:
    \begin{align}
        \label{eq:stable_power_ramp}
        \centering
        \forall u \in U^{th}, & \mkern9mu \forall T_{id}, \notag\\
        \textbf{if} \mkern9mu (P_{u,t_{id}^{start}-1,t_{m}^{ex}}^{plan} \neq P_{u,t_{id}^{start},t_{m}^{ex}}^{plan} \mkern9mu \& \mkern9mu P_{u,t_{id}^{end},t_{m}^{ex}}^{plan} \neq  & P_{u,t_{id}^{end}+1,t_{m}^{ex}}^{plan}) \mkern9mu \textbf{then} \quad \begin{cases}
            Q_{u,T_{id},t_{m}^{ex}}^{up,max} \gets 0\\
            Q_{u,T_{id},t_{m}^{ex}}^{dn,max} \gets 0
        \end{cases}
    \end{align}
    
    \item If $P_{u,t_{id}^{start}-1,t_{m}^{ex}}^{plan} > P_{u,t_{id}^{start},t_{m}^{ex}}^{plan}$, then the previous stable power level can be extended by an upward order only:
    \begin{align}
        \label{eq:stable_power_extension_before_up}
        \centering
        \forall u \in U^{th}, & \mkern9mu \forall T_{id}, \notag\\
        \textbf{if} \mkern9mu (P_{u,t_{id}^{start}-1,t_{m}^{ex}}^{plan} > 
        P_{u,t_{id}^{start},t_{m}^{ex}}^{plan}) \mkern9mu \textbf{then} & \quad \begin{cases}
            Q_{u,T_{id},t_{m}^{ex}}^{up,max} \gets P_{u,t_{id}^{start}-1,t_{m}^{ex}}^{plan} - P_{u,t_{id}^{start},t_{m}^{ex}}^{plan}\\
            Q_{u,T_{id},t_{m}^{ex}}^{up,min} \gets Q_{u,T_{id},t_{m}^{ex}}^{up,max}\\
            Q_{u,T_{id},t_{m}^{ex}}^{dn,max} \gets 0\\
            Q_{u,T_{id},t_{m}^{ex}}^{dn,min} \gets 0
        \end{cases}
    \end{align}

    \item If $P_{u,t_{id}^{start}-1,t_{m}^{ex}}^{plan} < P_{u,t_{id}^{start},t_{m}^{ex}}^{plan}$, then the previous stable power level can be extended by a downward order only:
    \begin{align}
    \label{eq:stable_power_extension_before_down}
        \centering
        \forall u \in U^{th}, & \mkern9mu \forall T_{id}, \notag\\ 
        \textbf{if} \mkern9mu (P_{u,t_{id}^{start}-1,t_{m}^{ex}}^{plan} < P_{u,t_{id}^{start},t_{m}^{ex}}^{plan}) \mkern9mu \textbf{then} & \quad \begin{cases}
            Q_{u,T_{id},t_{m}^{ex}}^{up,max} \gets 0\\
            Q_{u,T_{id},t_{m}^{ex}}^{up,min} \gets 0\\
            Q_{u,T_{id},t_{m}^{ex}}^{dn,max} \gets P_{u,t_{id}^{start},t_{m}^{ex}}^{plan} - P_{u,t_{id}^{start}-1,t_{m}^{ex}}^{plan}\\
            Q_{u,T_{id},t_{m}^{ex}}^{dn,min} \gets Q_{u,T_{id},t_{m}^{ex}}^{dn,max}
        \end{cases}
    \end{align}

    \item If $P_{u,t_{id}^{end},t_{m}^{ex}}^{plan} < P^{plan}_{u,t_{id}^{end}+1,t_{m}^{ex}}$, then the next stable power level can be extended by an upward order only:
    \begin{align}
        \label{eq:stable_power_extension_after_up}
        \centering
        \forall u \in U^{th}, & \mkern9mu \forall T_{id}, \notag\\
        \textbf{if} \mkern9mu (P_{u,t_{id}^{end},t_{m}^{ex}}^{plan} < P_{u,t_{id}^{end}+1,t_{m}^{ex}}^{plan}) \mkern9mu \textbf{then} & \quad \begin{cases}
            Q_{u,T_{id},t_{m}^{ex}}^{up,max} \gets P_{u,t_{id}^{end}+1,t_{m}^{ex}}^{plan} - P_{u,t_{id}^{end},t_{m}^{ex}}^{plan}\\
            Q_{u,T_{id},t_{m}^{ex}}^{up,min} \gets Q_{u,T_{id},t_{m}^{ex}}^{up,max}\\
            Q_{u,T_{id},t_{m}^{ex}}^{dn,max} \gets 0\\
            Q_{u,T_{id},t_{m}^{ex}}^{dn,min} \gets 0
        \end{cases}
    \end{align}

    \item If $P_{u,t_{id}^{end},t_{m}^{ex}}^{plan} > P^{plan}_{u,t_{id}^{end}+1,t_{m}^{ex}}$, then the next stable power level can be extended by a downward order only:
    \begin{align}
        \label{eq:stable_power_extension_after_down}
        \centering
        \forall u \in U^{th}, & \mkern9mu \forall T_{id}, \notag\\ 
        \textbf{if} \mkern9mu (P_{u,t_{id}^{end},t_{m}^{ex}}^{plan} > P_{u,t_{id}^{end}+1,t_{m}^{ex}}^{plan}) \mkern9mu \textbf{then} & \quad \begin{cases}
            Q_{u,T_{id},t_{m}^{ex}}^{up,max} \gets 0\\
            Q_{u,T_{id},t_{m}^{ex}}^{up,min} \gets 0\\
            Q_{u,T_{id},t_{m}^{ex}}^{dn,max} \gets P_{u,t_{id}^{end},t_{m}^{ex}}^{plan} - P_{u,t_{id}^{end}+1,t_{m}^{ex}}^{plan}\\
            Q_{u,T_{id},t_{m}^{ex}}^{dn,min} \gets Q_{u,T_{id},t_{m}^{ex}}^{dn,max}
        \end{cases}
    \end{align}
\end{itemize}

\subsubsection{Hydraulic- and Storage-specific constraints}
\label{sec:bsp_orders_hydraulic_storage_constraints}
Specific constraints applied to hydraulic and storage units include:
\begin{itemize}
    \setlength\itemsep{-0.2em}
    \item Maximum ramping, although it is usually set to infinite for this kind of unit.
    \item Maximum and minimum storage level.
    \item For pumped hydraulic storage units, the transition duration between pumping and turbining states.
\end{itemize}

In contrast with thermic units, there is no combinatorial index to be considered, and each time step of the market time frame is considered individually.

\paragraph{Maximum ramping constraint}
\label{sec:bsp_orders_hydro_storage_max_ramp}
The maximum ramping constraint is verified for both upward and downward orders for hydraulic and storage units, when the maximum ramping $\Delta P^{max}$ is not equal to $0$ (which indicates an infinite ramping limit by convention):
\begin{subequations}
    \label{eq:bsp_orders_h_st_max_ramping}
    \begin{align}
        \centering
        & \forall u \in U^{h} \cup U^{st}, \mkern9mu \forall t \in T_{m}, \notag\\  
        Q_{u,t,t_{m}^{ex}}^{up,max} \gets min(& Q_{u,t,t_{m}^{ex}}^{up,max}, \notag\\
        & \Delta P_{u,t}^{max} * \Delta t_{m} - (P_{u,t,t_{m}^{ex}}^{plan} - P_{u,t-1,t_{m}^{ex}}^{plan}), \notag \\ 
        & \Delta P_{u,t_{id}^{end}}^{max} * \Delta t_{m} - (P_{u,t+1,t_{m}^{ex}}^{plan} - P_{u,t,t_{m}^{ex}}^{plan})) \label{eq:bsp_orders_upward_max_ramping_h_st} \\ 
        Q_{u,t,t_{m}^{ex}}^{dn,max} \gets min(& Q_{u,t,t_{m}^{ex}}^{dn,max}, \notag\\
        & \Delta P_{u,t}^{max} * \Delta t_{m} - (P_{u,t-1,t_{m}^{ex}}^{plan} - P_{u,t,t_{m}^{ex}}^{plan}), \notag \\ 
        & \Delta P_{u,t}^{max} * \Delta t_{m} - (P_{u,t,t_{m}^{ex}}^{plan} - P_{u,t+1,t_{m}^{ex}}^{plan})) \label{eq:bsp_orders_downward_max_ramping_h_st}
    \end{align}
\end{subequations}

\paragraph{Maximum and minimal storage level}
\label{sec:bsp_orders_hydro_storage_res_level}
These constraints ensure that the activation of reserves (and the stored energy evolution resulting from it) does not violate the maximum and minimum storage levels of hydraulic and storage units. Any stored energy evolution resulting from activations in balancing markets is modifying not only storage levels within $T_{m}$, but also all subsequent time steps until the operational time frame of the next day-ahead market\footnote{The day-ahead market is here taken as the reference, where the main part of power outputs is decided, and consequently where the main decisions shaping the storage levels of hydraulic and storage units are taken.}. 2 different situations arise, depending on the comparison between $t_{m}^{ex}$ and the hour $h_{DA}^{ex}$ at which the daily day-ahead market is executed:

\begin{itemize}
    \item If the current balancing market is occurring after $h_{DA}^{ex}$, it means that storage levels are already largely planned for the following day. In that situation, balancing orders are constrained not to violate reservoir level limits on the following day, and the associated time frame $T_{storage}$ is given in Equation \ref{eq:2.3.1.7_storage_level_timeframe_1}.
    \begin{equation}
        \label{eq:2.3.1.7_storage_level_timeframe_1}
        \textbf{if} \mkern6mu (hour(t_{m}^{ex}) > h_{DA}^{ex}) \mkern9mu \textbf{then} \quad T_{storage} = \{t_{m}^{start}, \mkern6mu t_{m}^{start} + \Delta t_{m}, \mkern6mu \dots, \mkern6mu t_{DA+1}^{end}\}
    \end{equation}
    Where $t_{DA+1}^{end}$ corresponds to the end of the following day, and the function $hour(date)$ simply returns the hour of $date$.
    \item If the current balancing market is occurring before $h_{DA}^{ex}$, then the storage time frame lasts until the end of the current day ($t_{DA}^{end}$):
    \begin{equation}
        \label{eq:2.3.1.7_storage_level_timeframe_2}
        \textbf{if} \mkern6mu (hour(t_{m}^{ex}) < h_{DA}^{ex}) \mkern9mu \textbf{then} \quad T_{storage} = \{t_{m}^{start}, \mkern6mu t_{m}^{start} + \Delta t_{m}, \mkern6mu \dots, \mkern6mu t_{DA}^{end}\}
    \end{equation}
\end{itemize}

Once the storage time frame is defined, $Q^{up,max}$ and $Q^{dn,max}$ are updated, potentially reduced if they would induce a violation of $E_{u,t, t_{m}^{ex}}^{max}$ or $E_{u,t, t_{m}^{ex}}^{min}$ for any time $t$ in this time frame:
\begin{subequations}
    \label{eq:storage_levels}
    \begin{align}
        \centering
        \forall u \in U^{h} & \cup U^{st}, \mkern9mu \forall t \in T_{m}, \mkern9mu \forall t_{st} \in T_{storage}, \notag\\
        Q_{u,t,t_{m}^{ex}}^{up,max} \gets & \min \biggl(Q_{u,t,t_{m}^{ex}}^{up,max}, \frac{\min\limits_{t' \in T_{m}}(E_{u,t', t_{m}^{ex}}^{stored}) - E_{u,t_{st}, t_{m}^{ex}}^{min}}{\Delta T_{m}/60} \biggl) \label{eq:max_storage_level} \\
        Q_{u,t,t_{m}^{ex}}^{dn,max} \gets & \min\biggl(Q_{u,t,t_{m}^{ex}}^{dn,max}, \frac{E_{u,t_{st}, t_{m}^{ex}}^{max} - \max\limits_{t' \in T_{m}}(E_{u,t', t_{m}^{ex}}^{stored})}{\Delta T_{m}/60} \biggl) \label{eq:min_storage_level}
    \end{align}
\end{subequations}

Finally, if the constraint is limiting $Q^{max}$, then it means that orders of each time $t \in T_{m}$ should be linked by an \textit{Exclusion} coupling with all orders of other time steps of $T_{m}$. If not, simultaneous activations across different time steps may lead to a violation of said constraint (see section \ref{sec:bsp_orders_coupling_excl_explain} for more details about the \textit{Exclusion} coupling type).

\paragraph{Pumping / turbining transition duration constraint}
\label{sec:bsp_orders_phs_transition}
This constraint is only applied to PHS units, that are modeled as storage equipments in ATLAS. It represents the duration required for the unit to switch from a pumping mode (when the unit is consuming energy) to a turbining mode (when the unit is producing energy), or conversely. During this transition phase, the power output of the unit must be equal to 0. By hypothesis, we assume that this duration is symmetrical in both directions (from pumping to turbining, or from turbining to pumping).

Because of the limited number of time steps in the current design of balancing markets (at most 4, for the RR market), it is basically impossible for a unit to switch between modes within the time frame of these markets. To keep a reasonable complexity, the module only considers that, if the transition constraint is active and applied to a unit, then it cannot switch mode. Formally, this means:
\begin{subequations}
    \label{eq:transition_duration}
    \begin{align}
        \centering
        & \forall u \in U^{phs}, \mkern9mu \forall t \in T_{m}, \mkern9mu \textbf{if} \mkern9mu (d_{u}^{trans} > \Delta t_{m}) \mkern9mu \textbf{then} \notag\\
        & \begin{cases}
            \textbf{if} \mkern9mu (P^{plan}_{u,t,t_{m}^{ex}} < 0) \mkern9mu \textbf{then} \quad Q_{u,t,t_{m}^{ex}}^{up,max} \gets \min(Q_{u,t,t_{m}^{ex}}^{up,max}, \mkern9mu |P^{plan}_{u,t,t_{m}^{ex}}|)\\
            \textbf{if} \mkern9mu (P^{plan}_{u,t,t_{m}^{ex}} > 0) \mkern9mu \textbf{then} \quad 
            Q_{u,t,t_{m}^{ex}}^{dn,max} \gets \min(Q_{u,t,t_{m}^{ex}}^{dn,max}, \mkern9mu |P^{plan}_{u,t,t_{m}^{ex}}|)
        \end{cases}
    \end{align}
\end{subequations}

\subsection{Setting $q_o^{max}$ and $q_o^{min}$ for all orders $o$}
\label{sec:bsp_orders_qmax_qmin}
For any unit $u \in U$, once the available power has been computed following all steps in section \ref{sec:bsp_orders_available_power_determination}, market orders $o$ are created. No strategic behavior is modeled regarding the amount of reserve offered on the balancing markets, meaning that each unit offers all the energy available.

In a general fashion, the following criteria are applied to determine if an upward (resp. downward) order should be formulated for any $t \in T_{m}$: the upper bound of the upward (resp. downward) available power has to be strictly positive ($Q_{u,t,t_{m}^{ex}}^{\bullet,max} > 0$) and, and it has to be greater than the lower bound ($Q_{u,t,t_{m}^{ex}}^{\bullet,max} > Q_{u,t,t_{m}^{ex}}^{\bullet,min}$).

The general case is described in Section \ref{sec:bsp_orders_general_qmax_qmin}. Once again, specific unit types require to be treated separately from the others:
\begin{itemize}
    \setlength\itemsep{-0.2em}
    \item For thermal units, a specific treatment is required to take into account the different combinatorial indexes as well as possible startup and shutdown orders (Section \ref{sec:bsp_orders_thermic_qmax_qmin}).
    \item For hydraulic units, the convention of both day-ahead and intraday markets is taken into account (see \cite{little_atlas_nodate}, Section 4.2.4). The range $[0,P^{max}_{u}]$ is divided into 7 fragments, each one priced according to a spread around the water value $WV_{u,t,E^{stored}_{u,t,t^{ex}_{m}}}$ to avoid an all or nothing effect (Section \ref{sec:bsp_orders_hydraulic_qmax_qmin}).
\end{itemize}

\subsubsection{General case}
\label{sec:bsp_orders_general_qmax_qmin}
For all storage, wind, photovoltaic and flexible load units, the characteristics $q^{max}$ and $q^{min}$ of an upward order $o_{t}^{up}$ are directly given by the available upward power (Equation \ref{eq:bsp_general_up_order_qmax}). The same principle is applied for downward orders (Equation \ref{eq:bsp_general_down_order_qmax})

\begin{equation}
    \label{eq:bsp_general_up_order_qmax}
    \forall u \in U^{st} \cup U^{w} \cup U^{pv} \cup U^{l}, \mkern9mu \forall t \in T_{m}, \mkern9mu \forall o_{t}^{up} \in O_{up}, \quad \begin{cases}
        \sigma_{o_{t}^{up}} = 1\\
        q_{o_{t}^{up}}^{min} = Q_{u,t,t_{m}^{ex}}^{up,min}\\
        q_{o_{t}^{up}}^{max} = Q_{u,t,t_{m}^{ex}}^{up,max}
    \end{cases}
\end{equation}

\begin{equation}
    \label{eq:bsp_general_down_order_qmax}
    \forall u \in U^{st} \cup U^{w} \cup U^{pv} \cup U^{l}, \mkern9mu \forall t \in T_{m}, \mkern9mu \forall o_{t}^{dn} \in O_{dn}, \quad \begin{cases}
        \sigma_{o_{t}^{dn}} = 1\\
        q_{o_{t}^{dn}}^{min} = Q_{u,t,t_{m}^{ex}}^{dn,min}\\
        q_{o_{t}^{dn}}^{max} = Q_{u,t,t_{m}^{ex}}^{dn,max}
    \end{cases}
\end{equation}

\subsubsection{Thermic units specific case}
\label{sec:bsp_orders_thermic_qmax_qmin}
For a given thermic unit $u \in U^{th}$, and for each combinatorial index $id \in ID_{u}$, all market orders $o_{t}$ are created for every time $t \in T_{id}$, when possible. Compared to the general case, two specific types of orders can be formulated for thermal units: 
\begin{itemize}
    \setlength\itemsep{-0.2em}
    \item If the tracking variable $\delta_{u,T_{id},t_{m}^{ex}}^{SU} = 1$, then the upward orders on index $id$ will all be startup orders.
    \item If the tracking variable $\delta_{u,T_{id},t_{m}^{ex}}^{SD} = 1$, then shutdown orders will be formulated in addition to usual downward orders.
\end{itemize}

The characteristics $q_o^{max}$ and $q_o^{min}$ of every order $o$ can be deduced from $Q_{u,T_{id},t_{m}^{ex}}^{\sigma^{R},max}$ and $Q_{u,T_{id},t_{m}^{ex}}^{\sigma^{R},max}$. The relation is straightforward for downward orders $o_{id}^{dn} \in O^{dn}$ (Equation \ref{eq:bsp_orders_th_down_order_qmax_qmin}), and for upward orders $o_{id}^{up} \in O^{up}$ that do not induce or cancel a startup (Equation \ref{eq:bsp_orders_th_up_order_qmax_qmin}):
\begin{itemize}[label={}]
    \item \begin{equation}
        \label{eq:bsp_orders_th_up_order_qmax_qmin}
        \forall t \in T_{id}, \quad  \textbf{if}  \mkern9mu (\delta_{u,T_{id},t_{m}^{ex}}^{SU} = 0) \mkern9mu \textbf{then} \quad \begin{cases}
            \sigma_{o_{t}^{up}} = -1\\
            q_{o_{t}^{up}}^{min} = Q_{u,T_{id},t_{m}^{ex}}^{up,min}\\
            q_{o_{t}^{up}}^{max} = Q_{u,T_{id},t_{m}^{ex}}^{up,max}
        \end{cases}
    \end{equation}
    \item \begin{equation}
        \label{eq:bsp_orders_th_down_order_qmax_qmin}
        \forall t \in T_{id}, \quad \begin{cases}
            \sigma_{o_{t}^{dn}} = 1\\
            q_{o_{t}^{dn}}^{min} = Q_{u,T_{id},t_{m}^{ex}}^{dn,min}\\
            q_{o_{t}^{dn}}^{max} = Q_{u,T_{id},t_{m}^{ex}}^{dn,max}
        \end{cases}
    \end{equation}
\end{itemize}

For upward orders, if a startup generating is required after verifying all constraints ($\delta_{u,T_{id},t_{m}^{ex}}^{SU} = 1$), two orders are actually formulated for each time step to have a proper representation of startup costs:
\begin{itemize}
    \item An indivisible order $o_{id}^{s1}$ between 0 and minimum power. For this first order, $q_{o_{id}^{s1}}^{min}$ and $q_{o_{id}^{s1}}^{max}$ are equal and correspond to the minimum power.
    \begin{equation}
        \label{eq:startup_order_bottom_q}
        \centering
        \forall u \in U^{th}, \mkern9mu \forall t \in T_{id}, \quad \textbf{if} \mkern9mu (\delta_{u,T_{id},t_{m}^{ex}}^{SU} = 1) \mkern9mu \textbf{then} \quad \begin{cases}
            \sigma_{o_{t}^{s1}} = -1\\
            q_{o_{t}^{s1}}^{min} = P_{u,t_{id}^{start}}^{min}\\
            q_{o_{t}^{s1}}^{max} = P_{u,t_{id}^{start}}^{min}
        \end{cases}
    \end{equation}

    \item A divisible order $o_{id}^{s2}$ that covers the entire power range between minimum power and maximum power. As a reminder, it is assumed that the unit can start anywhere in this range, without considering ramping constraints. Consequently, $q_{o_{id}^{s2}}^{min}$ and $q_{o_{id}^{s2}}^{max}$ are set as followed:
    \begin{equation}
        \label{eq:startup_order_top_q}
        \centering
        \forall u \in U^{th}, \mkern9mu \forall t \in T_{id}, \quad \textbf{if} \mkern9mu (\delta_{u,T_{id},t_{m}^{ex}}^{SU} = 1) \mkern9mu \textbf{then} \quad \begin{cases}
            \sigma_{o_{t}^{s2}} = -1\\
            q_{o_{t}^{s2}}^{min} = 0\\
            q_{o_{t}^{s2}}^{max} = P_{u,t_{id}^{start}}^{max} - P_{u,t_{id}^{start}}^{min}
        \end{cases}
    \end{equation}
\end{itemize}
Both startup orders are linked with a \textit{Parent Children}, $o_{id}^{s1}$ being the parent and $o_{id}^{s2}$ being the child (more information in \ref{sec:bsp_orders_coupling_pc_creation}).

Finally, all feasible shutdown orders $o_{id}^{shut} \in O^{shut}$ are formulated as indivisible orders:
\begin{equation}
    \label{eq:shutdown_order_q}
    \centering
    \forall u \in U^{th}, \mkern9mu \forall t \in T_{id}, \quad \textbf{if} \mkern9mu (\delta_{u,T_{id},t_{m}^{ex}}^{SD} = 1) \mkern9mu \textbf{then} \quad \begin{cases}
        \sigma_{o_{id}^{shut}} = 1\\
        q_{o_{id}^{shut}}^{min} = P_{u,t,t_{m}^{ex}}^{plan}\\
        q_{o_{id}^{shut}}^{max} = P_{u,t,t_{m}^{ex}}^{plan}
    \end{cases}
\end{equation}

Downward orders and shutdown orders formulated over overlapping combinatorial index are linked with an \textit{Exclusion} to avoid simultaneous activations (see Section \ref{sec:bsp_orders_coupling_excl_diff_indexes}).

\subsubsection{Hydraulic units specific case}
\label{sec:bsp_orders_hydraulic_qmax_qmin}
For hydraulic units, the available upward power and the available downward power are divided into fragments to keep a relevant level of consistency with both day-ahead and intraday market results. In these markets, the entire range between $0$ and the maximum power of each unit $u \in U^{h}$ was divided into 7 fragments, each one leading to an individual market order.

\begin{figure}[H]
    \centering
    \begin{subfigure}[b]{0.48\textwidth}
        \centering
        \includegraphics[width = \textwidth]{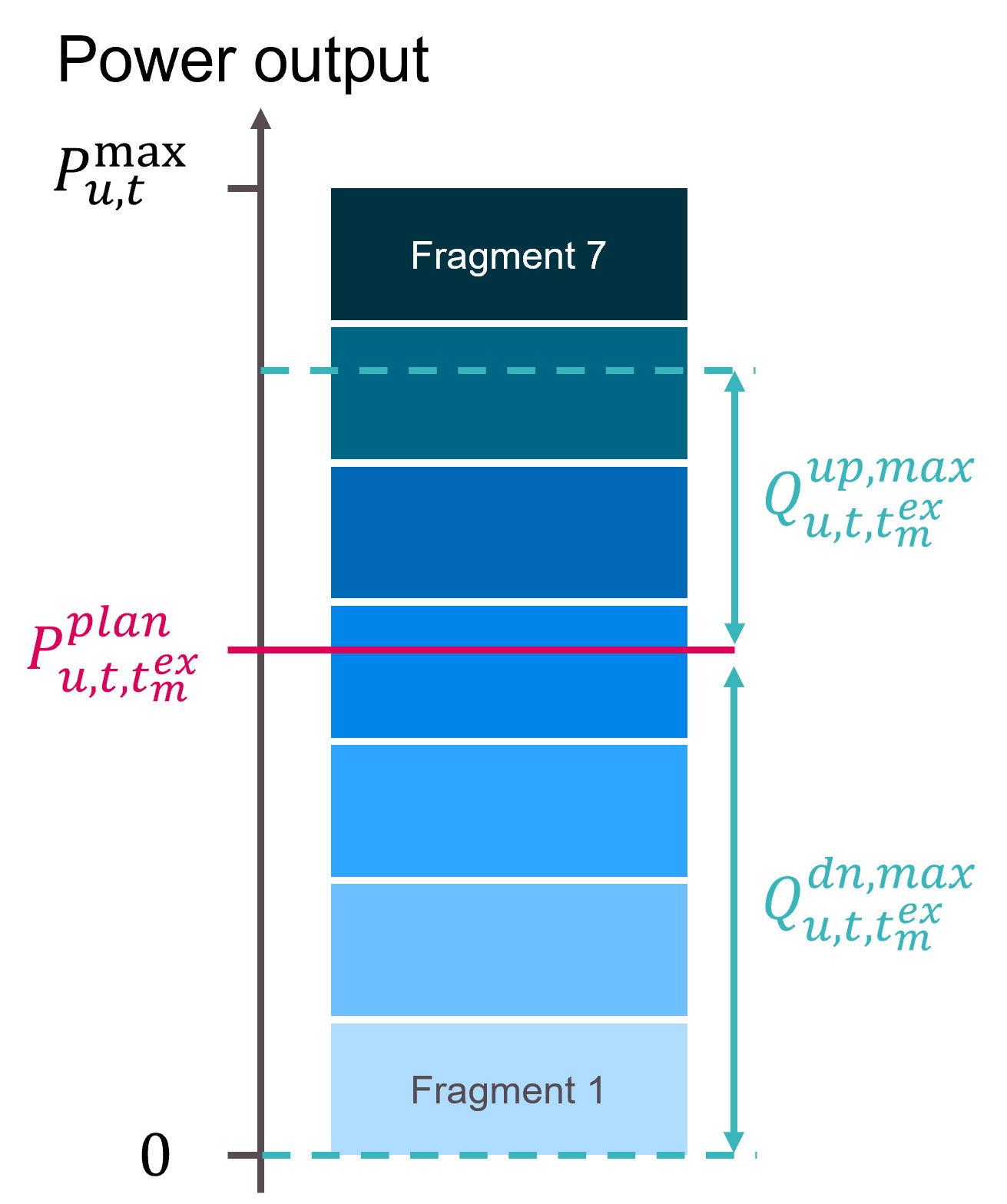}
        \caption{Identification of the power output state}
        \label{fig:bsp_orders_hydro_frag_qmax_1}
    \end{subfigure}
    \hfill
    \begin{subfigure}[b]{0.48\textwidth}
        \centering
        \includegraphics[width = \textwidth]{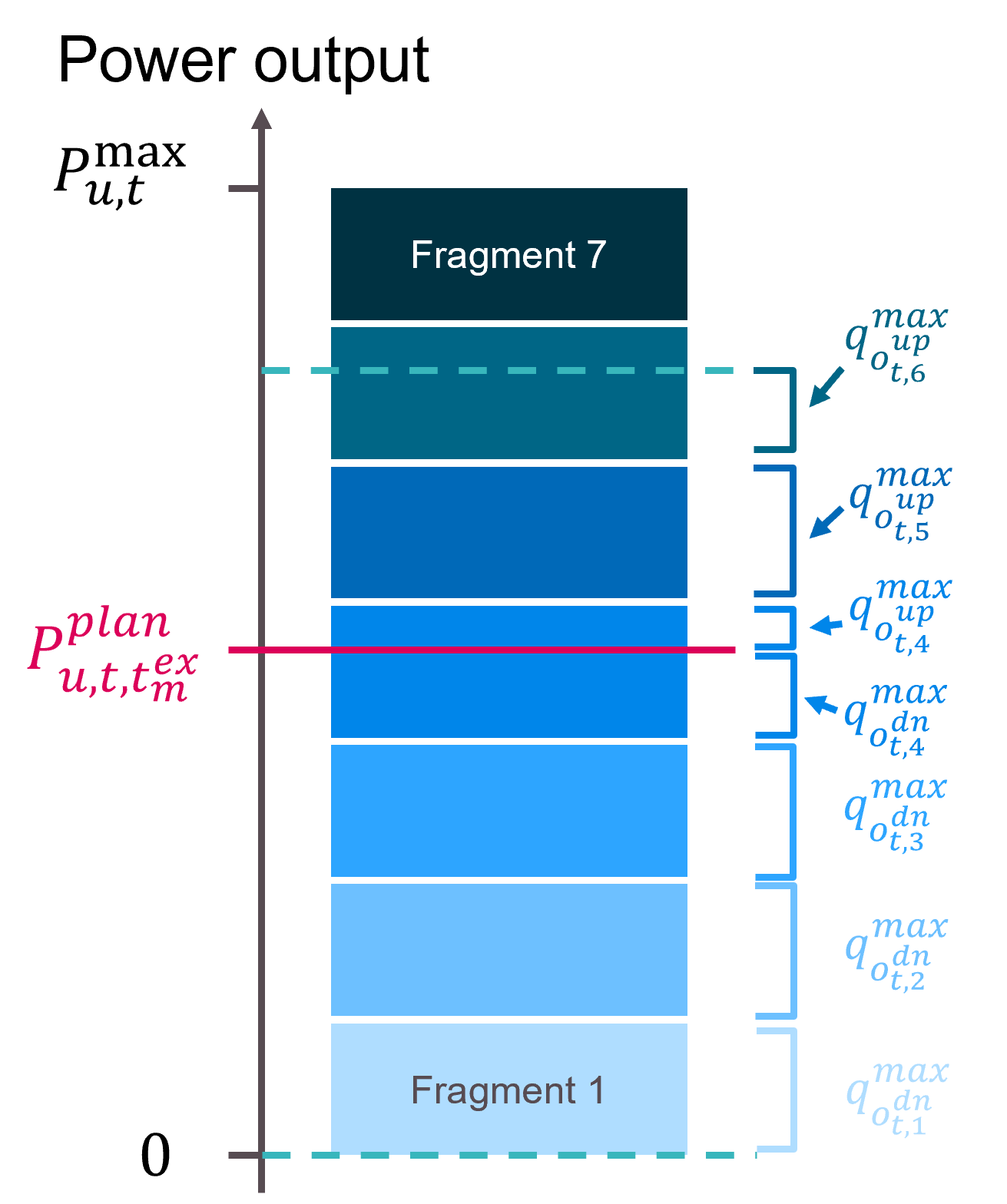}
        \caption{Computation of $q^{max}_{o}$}
        \label{fig:bsp_orders_hydro_frag_qmax_2}
    \end{subfigure}
    \caption{Example of the computation of orders volume for hydraulic units}
    \label{fig:bsp_orders_hydro_frag_qmax}
\end{figure}

In balancing markets, the same fragment division is kept. The last generation plan of the unit is used to identify where it stands in this division, and to deduce the subsequent upward and downward volumes that should be formulated. This computation is illustrated in Figures \ref{fig:bsp_orders_hydro_frag_qmax_1} and \ref{fig:bsp_orders_hydro_frag_qmax_2} in a general case, for a given time $t$.

In this example, the previously calculated available upward power does not allow the unit to reach its maximum power $P^{max}_{u,t}$, resulting in 3 upward orders of respective maximum quantities $q^{max}_{o^{up}_{t,i}}$, where $i$ is the fragment number associated with the order.

The minimum quantity $q^{min}_{o^{\bullet}_{t,\bullet}}$ of all of these orders is set to $0$.

\subsection{Order price determination}
\label{sec:bsp_orders_price_determination}
The price of the order $o^{\bullet}_{u}$ is determined based on the type of unit $u$, and the order direction $\bullet$.

\subsubsection{Thermic units}
\label{sec:bsp_orders_th_price}
For a thermal unit $u \in U^{th}$, 2 cases are possible depending on the value of the variable $\delta_{u,T_{id},t_{m}^{ex}}^{SU}$ tracking a startup order:
\begin{itemize}
    \item If the order is not part of a startup, then its price is computed as a combination of its variable costs $c_{u}^{var}$ and startup costs $c_{u}^{SU}$ (Equation \ref{eq:bsp_orders_price_th_classic}). Remark: a non startup order can still induce startup costs.
    \begin{align}
        \label{eq:bsp_orders_price_th_classic}
        \forall u \in U^{th}, \mkern9mu \forall o_{u,id} \in & O^{up} \cup O^{dn} \cup O^{shut}, \notag\\ 
        \textbf{if} \mkern9mu (\delta_{u,T_{id},t_{m}^{ex}}^{SU} = 0) \mkern9mu \textbf{then} & \quad p_{o_{u,id}} = c_u^{var} + \sigma_{u,T_{id}}^{SU} \mkern3mu * \mkern3mu \frac{c_u^{SU}}{Q_{u,T_{id},t_{m}^{ex}}^{max} * \Delta T_{id}}
    \end{align}

    \item If the order creates a startup, then it is separated into 2 parts as explained in Section \ref{sec:bsp_orders_thermic_startup}. The price of the first order $o_{id}^{s1}$, which corresponds to the start of the unit from 0 to its minimum power, includes the startup cost and is calculated in Equation \ref{eq:bsp_orders_price_th_su1}:
    \begin{align}
        \centering
        \label{eq:bsp_orders_price_th_su1}
        \forall u \in U^{th}, & \mkern9mu \forall T_{id}, \notag\\ 
        \textbf{if} \mkern9mu (\delta_{u,T_{id},t_{m}^{ex}}^{SU} = 1) \mkern9mu \textbf{then} \quad p_{o_{u,id}^{s1}} & = c_u^{var} + \sigma_{u,T_{id}}^{SU} \mkern3mu * \mkern3mu \frac{c_u^{SU}}{q_{o_{u,id}^{s1}}^{max} * \Delta T_{id}/60}
    \end{align}
    Here, startup costs are spread across all startup orders of the combinatorial index $id$, as is indicated by the division by $\Delta T_{id}/60$.

    The price of the second order $o_{id}^{s2}$ is directly given by the variable cost, as Equation \ref{eq:bsp_orders_price_th_su2} indicates:
    \begin{align}
        \centering
        \label{eq:bsp_orders_price_th_su2}
        \forall u \in U^{th}, & \mkern9mu \forall T_{id}, \mkern9mu \forall o \in O^{up}, \notag\\ 
        \textbf{if} \mkern9mu (\delta_{u,T_{id},t_{m}^{ex}}^{SU} = 1) & \mkern9mu \textbf{then} \quad p_{o_{u,id}^{s2}} = c_u^{var}
    \end{align}
\end{itemize}

\subsubsection{Hydraulic units}
\label{sec:bsp_orders_hydro_price}
For hydraulic units, the price of an order is computed using water value $WV_{u,t,E_{u,t,t_{m}^{ex}}^{stored}}$, which indicates the value of the current energy stored in the unit reservoir by estimating the opportunity cost of using it later over a certain period. In ATLAS, water values are preemptively computed over an entire year, for each hydraulic unit in the input dataset. This module then performs linear interpolations on both the storage level and the temporal axis to extract the storage marginal value corresponding to the situation at time $t$, for stored energy $E_{u,t,t_{m}^{ex}}^{stored}$.

In Section \ref{sec:bsp_orders_hydraulic_qmax_qmin}, the available upward and downward power was divided into orders corresponding to power fragments. In the day-ahead and intraday market processes, each fragment is associated with a given spread that is applied to the water value, and that will be noted $Spread_{i}$, with $i$ the fragment number. The same spreads are used to compute the price of balancing market orders:
\begin{equation}
    \label{eq:price_hyd}
    \forall u \in U^{h}, \mkern9mu \forall t\in T_{m}, \mkern9mu \forall o_{t,i} \in O^{up} \cup O^{dn}, \quad p_{o_{t,i}} = WV_{u,t,E_{u,t,t_{m}^{ex}}^{stored}} + Spread_{i}
\end{equation}

\subsubsection{Storage units}
\label{sec:bsp_orders_st_price}
For storage units, a daily variable cost is estimated during the day-ahead market (in the order formulation stage, see \cite{little_atlas_nodate}), and is used as a price reference for the balancing market orders. The price is then given by:
\begin{align}
    \label{eq:price_stor}
    \forall u \in U^{st}, \mkern9mu \forall t & \in T_{m},  \mkern9mu \forall o_{u} \in O^{up} \cup O^{dn}, \notag\\ 
    & p_{o_{u,t}} = c_{u,t}^{var}
\end{align}

\subsubsection{Wind, photovoltaic and flexible load units}
\label{sec:bsp_orders_w_pv_l_price}
Finally, for wind, photovoltaic and flexible load units units, the price of both upward and downward orders is directly given by the variable cost associated with each unit (Equation \ref{eq:bsp_orders_price_wind_pv_l}). 
\begin{align}\label{eq:bsp_orders_price_wind_pv_l}
    \forall u \in U^{w} \cup U^{pv} \cup U^{l}, \mkern9mu & t \in T_{m}, \mkern9mu \forall o_{u,t} \in O^{up} \cup O^{dn}, \notag\\
    p_{o_{u,t}} & = c_u^{var}
\end{align}

For wind and photovoltaic units, variable costs are usually set to 0, although the user can choose to set a different value at will. This can be used to represent specific market incentives for instance. For flexible load units, variable costs are often set exogenously. For example, power-to-gas units usually take as a reference the price of the gas market.

\section{Coupling links between orders} \label{sec:bsp_orders_coupling_links}
Once all their characteristics are determined by previous steps, market orders are formulated by creating instances of the Order class. However, one last phase is required to complete the balancing market orders formulation. As the aforementioned characteristics are not sufficient to accurately translate operating constraints in market orders, coupling links must be created between them in many cases. The types of coupling available in ATLAS are described in Section \ref{sec:bsp_orders_coupling_types}, and Section \ref{sec:bsp_orders_coupling_creation_general} explains how they are used in the module.\\

\subsection{Types of coupling}
\label{sec:bsp_orders_coupling_types}
This module uses 3 different types of coupling instances, which are explained hereafter.

\subsubsection{\textit{Identical ratio}}
\label{sec:bsp_orders_coupling_idr_explain}
If several orders are coupled by \textit{Identical ratio}, it will force the Clearing stage to activate them with the same acceptance ratio. This ratio is calculated based on $q_{o_{i}}^{max}$ and $q_{o_{i}}^{min}$ of each order $o_{i}$, and is shown in equation \ref{eq:id_ratio_example} in a simplified example with 2 orders:
\begin{equation}
    \label{eq:id_ratio_example}
    \centering
    \forall c^{idr} \in C^{idr}, \mkern9mu \forall \{o_{1},o_{2}\} \in \mkern3mu c^{idr}, \quad
    \frac{q_{o_{1}}^{acc}-q_{o_{1}}^{min}}{q_{o_{1}}^{max}-q_{o_{1}}^{min}} = \mkern9mu \frac{q_{o_{2}}^{acc}-q_{o_{2}}^{min}}{q_{o_{2}}^{max}-q_{o_{2}}^{min}}
\end{equation}

\subsubsection{\textit{Exclusion}}
\label{sec:bsp_orders_coupling_excl_explain}
If several orders are linked by an \textit{Exclusion} coupling, the Clearing will only be able to activate at most one of these orders (even if it is partially accepted). If we note $\delta_{o_{i}}^{acc}$ the boolean indicating if an order $o_{i}$ is accepted by the Clearing (meaning that $\delta_{o_{i}}^{acc} = 1 \mkern9mu if  \mkern9mu q_{o_{i}}^{acc} > 0$ and $\delta_{o_{i}}^{acc} = 0 \mkern9mu if  \mkern9mu q_{o_{i}}^{acc} = 0$), equation \ref{eq:exclusion_example} gives an example of this coupling link for two orders:
\begin{equation}
    \label{eq:exclusion_example}
    \centering
    \forall c^{excl} \in C^{excl}, \mkern9mu \forall \{o_{1},o_{2}\} \in c^{excl}, \quad \delta_{o_{1}}^{acc} + \delta_{o_{2}}^{acc} \leq 1 
\end{equation}

\subsubsection{\textit{Parent Children}}
\label{sec:bsp_orders_coupling_pc_explain}
A coupling of type \textit{Parent Children} contains a list of orders $\{o_{1}, ..., o_{N}\}$. The first order $o_{1}$ is the parent by convention, and the rest are the children. This coupling type forces the Clearing to activate (at least partially) the parent order if at least one child is activated, which is described by equation \ref{eq:pc_example}:
\begin{equation}
    \label{eq:pc_example}
    \centering
    \forall c^{pc} \in C^{pc}, \mkern9mu \forall \{o_{1}, ..., o_{N}\} \in c^{pc}, \quad \delta_{o_{1}}^{acc} \geq \min(1, \sum \limits_{n=2}^{N} \delta_{o_{n}}^{acc})
\end{equation}

\subsection{Creation of coupling links}
\label{sec:bsp_orders_coupling_creation_general}

\subsubsection{\textit{Parent Children} coupling between both parts of a startup order}
\label{sec:bsp_orders_coupling_pc_creation}
The two orders $o^{s1}$ and $o^{s2}$ that form a startup are linked with a \textit{Parent Children} coupling instance in the following way: the indivisible order $o^{s1}$ between 0 and the minimum power of the unit is the parent order, and the other is the child (Equation \ref{eq:bsp_orders_th_startup_pc}). 
\begin{equation}
    \label{eq:bsp_orders_th_startup_pc}
    \centering
    \forall u \in U^{th}, \mkern9mu \forall T_{id} \mkern6mu | \mkern6mu \delta_{u,T_{id},t_{m}^{ex}}^{SU} = 1, \mkern9mu \forall t \in T_{id}, \quad \exists \mkern3mu c^{pc} \in C^{pc} \mkern6mu | \mkern6mu \{o_{u,t}^{s1}, o_{u,t}^{s2}\} \in c^{pc}
\end{equation}

Consequently, the Clearing has to activate the bottom order (entirely, because of its indivisibility--$q^{min}_{o_{u,t}^{s1}} = q^{max}_{o_{u,t}^{s1}}$) in order to activate the top one.

\subsubsection{\textit{Identical Ratio} coupling between orders of the same combinatorial index}
\label{sec:bsp_orders_coupling_idr_same_index}
For thermic units, all orders from the same combinatorial index $id$ are linked with an \textit{Identical Ratio} coupling, to represent the fact that they are meant to be a single order over several time steps (Equation \ref{eq:bsp_orders_th_combi_idr}). The only exception is the bottom part $o^{s1}$ of startup orders. Since all bottom parts are indivisible, there is no need for an additional coupling between them as they are all fully activated, and only top orders $o^{s2}$ are linked (Equation \ref{eq:bsp_orders_th_combi_idr_startup}).
\begin{equation}
    \label{eq:bsp_orders_th_combi_idr}
    \centering
    \forall u \in U^{th}, \mkern9mu \forall T_{id} \mkern6mu | \mkern6mu \delta_{u,T_{id},t_{m}^{ex}}^{SU} = 0, \quad \exists \mkern3mu c^{idr} \in C^{idr} \mkern6mu | \mkern9mu \forall o_{u} \in O_{u,id}, \mkern9mu o_{u} \in c^{idr}
\end{equation}

\begin{equation}
    \label{eq:bsp_orders_th_combi_idr_startup}
    \centering
    \forall u \in U^{th}, \mkern9mu \forall T_{id} \mkern6mu | \mkern6mu \delta_{u,T_{id},t_{m}^{ex}}^{SU} = 1, \quad \exists \mkern3mu c^{idr} \in C^{idr} \mkern6mu | \mkern9mu \forall o_{u}^{s2} \in O_{u,id}, \mkern9mu o_{u}^{s2} \in c^{idr}
\end{equation}

\subsubsection{\textit{Exclusion} between orders of different combinatorial indexes formulated on the same time period}
\label{sec:bsp_orders_coupling_excl_diff_indexes}
The order formulation on thermic units can create several market orders of the same direction in the same time period that belong to different combinatorial indexes. These orders are meant to provide flexibility regarding operating constraints, but should never be activated together by the Clearing. This is enforced by creating \textit{Exclusion} couplings between them (Equation \ref{eq:bsp_orders_th_combi_excl}), while excluding startup orders linked with a \textit{Parent Children} coupling.
\begin{equation}
    \label{eq:bsp_orders_th_combi_excl}
    \centering
    \forall u \in U^{th}, \mkern9mu t \in T_{m}, \quad \begin{cases}
        \exists \mkern3mu c^{excl}_{up} \in C^{excl} \mkern6mu | \mkern9mu \biggl(\forall o_{u,t} \in O_{u,t} \mkern6mu | \mkern9mu (\sigma_{o_{u,t}} = -1) \mkern6mu \& \mkern6mu (o_{u,t} \notin c^{pc}), \mkern9mu o_{u,t} \in c^{excl}_{up}\biggl)\\
        \exists \mkern3mu c^{excl}_{dn} \in C^{excl} \mkern6mu | \mkern9mu \biggl(\forall o_{u,t} \in O_{u,t} \mkern6mu | \mkern9mu (\sigma_{o_{u,t}} = 1) \mkern6mu \& \mkern6mu (o_{u,t} \notin c^{pc}), \mkern9mu o_{u,t} \in c^{excl}_{dn}\biggl)
    \end{cases}
\end{equation}

Remark: this does not concern orders of opposite directions. If their price is set correctly, it should prevent the Clearing from activating them simultaneously.

\subsubsection{\textit{Exclusion} between consecutive orders of opposite directions}
\label{sec:bsp_orders_coupling_excl_consec_up_down}
\begin{figure}[H]
    \centering
    \includegraphics[width = 0.8\textwidth]{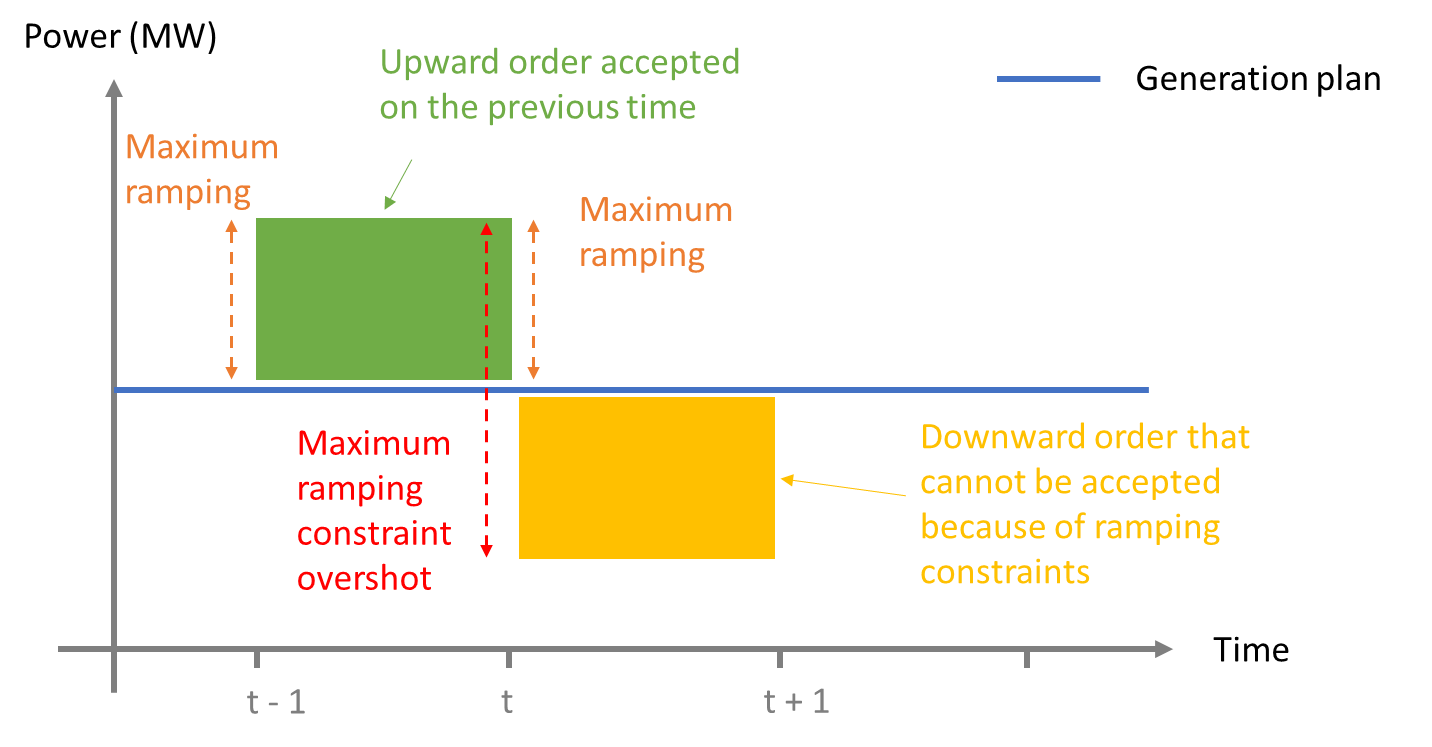}
    \caption{Consecutive opposite orders violating gradient constraints}
    \label{fig:consec_opp_orders_violating_constraint}
\end{figure}

The maximum ramping constraints may be violated if the Clearing activates consecutive orders of opposite directions (see Figure \ref{fig:consec_opp_orders_violating_constraint}). Indeed, this module tries to maximize the quantity offered by each unit, it can frequently correspond to the quantity directly feasible by its maximum ramping constraints.

Consequently, an \textit{Exclusion} coupling is created between the first order of a combinatorial index (so at $t_{id}^{start}$) and every order of the opposite direction formulated on the previous time period (at $t_{id}^{start}- 1$). The same operation is done for the last order (at $t_{id}^{end}$) and orders of the next time period ($t_{id}^{end} + 1$).

\begin{equation}
    \label{eq:bsp_orders_th_successive_excl}
    \centering
    \forall u \in U^{th}, \mkern9mu \forall T_{id}, \quad \begin{cases}
        \forall o_{u} \in O_{u} \mkern6mu | \mkern9mu \biggl((t^{start}_{o_{u}} = t^{start}_{id} - 1) \mkern6mu \& \mkern6mu (\sigma_{o_{u}} \neq \sigma_{o_{u,id}^{start}}) \biggl), \quad \exists \mkern3mu c^{excl} \in C^{excl} \mkern6mu | \mkern9mu \{o_{u,id}^{start}, o_{u}\}  \in c^{excl}\\
        \forall o_{u} \in O_{u} \mkern6mu | \mkern9mu \biggl((t^{start}_{o_{u}} = t^{end}_{id} + 1) \mkern6mu \& \mkern6mu (\sigma_{o_{u}} \neq \sigma_{o_{u,id}^{end}}) \biggl), \quad \exists \mkern3mu c^{excl} \in C^{excl} \mkern6mu | \mkern9mu \{o_{u,id}^{end}, o_{u}\}  \in c^{excl}
    \end{cases} 
\end{equation}

Where $o_{u,id}^{start}$ is the order in $id$ with the start date equal to the first date of $id$ ($t^{start}_{o_{u,id}^{start}} = t^{start}_{id}$), and $o_{u,id}^{end}$ the corresponding order for the last date of $id$ ($t^{start}_{o_{u,id}^{end}} = t^{end}_{id}$).

\subsubsection{Orders completely exclusive}
\label{sec:bsp_orders_coupling_excl_completely}
Certain orders are considered to be completely exclusive, which means that they need to be exclusive to all other orders of the same direction, regardless of the time period. These orders are those for which:

\begin{itemize}
    \item A storage capacity constraint is limiting the quantity offered. Since the orders of the combinatorial index are formulated so that the constraint is respected on this index exactly, and not on other time periods of the balancing time frame, it would conflict with another activation (in the same direction) within the balancing time frame.
    \item The minimum stable power duration constraint is active. For a given index, the unit power must remain unchanged to be sure that the constraint is respected.
\end{itemize}

Eventually, \textit{Exclusion} coupling instances are created between every order completely exclusive and all other orders of the same direction.

\chapter{Balancing Market - TSO Orders} 
\label{ch:Balancing_Market_TSO_Orders}

\section{Overview}
\label{sec:TSO Orders - Overview}

The objective of this module is to formulate TSO balancing energy market orders, for RR or mFRR markets. Submitted market orders are required to display all information indicated in Table \ref{tab:tso_orders_nomenclature_orders}. The general process described in this chapter is repeated for each individual TSO in the power system, and consists of two main steps: the computation of the balancing need in the control area $ca$ of the studied TSO (Section \ref{sec:TSO needs}), followed by the computation of orders prices and volumes depending on the volume of need previously computed and on the bidding strategy parameter chosen for the TSO (Section \ref{sec:Pricing strategies}). The bidding step is designed to be flexible in multiple aspects.

This module and the bidding strategies previously presented can be used for formulating both RR or mFRR market orders, depending on the parameters chosen by the user. Note that if the simulated market is mFRR, the alternative chosen should logically not be the mFRR market (but the aFRR market is still a relevant alternative). 

Finally, as the process is the same for every TSO, it is assumed in all following sections that a specific TSO is chosen, along with its control area $ca \in CA$. This control area may include several market areas $z \in Z_{ca}$. 

\subsection{Nomenclature}
\label{sec:tso_orders_nomenclature}
The following notations are used in this chapter. Some elements correspond to a specific type, which is indicated in bold and italics:
\begin{itemize}
    \item \textbf{\textit{Parameter}} refers to a parameter, which is indicated by the user before the execution of the module.
    \item \textbf{\textit{Input data}} refers to an element that is extracted from the input dataset.
    \item \textbf{\textit{Variable}} refers to an optimization variable.
\end{itemize}

Remark: For sets, the notation $A_{b}$ refers to the subset of $A$ linked with variable $b$. For instance, $Z_{ca}$ indicates the subset of market areas belonging to the control area $ca$.

\renewcommand{\arraystretch}{1.3} 
\begin{longtable}[H]{!{\color{Grey}\vrule}>{\centering\arraybackslash}p{3cm} !{\color{Grey}\vrule} p{13cm}!{\color{Grey}\vrule} }
    \arrayrulecolor{Grey} \hline
    \rowcolor{Grey} \multicolumn{2}{|c|}{\large{\textbf{Sets and global market notations}}} \label{tab:tso_orders_nomenclature_sets}\\ 
    \hline
    \rowcolor{Grey} 
    \textbf{Notation} & \textbf{Description}\\
    \hline
    $CA$ & Set of all control areas (or control blocks)\\
    \hline
    $Z$ & Set of all market areas. $Z_{ca}$ indicates all market areas within control area $ca \in CA$\\
    \hline
    $PF$ & Set of all portfolios. $PF_{z}$ indicates all portfolios in market area $z \in Z$\\
    \hline 
    $U$ & Set of all units. $U_{z}$ indicates units in market area $z \in Z$, and $U_{pf}$ indicates units in portfolio $pf \in PF$\\
    \hline 
    $U^{unit\_type}$ & Set of all units of type $unit\_type \in [g, l, th, h, st, w, pv]$. ($g$ = generation, $l$ = flexible load, $th$ = thermal, $h$ = hydraulic, $st$ = storage, $w$ = wind, $pv$ = photovoltaic)\\
    \hline 
    $O$ & Set of all market orders\\
    \hline 
    $O^{up}$ & Set of all upward market orders\\
    \hline 
    $O^{dn}$ & Set of all downward market orders\\
    \hline 
    $C^{excl}$ & Set of all coupling instances of type $Exclusion$\\
    \hline
    $m$ & Type of market simulation, amongst $\{RR, \mkern6mu mFRR\}$\\
    \hline
\end{longtable}

\begin{longtable}[H]{!{\color{Grey}\vrule}>{\centering\arraybackslash}p{3cm} !{\color{Grey}\vrule} p{10.5cm}!{\color{Grey}\vrule} >{\centering\arraybackslash} p{2cm}!{\color{Grey}\vrule}}
    \arrayrulecolor{Grey} \hline
    \rowcolor{Grey} \multicolumn{3}{|c|}{\large{\textbf{Temporal variables}}} \label{tab:tso_orders_nomenclature_temporal}\\
    \hline
    \rowcolor{Grey} 
    \textbf{Notation} & \textbf{Description} & \textbf{Units}\\
    \hline
    $t_{m}^{ex}$ & Execution date of the market $m \in \{RR, mFRR\}$. \textbf{\textit{Parameter}} & -\\
    \hline
    $t_{m}^{start}$ & Start date of the effective period of the market $m$. \textbf{\textit{Parameter}} & -\\
    \hline
    $t_{m}^{end}$ & End date of the effective period of the market $m$. \textbf{\textit{Parameter}} & -\\
    \hline
    $\Delta t_{m}$ & Time step of the market $m$. \textbf{\textit{Parameter}} & min\\
    \hline
    $T_{m}$ & Effective time frame of the market $m$, i.e. $T_{m} = [t^{start}_m, t^{start}_m + \Delta t_{m},\, \dots \,, t^{end}_m - \Delta t_{m}]$ & -\\
    \hline
\end{longtable}

\begin{longtable}[H]{!{\color{Grey}\vrule}>{\centering\arraybackslash}p{3cm} !{\color{Grey}\vrule} p{10.5cm}!{\color{Grey}\vrule} >{\centering\arraybackslash} p{2cm}!{\color{Grey}\vrule}}
    \arrayrulecolor{Grey} \hline
    \rowcolor{Grey} \multicolumn{3}{|c|}{\large{\textbf{TSO-specific characteristics}}} \label{tab:tso_orders_nomenclature_tso_charac}\\
    \hline
    \rowcolor{Grey} 
    \textbf{Notation} & \textbf{Description} & \textbf{Units}\\
    \hline
    $bn_{m,ca,t,t_{m}^{ex}}$ \newline (short: $\mkern3mu bn_{t}$) & Balancing needs for market $m$ in control area $ca \in CA$ for time $t \in T_{m}$, seen from $t_{m}^{ex}$. For readability, it is shortened as $bn_{t}$ & MW\\
    \hline
    $\sigma_{ca,t,t_{m}^{ex}}^{bn}$ \newline (short: $\mkern3mu \sigma^{bn}_{t}$) & Binary parameter indicating the direction of TSO (of area $ca$) balancing needs at time $t$ seen from $t_{m}^{ex}$. $\sigma_{ca,t,t_{m}^{ex}}^{bn} = up \mkern6mu if \mkern6mu bn_{m,ca,t,t_{m}^{ex}} > 0$, and $\sigma_{ca,t,t_{m}^{ex}}^{bn} = dn$ otherwise. For readability, it is shortened as $\sigma^{bn}_{t}$& - \\
    \hline
    $\delta_{ca}^{for}$ & Binary parameter indicating if TSO associated with control area $ca \in CA$ takes into account forecasts errors in its imbalance needs computation. \textbf{\textit{Parameter}} & - \\
    \hline
    $\delta_{ca}^{elas}$ & Binary parameter indicating if the demand of TSO associated with control area $ca \in CA$ is formulated as elastic ($\delta_{ca}^{elas} = 1$) or not ($\delta_{ca}^{elas} = 0$). \textbf{\textit{Parameter}} & - \\
    \hline
    $alt_{ca}$ & Alternative chosen to compute prices and volumes of the demand of TSO associated with control area $ca \in CA$. Possible values are $\{"mFRRalt", "FrBMalt"\}$. \textbf{\textit{Parameter}}. & - \\
    \hline
    $V^{s}$ & Maximum quantity of each slice of TSO demand formulated on markets. Only relevant if $\delta_{ca}^{elas} = 1$. \textbf{\textit{Parameter}} & MW \\
    \hline
    $\delta_{ca}^{risk}$ & Binary parameter indicating if the demand of TSO associated with control area $ca \in CA$ is taking into account volume-based risk aversion ($\delta_{ca}^{risk} = 1$) or not ($\delta_{ca}^{risk} = 0$). Only relevant if $\delta_{ca}^{elas} = 1$. \textbf{\textit{Parameter}} & - \\
    \hline
    $\epsilon_{ca}^{alt_{ca}}$ & List of quantiles of the distribution of the TSO forecast error function of area $ca$ associated with alternative $alt_{ca}$. Only relevant if $\delta_{ca}^{risk} = 1$. \textbf{\textit{Parameter}} & - \\
    \hline
\end{longtable}

\begin{longtable}[H]{!{\color{Grey}\vrule}>{\centering\arraybackslash}p{3cm} !{\color{Grey}\vrule} p{10.5cm}!{\color{Grey}\vrule} >{\centering\arraybackslash} p{2cm}!{\color{Grey}\vrule}}
    \arrayrulecolor{Grey} \hline
    \rowcolor{Grey} \multicolumn{3}{|c|}{\large{\textbf{Zonal-specific characteristics}}} \label{tab:tso_orders_nomenclature_zonal_charac}\\
    \hline
    \rowcolor{Grey} 
    \textbf{Notation} & \textbf{Description} & \textbf{Units}\\
    \hline
    $q_{m,ca,t}^{\sigma^{bn}_{t},max}$ & Overall quantity for orders in direction opposite to $\sigma^{bn}_{t}$ formulated by BSPs included in control area $ca$ on market $m$, at time $t$ & MW\\
    \hline
    $(\Delta q)^{bal}_{z,t}$ & Commercial (power) balance of area $z \in Z$ at time $t$, equal to the sum of all power exports minus the sum of all power imports & MW \\
    \hline
    $\rho_{m, ca}^{FrBMalt}$ & Percentage of BSP available capacity in area $ca$ that is not submitted on the market $m$, and is consequently directly sent to the FrBM market. Only relevant if $\delta_{ca}^{elas} = 1$, with $alt_{ca} = "FrBMalt"$. \textbf{\textit{Parameter}} & - \\
    \hline
\end{longtable}

\begin{longtable}[H]{!{\color{Grey}\vrule}>{\centering\arraybackslash}p{3cm} !{\color{Grey}\vrule} p{10.5cm}!{\color{Grey}\vrule} >{\centering\arraybackslash} p{2cm}!{\color{Grey}\vrule}}
    \arrayrulecolor{Grey} \hline
    \rowcolor{Grey} \multicolumn{3}{|c|}{\large{\textbf{Global unit characteristics}}} \label{tab:tso_orders_nomenclature_unit_charac}\\
    \hline
    \rowcolor{Grey} 
    \textbf{Notation} & \textbf{Description} & \textbf{Units}\\
    \hline
    $P_{u,t,t_{m}^{ex}}^{plan}$ & Power output of unit $u \in U$ at time $t \in T_{m}$, seen from time $t_{m}^{ex}$ & MW\\
    \hline
    $P_{u,t}^{max}$ & Maximum power output of unit $u \in U$ at time $t \in T_{m}$ & MW\\
    \hline
    $P_{u,t}^{min}$ & Minimum power output of unit $u \in U$ at time $t \in T_{m}$ & MW\\
    \hline
    $R_{ut,t_{m}^{ex}}^{m^{R}, \sigma^{R}}$ & Procured reserves of type $m^{R} \in [FCR, aFRR, mFRR, RR]$ in direction $\sigma^{R} \in [up, dn]$  on unit $u$ at time $t$, seen from $t_{m}^{ex}$ & -\\
    \hline
\end{longtable}

\begin{longtable}[H]{!{\color{Grey}\vrule}>{\centering\arraybackslash}p{3cm} !{\color{Grey}\vrule} p{10.5cm}!{\color{Grey}\vrule} >{\centering\arraybackslash} p{2cm}!{\color{Grey}\vrule}}
    \arrayrulecolor{Grey} \hline
    \rowcolor{Grey} \multicolumn{3}{|c|}{\large{\textbf{Hydraulic- and Storage-specific unit characteristics}}} \label{tab:tso_orders_nomenclature_hydro_storage}\\
    \hline
    \rowcolor{Grey} 
    \textbf{Notation} & \textbf{Description} & \textbf{Units}\\
    \hline
    $E_{u,t,t_{m}^{ex}}^{stored}$ & Stored energy in the reservoir of unit $u \in U^{h} \, \cup \, U^{st}$ at time $t \in T_{m}$, seen from time $t_{m}^{ex}$ & MWh\\
    \hline
    $E_{u,t}^{max}$ & Maximum storage level in the reservoir of unit $u \in U^{h} \, \cup \, U^{st}$ at time $t \in T_{m}$ & MWh\\
    \hline
    $E_{u,t}^{min}$ & Minimum storage level in the reservoir of unit $u \in U^{h} \, \cup \, U^{st}$ at time $t \in T_{m}$ & MWh\\
    \hline
    $d_{u}^{tran}$ & Transition duration between pumping and turbining of unit $u$ of type Pumped Hydraulic Storage (PHS), that are modeled as storage units in ATLAS (i.e. $u \in U^{st}$) & min\\
    \hline
    $WV_{u,t,E_{u,t,t_{m}^{ex}}^{stored}}$ & Marginal storage value of unit $u  \in U^{h}$ at time $t$, for stored energy level $E_{u,t,t_{m}^{ex}}^{stored}$ & €/MWh\\
    \hline
\end{longtable}

\begin{longtable}[H]{!{\color{Grey}\vrule}>{\centering\arraybackslash}p{3cm} !{\color{Grey}\vrule} p{10.5cm}!{\color{Grey}\vrule} >{\centering\arraybackslash} p{2cm}!{\color{Grey}\vrule}}
    \arrayrulecolor{Grey} \hline
    \rowcolor{Grey} \multicolumn{3}{|c|}{\large{\textbf{Load-, Photovoltaic- and Wind-specific unit characteristics}}} \label{tab:tso_orders_nomenclature_pv_wind}\\
    \hline
    \rowcolor{Grey} 
    \textbf{Notation} & \textbf{Description} & \textbf{Units}\\
    \hline
    $P_{u,t,t_{m}^{ex}}^{for}$ & Forecast of the maximum power output of unit $u \in \{U^{w}, U^{pv}, U^{l}\}$ at time $t \in T_{m}$ & MW\\
    \hline
    $Curt_{u}$ & Percentage of curtailed power allowed on unit $u \in \{U^{w}, U^{pv}\}$ & - \\
    \hline
\end{longtable}

\begin{longtable}[!ht]{!{\color{Grey}\vrule}>{\centering\arraybackslash}p{3cm} !{\color{Grey}\vrule} p{13cm}!{\color{Grey}\vrule} }
    \arrayrulecolor{Grey} \hline
    \rowcolor{Grey} \multicolumn{2}{|c|}{\large{\textbf{Market order characteristics}}} \label{tab:tso_orders_nomenclature_orders}\\ 
    \hline
    \rowcolor{Grey} 
    \textbf{Notation} & \textbf{Meaning}\\ \hline
    $p_o$ & Price of order $o$ \\
    \hline
    $q_o^{min}$ & Minimum quantity of power offered for order $o$ \\
    \hline
    $q_o^{max}$ & Maximum quantity of power offered for order $o$ \\
    \hline
    $t_o^{start}$ & Start date of order \\
    \hline
    $t_o^{end}$ & End date of order \\
    \hline
    $t_o^{ex}$ & Execution date of order \\
    \hline
    $\sigma_{o}$ & Sale/Purchase indicator, $\sigma = 1$ for purchase, -1 for sale \\ 
    \hline
    $\delta_o^{TSO}$ & Binary variable indicating if order $o$ is from a TSO ($\delta_o^{TSO} = 1$) or not. \\
    \hline
    $\delta_{o}^{acc}$ & Binary variable indicating if order $o$ is activated by the Clearing stage ($\delta_{o}^{acc} = 1$ if $q_o^{acc} > 0$)\\
    \hline
\end{longtable}

\begin{longtable}[H]{!{\color{Grey}\vrule}>{\centering\arraybackslash}p{3cm} !{\color{Grey}\vrule} p{10.5cm}!{\color{Grey}\vrule} >{\centering\arraybackslash} p{2cm}!{\color{Grey}\vrule}}
    \arrayrulecolor{Grey} \hline
    \rowcolor{Grey} \multicolumn{3}{|c|}{\large{\textbf{Other notations}}} \label{tab:tso_orders_nomenclature_others}\\
    \hline
    \rowcolor{Grey} 
    \textbf{Notation} & \textbf{Description} & \textbf{Units}\\
    \hline
    $VoLL$ & Value of loss load & €/MWh\\
    \hline
    $E^{VoLL}_{ca,t}$ & Loss load energy in control area $ca$ at time $t$ & MWh\\
    \hline
    $E_{ca,t}^{spill}$ & Spillage power in control area $ca$ at time $t$ & MW\\
    \hline
\end{longtable}

\renewcommand{\arraystretch}{1.1} 

\section{TSO balancing needs computation}
\label{sec:TSO needs}
The module studies independently each TSO in the power system extracted from the input marker, and aims at formulating its balancing market orders. To do so, the first step is to compute its balancing needs $bn_{t}$ for every time $t$ in the entire effective time frame $T_{m}$ of the market $m$, seen from $t_{m}^{ex}$. They correspond to the forecasted imbalance in its control area $ca \in CA$ (that may be composed of several market areas), computed based on the planned power output of all generation and consumption units in $ca$. This calculation is not trivial for several reasons: 
\begin{itemize}
    \item A specific restriction exists for balancing markets: in a nominal state, the volume of TSO needs submitted to the market in a given direction $\sigma_{t}^{bn}$ cannot exceed $q_{m,ca,t}^{\sigma_{t}^{bn},max}$ the sum of reserves offered by BSPs in $ca$ in the opposite direction to $\sigma_{t}^{bn}$. This models the actual RR market rule stating that a TSO cannot submit a larger volume of needs than what market orders in its area could compensate for\footnote{This is indicated in the document https://eepublicdownloads.entsoe.eu/clean-documents/events/2018/terre/20180319\_TERRE\_Stakeholders\_presentation.pdf}. In practice, TSOs can ask for a specific exemption to this rule for a given time step if the security of supply or their network is compromised, which has to be justified thoroughly (this feature is not implemented in ATLAS). The estimation of $q_{m,ca,t}^{\sigma_{t}^{bn},max}$ is done in Section \ref{sec:TSO needs - overall BSP volume}.
    \item The type of power output that should be taken into account when computing TSO needs is not straightforward. It challenges the role of TSOs in these new balancing markets, by questioning the integration of forecasts in the computation process. This issue is discussed in Section \ref{sec:TSO needs - imbalance}
\end{itemize}

\subsection{Computation of the overall volume submitted by BSPs in each control area}
\label{sec:TSO needs - overall BSP volume}
For time step $t \in T_{m}$, the overall volume of power offered by BSPs $q_{m,ca,t}^{\sigma_{t}^{bn},max}$ is computed for depending on $\sigma_{t}^{bn}$, the direction of TSO balancing needs. Only BSP orders in the opposite direction are taken into account (BSP shutdown orders explained in Section \ref{sec:bsp_orders_thermic_shutdown} are included in the $dn$ direction). Coupling links need to be considered in the calculation. Indeed, multiple orders of the same time step ($t_{o}^{start} = t$, that is noted $o_t$ in the equation below) linked with an \textit{Exclusion} coupling should not be simultaneously participating in the overall volume. In that situation, only the order with the largest $q_o^{max}$ is included in the computation:
\begin{align}
    \centering
    \label{eq:bsp_overall_volume}
    \forall  u \in U_{ca}, \mkern9mu \forall c \in C^{excl}, \mkern9mu & \forall t \in T^{m},\notag\\ q_{m,ca,t}^{\sigma_{t}^{bn},max} = \sum_{o_t \in O_{u} \mkern3mu | \mkern3mu (\sigma_{o_t} \mkern3mu \neq \mkern3mu \sigma_{t}^{bn}) \mkern3mu \& \mkern3mu (o_t \notin c)} q_{o_t}^{max} \quad + &  \max_{o_t \in c \mkern3mu | \mkern3mu (\sigma_{o_t} \mkern3mu \neq  \mkern3mu\sigma_{t}^{bn})}q_{o_t}^{max}
\end{align} 

\subsection{Imbalance computation}
\label{sec:TSO needs - imbalance}
As hinted at before, the balancing needs computation depends on the role of TSOs in balancing markets. The main question revolves around whether TSOs or Balancing Responsible Parties (BRPs)\footnote{BRPs possess a portfolio of generation and consumption units and face financial penalties if their portfolio is imbalanced compared to the energy they sold/bought on markets} should take into account imbalances that arise because of forecasts errors or even because of unexpected shutdown of units. To be as flexible as possible, both options exist in ATLAS for a given $ca$, depending on the parameter $\delta_{ca}^{for}$.
\begin{align}
    \centering
    \label{eq:tso_need_computation}
    & \forall ca \in \mkern3mu CA, \mkern9mu \forall t \in T_{m},\notag\\
    if \mkern9mu \delta_{ca}^{for} = 1, \quad bn_{t} = \sum_{u \in U^{l}} | P_{u,t, t_{m}^{ex}}^{for} | &  - (\sum_{u \in U^{th} \cup U^{h} \cup U^{st}}  P_{u,t, t_{m}^{ex}}^{plan} + \sum_{u \in U^{w} \cup U^{pv}}  P_{u,t, t_{m}^{ex}}^{for}) + \sum_{z \in Z_{ca}} (\Delta q)^{bal}_{z,t}\\
    if \mkern9mu \delta_{ca}^{for} = 0, \quad bn_{t} & = \sum_{u \in U^{l}} | P_{u,t, t_{m}^{ex}}^{plan} |  - \sum_{u \in U^{g}}  P_{u,t, t_{m}^{ex}}^{plan} + \sum_{z \in Z_{ca}} (\Delta q)^{bal}_{z,t}
\end{align}

Where $(\Delta q)^{bal}_{z,t}$ is the power balance of area $z$ at time $t$, meaning the sum of all exports minus the sum of all imports. Absolute values of power outputs and power forecasts of load units are considered in the computation, since by convention in ATLAS a negative power output corresponds to a consumption.

Finally, balancing needs are capped by the overall volume submitted by BSPs, which was just computed before:
\begin{align}
    \centering
    \label{eq:tso_need_capped}
    if \mkern9mu bn_{t} < 0, &  \quad bn_{t} \gets \min (bn_{t}, \mkern6mu q_{m,ca,t}^{dn,max})\\
    if \mkern9mu bn_{t} > 0, &  \quad bn_{t} \gets \min (bn_{t}, \mkern6mu q_{m,ca,t}^{up,max})
\end{align}

\section{Bidding strategies}
\label{sec:Pricing strategies}
\begin{figure}[H]
    \centering
    \includegraphics[width = 0.8\textwidth]{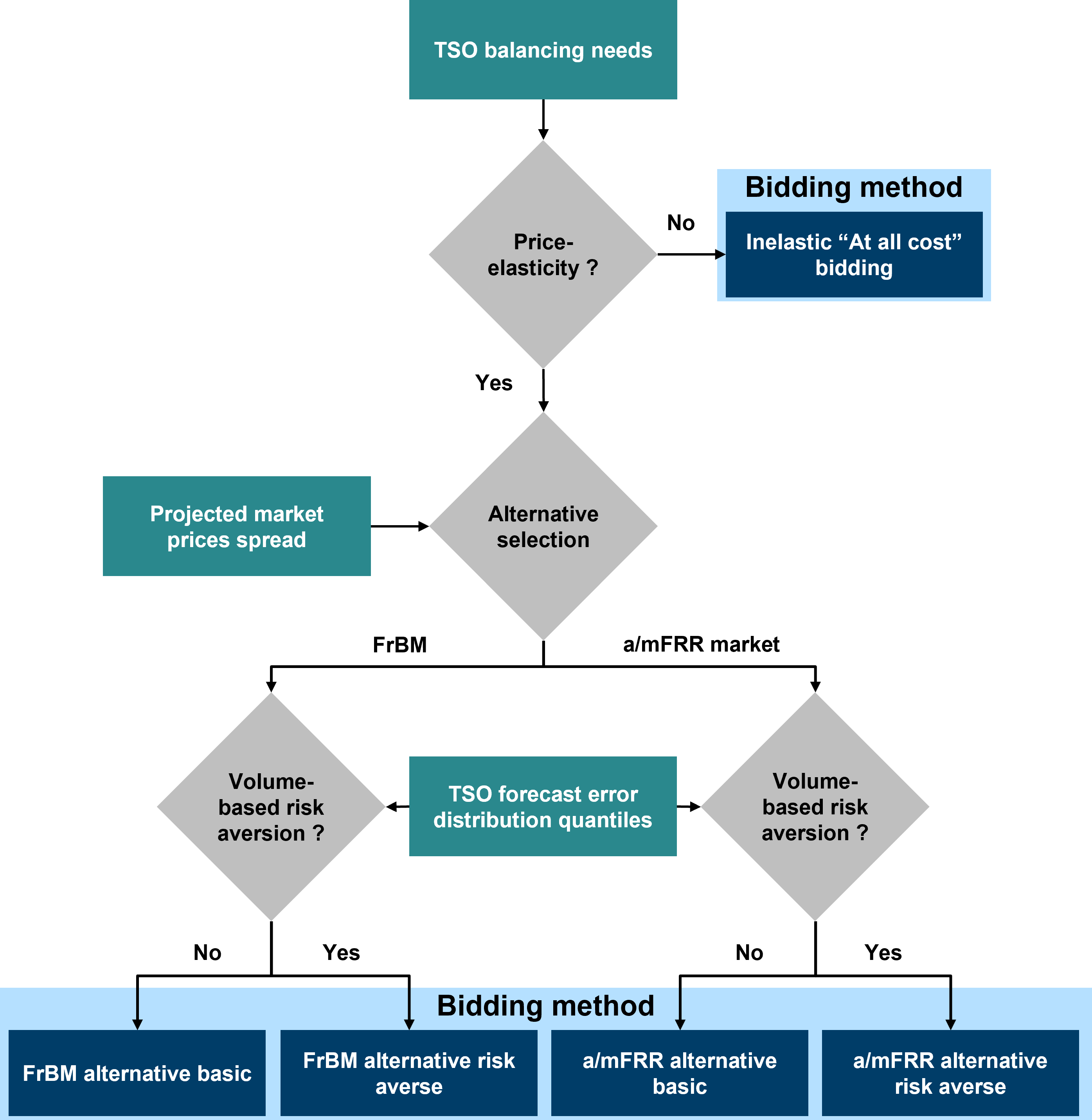}
    \caption{TSO Orders Formulation Strategies Framework}
    \label{fig:tso_orders_bidding_framework}
\end{figure}

Once overall balancing needs are defined, they are translated into market orders depending on the set of bidding strategy parameters chosen for the TSO of control area $ca$, which are illustrated in the diagram of Figure \ref{fig:tso_orders_bidding_framework}. The current section details all bidding strategies available in ATLAS, including the final characteristics of all market orders formulated (Table \ref{tab:tso_orders_nomenclature_orders}. Filling most characteristics is straightforward, as all orders $o$ formulated by a TSO at time $t$ share the following attributes: 

\begin{equation}
    \label{eq:tso_orders_general_attributes}
    \forall o \in O_{TSO}, \mkern9mu \forall t \in T_{m}, \quad \begin{cases}
        t_{o}^{ex} = t_{m}^{ex}\\
        t_{o}^{start} = t\\
        t_{o}^{end} = t + \Delta t_{m}\\
        q_{o}^{min} = 0\\
        \delta_{o}^{TSO} = 1
    \end{cases}
\end{equation}

The $q^{min}$ is set to 0 as the regulation of balancing markets requires TSO orders to be entirely divisible. The only characteristics that need to be determined by the bidding strategies are then the maximum quantity $p_{o}$, the price $p_{o}$ and the order direction $\sigma_{o}$.

\subsection{Price-inelastic bidding strategy}
\label{sec:Pricing - inelastic}
With the inelastic bidding strategy, the TSO is willing to accept any price for its entire balancing needs to be fulfilled by the market. Consequently, this strategy leads to the formulation of a single market order $o$, whose quantity, price and direction are described in Equation \ref{eq:inelastic_pricing_formulation}. Notably, the quantity corresponds to the entire balancing need of the given TSO, and the price to the maximum (resp. minimum) price authorized by the market for upward (resp. downward) balancing needs:
\begin{align}
    \centering
    \label{eq:inelastic_pricing_formulation}
    \forall ca \in CA, \mkern9mu \forall t & \in T_m, \mkern9mu if \mkern9mu \delta_{ca}^{elas} = 0, \notag\\ 
    if \mkern9mu bn_{t} > 0, & \quad \begin{cases}
        q_{o}^{max} = bn_{t}\\
        p_{o} = 10,000\\
        \sigma_{o} = -1
    \end{cases}\\
    if \mkern9mu bn_{t} < 0, & \quad \begin{cases}
        q_{o}^{max} = bn_{t}\\
        p_{o} = -10,000\\
        \sigma_{o} = 1
    \end{cases}
\end{align}

Notably, it is required for all TSO orders to be completely divisible ($q^{min}$ is always set to 0). This will be consistent across all other types of bidding strategies.

\subsection{Elastic bidding strategies}
\label{sec:Pricing - Basic elastic}

\subsubsection{Key concepts}
\label{sec:deter_key_principles}
In ATLAS, price-elastic bidding strategies result in the creation of bidding curves for TSO balancing needs. In terms of market orders at a given time $t$, it implies the creation of a finite set of orders $o_{i}, \quad \forall i \in {1, \dots, n}$ where $n$ is the number of slices into which the balancing needs are divided. All orders share the same general characteristics described at the beginning of Section \ref{sec:Pricing strategies}, and each order $o_{i}$ has its specific quantity-price-direction characteristics that we will note $q_{t,i}$, $p_{t,i}$ and $\sigma_{t,i}$. 

The price-elastic bidding process depends on two main parameters:
\begin{itemize}
    \item The type of alternative that is used as a reference to estimate opportunity costs, from which are derived the price of TSO market orders. That alternative can either be a closer-to-real-time market (such as the mFRR market when formulating orders on the RR market), or the local balancing process in the TSO area. These two categories of alternatives have their own specific features, that are detailed in Section \ref{sec:elastic_alternatives}.
    \item The integration of volume-based risk aversion in the bidding process, which can influence both the quantities and prices of TSO market orders. It is detailed in Section \ref{sec:elastic_risk_aversion_scenarios}.
\end{itemize}

\subsubsection{Alternative types and their features}
\label{sec:elastic_alternatives}

\paragraph{Local balancing process}
\label{sec:elastic_FrBM}
With this alternative selection, the TSO considers that its local balancing process is a viable alternative to the simulated market. In other words, the TSO assumes that it can rely on its local process after the balancing market, and uses this assumption to compute the prices of its market orders. In ATLAS, the local balancing process is modeled by the Balancing Mechanism module, detailed in Chapter \ref{ch:4. - BalancingMechanism}. It is based on the historical French Balancing Mechanism process, and will consequently be referred to as FrBM.

A key hypothesis is required when considering this type of alternative. Since the TSO uses an estimate of its local balancing process costs, it has to predict that a certain volume of power (either upward or downward, depending on its needs) will be available on the BM. This prediction is made before the clearing of balancing markets, for a BM that occurs after said balancing markets, which means that the TSO has to predict the power that is either not submitted to the market or not activated on it. The prediction process currently used in ATLAS revolves around the parameter $\rho_{m, ca}^{FrBMalt}$ which indicates the percentage of available capacity that is not submitted on the balancing market $m$ by BSPs, and goes directly on the FrBM. This process then consists of the following steps, according to the unit type: 
\begin{itemize}
    \item For thermic units, a random draw is performed for each subset of fuel types in the area of the TSO, selecting units that are only submitting capacity to the FrBM to reach $\rho_{m, ca}^{FrBMalt}$ with a tolerance band of 10\%. For instance, if an area has both CCGT and nuclear assets with $\rho_{m, ca}^{FrBMalt} = 50\%$, then each type of fuel is looked at individually. Units belonging to the CCGT group are randomly selected so that their combined capacity is within the range [40\%, 60\%] of the total CCGT capacity in the area. The same selection is applied to nuclear units. 
    Performing the selection by fuel type allows to avoid randomly selecting all units of a certain type, which could have an important impact on cost estimations.
    \item Hydraulic and storage units are usually modeled by a single asset for each zone in ATLAS. Consequently, it is assumed that the percentage $\rho_{m, ca}^{FrBMalt}$ of the available capacity of each unit is available on the FrBM.
\end{itemize}

The calibration of the capacity percentage parameter then plays a key role. An empirical analysis carried out on data published by RTE over 2022 was used for this calibration, leading to the following estimation (using a conservative approach that tends to favor the volume submitted on balancing markets): 50\% of the available upward (resp. downward) power is not formulated on the RR market, and is always available on the French local balancing process.

We are looking to refine this estimation in the future. First, the empirical study only provides aggregated data, and in particular does not distinguish unit types. It is consequently difficult to evaluate if the volume not submitted to the RR market is mostly associated with a specific type, or spread homogeneously (the latter is chosen as a basis in ATLAS). In addition, the study was only conducted in the French area for data availability reasons, and this estimator could be very different in other areas. Nonetheless, as this is the only value at our disposal, it is consequently applied to all areas in ATLAS.

Once the determination of available units on the projected FrBM is done, simulations of the alternative using the Balancing Mechanism module are performed. The cost $C^{FrBMalt}(q)$ of balancing a given volume of TSO needs $q$ is a direct output of the simulation process.

\begin{figure}[H]
    \centering
    \includegraphics[width = 0.9 \textwidth]{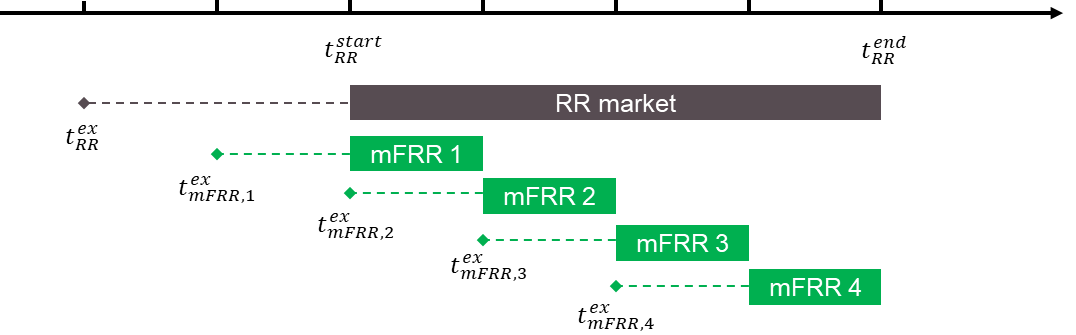}
    \caption{Schematic representation of RR and mFRR markets overlays used in the mFRR alternative pricing}
    \label{fig:mfrr_alternative_rr_and_mfrr_markets}
\end{figure}

\paragraph{Alternative market process}
\label{sec:elastic_mFRR}
A TSO can consider that a closer-to-real-time market is an alternative to its current studied market $m$, as illustrated by the timeline in Figure \ref{fig:mfrr_alternative_rr_and_mfrr_markets}. While the aFRR market could be a relevant alternative to both RR and mFRR markets, it is currently not implemented in ATLAS. This means that the alternative market process currently works in a single configuration: $m = RR$, and $alt_{ca} = mFRRalt$. However, the following description is sufficiently generic that it could be applied in the case $alt_{ca} = aFRRalt$, should the aFRR market be implemented in ATLAS in the future.

In contrast with the FrBM alternative, there is no need for an assumption of available units. The alternative market is not simulated, as a simulation of future balancing markets would require the TSO to make a significant number of assumptions about the state of the power system, especially given that balancing markets include multiple European areas. We considered that it was not a realistic approach. Instead, the projected market prices are estimated based on fixed input data.

Several options are possible for this estimation. The first one would be to use historical data of the alternative. However, actual mFRR markets have only been operational for a year\footnote{The Go-Live of the MARI platform was announced by ENTSO-E the 2022/10/05, c.f. https://www.entsoe.eu/news/2022/10/06/go-live-of-mari-the-european-implementation-project-for-the-creation-of-the-european-manual-frequency-restoration-reserves-mfrr-platform/}. Based on past experiences with the operational RR market, it seems that a year is not enough time to reach a stable state for the market, as both BSPs and TSOs are still progressively entering it and experimenting with it. Another basis could be to preemptively simulate mFRR markets in ATLAS and gather the output clearing prices. However, this would require large numbers of simulations of mFRR markets to generate sufficient data, and this option was currently ruled out for computational time reasons.

Consequently, the data we chose to use for the estimation of mFRR prices was extrapolated from the historical FrBM over 6 years, between 2018 and 2023\footnote{This data is public and was downloaded from the RTE services website https://www.services-rte.com/en/home.html, categorized in Market/Balancing Energy/Prices}. First, offers made by units that could have participated in mFRR markets were extracted. These offers were then used to generate a table giving the projected ratio between the day-ahead market price and the mFRR price, both for upward and downward regulations, for several ranges of imbalance to compensate \ref{tab:mFRRalt_price_spreads}.

\begin{table}[H]
    \centering
    \begin{tabular}{|c|c|c|}
    \hline
    \textbf{Imbalance volumes ranges (MW)} & \textbf{$\mathbf{Ratio^{dn}}$}  & \textbf{$\mathbf{Ratio^{up}}$}\\
    \hline
     0 - 300 & 0.81 & 1.28 \\
    \hline
    300 - 600 & 0.76 & 1.33 \\
    \hline
    600 - 900 & 0.73 & 1.36 \\
    \hline
    900 - 1200 & 0.7 & 1.37 \\
    \hline
    1200 - 1500 & 0.7 & 1.38\\
    \hline
    1500 - $\infty$ & 0.59 & 1.47\\
    \hline
    \end{tabular}
    \caption{Price ratio in France between day-ahead and mFRR prices over 2018-2023}
    \label{tab:mFRRalt_price_spreads}
\end{table}

Eventually, the price estimation $\Tilde{\lambda}^{\sigma^{R}}_{mFRR}$ of the mFRR for an imbalance $q$ in the direction $\sigma^{R} \in \{up, dn\}$ is given by:
\begin{equation}
    \label{eq:mFRR_price_estimation}
    \Tilde{\lambda}^{\sigma^{R}}_{mFRR}(q) = \lambda_{DA} * Ratio^{\sigma^{R}}(q)
\end{equation}
Where $\lambda_{DA}$ is the day-ahead price endogenously obtained within the ATLAS simulation process\footnote{In ATLAS, balancing markets are assumed to be simulated following at least a day-ahead market, and not entirely alone.}.\\

This leads to the following cost estimation for the mFRR market alternative for a given imbalance quantity $q$, where $\sigma^{R}$ is here given by the direction of $q$ ($\sigma^{R} = 1$ if $q >= 0$, and $\sigma^{R} = -1$ otherwise):

\begin{equation}
    \label{eq:2.5_mFRR_post_setimation}
    C^{mFRRalt}(q) = \sigma_q * q * \Tilde{\lambda}_{mFRR}^{\sigma^{R}}(q)
\end{equation}

\subsubsection{Risk aversion integration}
\label{sec:elastic_risk_aversion_scenarios}
Two different sources of risk aversion can be distinguished when looking at the TSO order formulation process:
\begin{itemize}
    \setlength\itemsep{-0.1em}
    \item Uncertainties associated with the cost estimation of the alternative, which we will call "price-based" risk aversion.
    \item Uncertainties regarding the volume of TSO balancing needs. Indeed, pro-active TSOs try to estimate this volume at the execution date of the studied market, and it can consequently be under- or over-estimated. These forecast errors would then induce subsequent balancing needs that should be resolved with the alternative. We call this type of risk aversion "volume-based".
\end{itemize}

The volume-based risk aversion has been implemented in ATLAS, while the price-based risk aversion is still currently not in the model. 
Two degrees relative to risk aversion can be chosen in ATLAS: a basic approach that considers estimated balancing needs to be exact (Section \ref{sec:elastic_basic}), and a risk-averse approach that takes into account volume-based uncertainties previously described (Section \ref{sec:elastic_risk_averse}). 

\begin{figure}[H]
    \centering
    \begin{subfigure}[b]{0.48\textwidth}
        \centering
        \includegraphics[width = \textwidth]{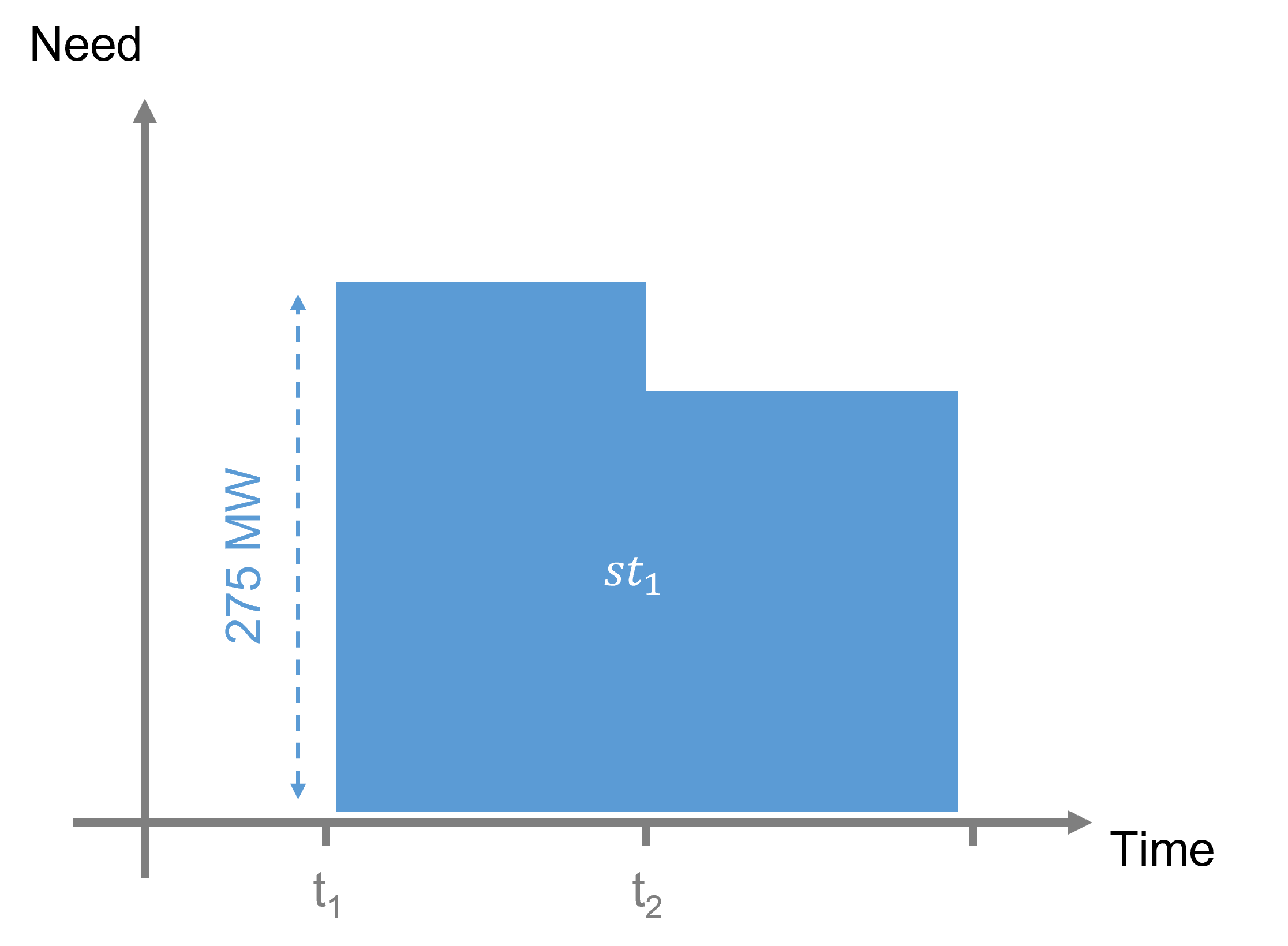}
        \caption{Total TSO need}
        \label{fig:deter_price_total_need}
    \end{subfigure}
    \hfill
    \begin{subfigure}[b]{0.48\textwidth}
        \centering
        \includegraphics[width = \textwidth]{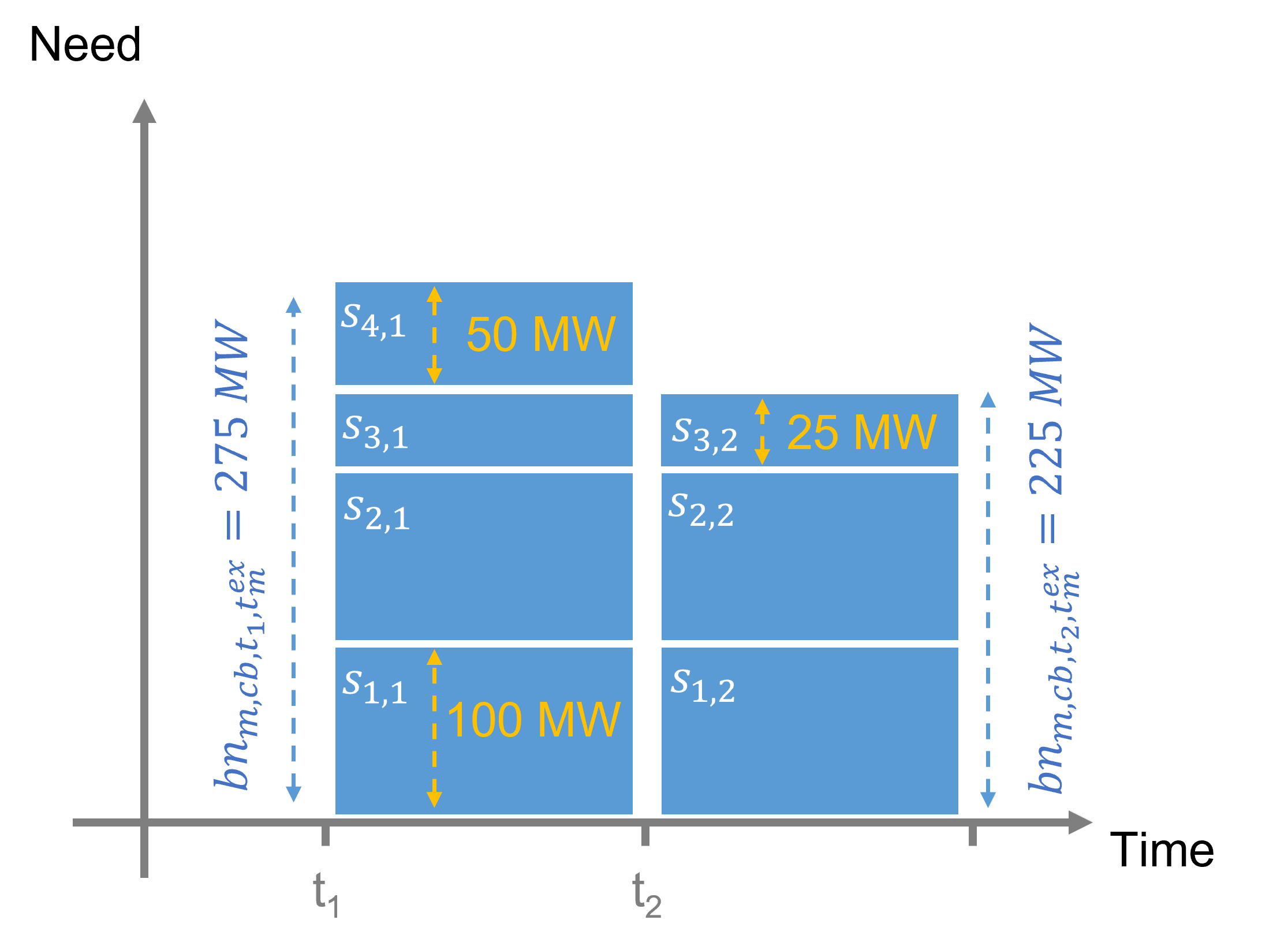}
        \caption{Division into slices}
        \label{fig:deter_price_slices_1}
    \end{subfigure}
    \caption{Deterministic Pricing Slicing Process Illustration}
    \label{fig:deter_pricing_slicing}
\end{figure}

\paragraph{Basic formulation}
\label{sec:elastic_basic}
With the basic formulation, the division of TSO balancing needs into slices follows a process based on the parameter $V^{s}$, which indicates the maximum quantity of each order wanted by the user. When possible, the quantity $q_{o_{i}}$ of order $i$ will be equal to $V^{s}$. However, it is required to modify this quantity in some cases, mainly to improve computational performances. For instance, the simulation of the FrBM alternative is performed over the complete time frame $T_{m}$, comprised of several time steps. For that reason, the size of certain slices is reduced to harmonize as much as possible the different time steps. This example is illustrated in Figure \ref{fig:deter_pricing_slicing}, which indicates the balancing needs division over two consecutive time steps for $V^{s} = 100 MW$.

The two first slices of need reach the maximum quantity $V_{s}$. To divide the entire balancing needs of both time steps, the quantities of the two top slices are set at a lower value, respectively 25 MW and 50 MW in the example. 

It is relevant to note that balancing needs can be positive for some time steps of a given time frame $T_{m}$, and negative for others. If this situation arises, the division is performed separately for positive and negative needs. The division process is formally described in Algorithm \ref{alg:elastic_slice_division_pos} for time steps with positive needs, and in Algorithm \ref{alg:elastic_slice_division_neg} for time steps with negative needs. In these algorithms, $V_{i}$ is the relative quantity of slice $i$\footnote{For all time steps $t$ for which slice $i$ is reached, $q_{t,i}$ is set to the absolute value of $V_{i}$ in the algorithm, and $\sigma_{o_{i}}$ is set in accordance to the sign of $V_{i}$.}. 

\begin{algorithm}[H]
    \caption{Basic elastic formulation slice division process (positive balancing needs)}\label{alg:elastic_slice_division_pos}
    \begin{algorithmic}
        \State \textbf{\textit{Initialization of the set of time steps}}
        \State $T_{div} \gets {t \in T_{m} \mkern9mu | \mkern9mu bn_{t} > 0}$\\
        
        \State \textbf{\textit{Initialization of the first slice $i = 1$}}
        \State $V_{1} \gets \min(V^{s}, \max\limits_{t \in T_{div}} bn_{t})$\\

        \State \textbf{\textit{Recursion}}
        \For{$i > 1$}
            \While{$\mkern9mu \sum\limits_{1 \leq j < i} (V_{j}) < \max\limits_{t \in T_{div}} bn_{t}$}
                \State $V_{i} \gets \min(V^{s}, \min\limits_{t \in T_{div}} (bn_{t} - \sum\limits_{1 \leq j < i} (V_{j})))$\Comment{Compute the maximum size of $V_{i}$}\\

                \For{$t \in T_{div}$}
                    \State $q_{t,i} \gets |V_{i}|$\Comment{Extract the order quantity for relevant time steps}
                    \State $\sigma_{t,i} \gets -1$\Comment{Set the order direction}
                    \If{$bn_{t} - \sum\limits_{1 \leq j \leq i} (V_{j}) = 0$}\Comment{Remove now "empty" time steps from $T_{div}$}
                        \State $T_{div} \gets T_{div} - \{t\}$
                    \EndIf
                \EndFor
                \State $i \gets i+1$
            \EndWhile
        \EndFor
    \end{algorithmic}
\end{algorithm}

\begin{algorithm}[H]
    \caption{Basic elastic formulation slice division process (negative balancing needs)}\label{alg:elastic_slice_division_neg}
    \begin{algorithmic}
        \State \textbf{\textit{Initialization of the set of time steps}}
        \State $T_{div} \gets {t \in T_{m} \mkern9mu | \mkern9mu bn_{t} < 0}$\\
        
        \State \textbf{\textit{Initialization of the first slice $i = 1$}}
        \State $V_{1} \gets \max(-V^{s}, \max\limits_{t \in T_{div}} bn_{t})$\\

        \State \textbf{\textit{Recursion}}
        \For{$i > 1$}
            \While{$\mkern9mu \sum\limits_{1 \leq j < i} (V_{j}) > \min\limits_{t \in T_{div}} bn_{t}$}
                \State $V_{i} \gets \max(-V^{s}, \max\limits_{t \in T_{div}} (bn_{t} - \sum\limits_{1 \leq j < i} (V_{j})))$\Comment{Compute the maximum size of $V_{i}$}\\

                \For{$t \in T_{div}$}
                    \State $q_{t,i} \gets |V_{i}|$\Comment{Extract the order quantity for relevant time steps}
                    \State $\sigma_{t,i} \gets 1$\Comment{Set the order direction}
                    \If{$bn_{t} - \sum\limits_{1 \leq j \leq i} (V_{j}) = 0$}\Comment{Remove now "empty" time steps from $T_{div}$}
                        \State $T_{div} \gets T_{div} - \{t\}$
                    \EndIf
                \EndFor
                \State $i \gets i+1$
            \EndWhile
        \EndFor
    \end{algorithmic}
\end{algorithm}

Eventually, the price of each slice $i$ is computed according to the cost of the complete stack of needs $\sum\limits_{1 \leq j \leq i} (V_{j})$ using the selected alternative $alt$, as Equation \ref{eq:tso_elastic_basic_price} indicates:

\begin{equation}
    \label{eq:tso_elastic_basic_price}
    \forall t \in T_{m}, \mkern9mu \forall i \in \{1, \dots, n\}, \quad p_{t,i} = \frac{C^{alt}(\sum\limits_{1 \leq j \leq i} (q_{t,j}))}{\sum\limits_{1 \leq j \leq i} (q_{t,j})} 
\end{equation}

With $C^{alt}(q)$ the cost estimation of balancing quantity $q$ using the alternative $alt$, as explained in Section \ref{sec:elastic_alternatives}.

\paragraph{Risk-averse formulation}
\label{sec:elastic_risk_averse}
The general idea of the volume-based risk-averse formulation is to create a bidding curve in which both the size of each slice $i$ and its price depend on its probability of occurrence. The method implemented in ATLAS is comprised of the following steps: 
\begin{enumerate}
    \setlength\itemsep{0em}
    \item Determining the distribution of the error made by the TSO when forecasting its balancing needs between two different dates: the execution date of the RR market, and the execution date of its chosen alternative. The user needs to perform this computation before running balancing modules, and indicate the list of quantiles $\epsilon_{ca}^{alt_{ca}}$ of this distribution as a module parameter.
    \item Using the list of quantiles to compute the size of each slice and the associated probability of occurrence. This step is illustrated in Figure \ref{fig:2.4_norm_distrib_example}, using an arbitrary normal distribution. Quantiles $\epsilon_i$ of probability $\alpha_i \in \{0.1, \dots, 0.9\}$ are displayed (for readability concerns, we will note $i \in \{1, \dots, 9\}$), bounded by extreme quantiles $\epsilon_{min}$ of probability $\alpha_{min} = 0.01$ and $\epsilon_{max}$ of probability $\alpha_{max} = 0.99$ that will be used as references for outer bounds.

    \begin{figure}[H]
        \centering
        \includegraphics[width = 0.7\textwidth]{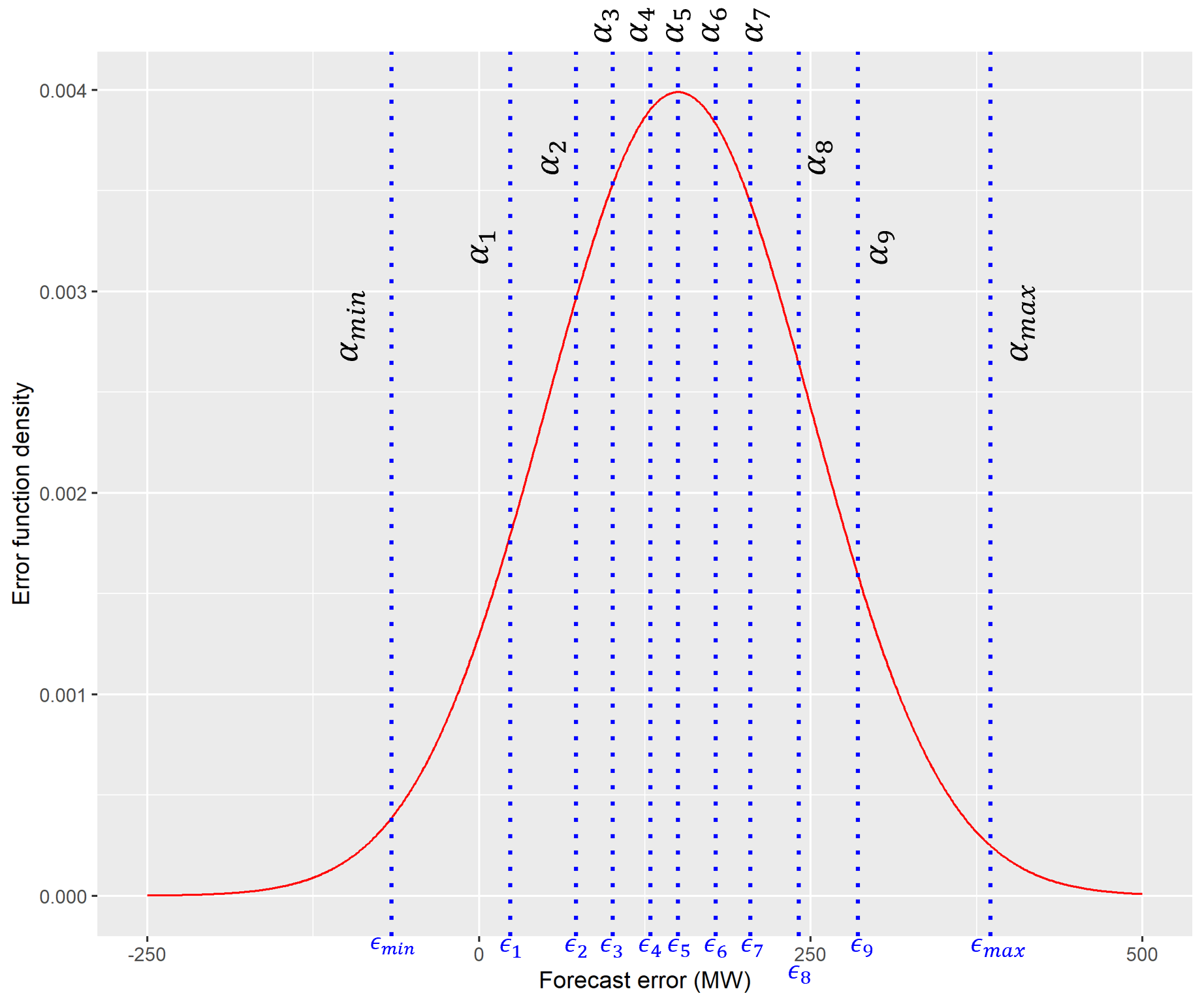}
        \caption{TSO forecast error distribution example}
        \label{fig:2.4_norm_distrib_example}
    \end{figure}

    The quantile $i$ indicates that the real forecast error $\epsilon$ made by the TSO has a probability $\alpha_i$ of being below the associated quantity $\epsilon_i$, and a probability $(1 - \alpha_i)$ of being above this value: 

    \begin{equation}
        \forall i \in \{1, \dots, 9\}, \quad P[\epsilon \leq \epsilon_i] = \alpha_i
    \end{equation}

    Information carried by quantiles can then be translated into the quantity of each slice of need submitted to the RR market. If we look at a given forecasted balancing need in the upward direction ($bn_{t} > 0$), then $bn_{t} + \epsilon_i$ should give the imbalance that is likely to be reached with an $\alpha_i$ probability. However, depending on the values of $bn_{t}$ and $\epsilon_i$, this calculated imbalance volume can be in the opposite direction as $bn_{t}$. This makes sense from a practical point of view: if the forecasted imbalance is low (and/or if the margin of error is high), then there is a decent chance that the actual balancing need is eventually in the opposite direction. But it also introduces another layer of complexity when determining the volume of each slice, as it leads to 3 different cases:

    \begin{itemize}
        \item If $bn_{t} > 0$ and $bn_{t} + \epsilon_1 \geq 0$, then we are in the simple case were the balancing need submitted is always positive, even when taking into account the possible forecast errors. In that situation, the first slice of need ranges from $0$ up until $bn_{t} + \epsilon_1$. The quantity of every subsequent slice of need $i$ is equal to the quantity between the two successive quantiles $\epsilon_{i-1}$ and $\epsilon_i$ (Equation \ref{eq:tso_orders_risk_quantity_pos_pos}). The direction of each order is, as explained before, always upward (Equation \ref{eq:tso_orders_risk_direction_pos_pos}).
        \begin{align}
            \label{eq:tso_orders_risk_quantity_pos_pos}
            \forall t \in T_{RR}, \mkern9mu if & \mkern6mu (bn_{t} > 0) \mkern9mu \& \mkern9mu (bn_{t} + \epsilon_1 \geq 0), \notag\\
            & \sigma_{t,i} = -1, \quad \forall i\\
            & \begin{cases}
                q_{t,1} = bn_{t} + \epsilon_1\\
                q_{t,i} = \epsilon_i - \epsilon_{i-1}, \quad \forall i \in \{2, \dots, 9\}
                \end{cases} 
        \end{align}

        \begin{equation}
            \label{eq:tso_orders_risk_direction_pos_pos}
            \forall t \in T_{RR}, \mkern9mu if \mkern6mu (bn_{t} < 0) \mkern9mu \& \mkern9mu (bn_{t} + \epsilon_9 \leq 0), \mkern9mu \forall i \in \{1, \dots, 9 \},
            \quad \sigma_{t,i} = -1
        \end{equation}

        \item If $bn_{t} < 0$ and $bn_{t} + \epsilon_9 \leq 0$, then this is the mirror case of the first one, with submitted needs being always negative.
        \begin{align}
            \label{eq:tso_orders_risk_quantity_neg_neg}
            \forall t \in T_{RR}, \mkern9mu if & \mkern6mu (bn_{t} < 0) \mkern9mu \& \mkern9mu (bn_{t} + \epsilon_9 \leq 0), \notag\\
            &  \begin{cases}
                q_{t,i} = |\epsilon_i - \epsilon_{i+1}|, \quad \forall i \in \{1, \dots, 8\}\\
                q_{t,9} = |bn_{t} + \epsilon_9|
                \end{cases}
        \end{align}

        \begin{equation}
            \label{eq:tso_orders_risk_direction_neg_neg}
            \forall t \in T_{RR}, \mkern9mu if \mkern6mu (bn_{t} < 0) \mkern9mu \& \mkern9mu (bn_{t} + \epsilon_9 \leq 0), \mkern9mu \forall i \in \{1, \dots, 9 \},
            \quad \sigma_{t,i} = 1
        \end{equation}

        \item If $bn_{t} > 0$ and $bn_{t} + \epsilon_1 < 0$, or if $bn_{t} < 0$ and $bn_{t} + \epsilon_9 > 0$, then some need slices submitted should be negative and others should be positive. It is then important to identify where submitted balancing needs shift from being negative to being positive when browsing through values of $i$, and between which consecutive slices it happens. Let's note these consecutive slices $i_{s1}$ and $i_{s2}$, such that:
        \begin{equation}
            \label{2.4_is1_is2_determination}
            \begin{cases}
                \nexists \mkern6mu i \in [i_{s1}, i_{s2}]\\
                bn_{t} + \epsilon_{i_{s1}} < 0 \mkern9mu \& \mkern9mu bn_{t} + \epsilon_{i_{s2}} > 0
            \end{cases}
        \end{equation}

        As Equations \ref{eq:tso_orders_risk_quantity_pos_neg} and \ref{eq:tso_orders_risk_direction_pos_neg} describe, for slices lower than $i_{s1}$, submitted balancing needs are always negative. For slices higher than $i_{s2}$, submitted balancing needs are always positive. Finally, slices $i_{s1}$ and $i_{s2}$ are dealt with in specific cases:
        \begin{align}
            \label{eq:tso_orders_risk_quantity_pos_neg}
            \forall t \in T_{RR}, \mkern9mu if \mkern6mu &(bn_{t}  > 0) \mkern9mu \& \mkern9mu (bn_{t} + \epsilon_1 < 0),\notag\\
            & \begin{cases}
                q_{t,i} = |\epsilon_i - \epsilon_{i+1}|, \quad \forall i \in [1, \dots ,i_{s1} - 1]\\
                q_{t,i_{s1}} = |\epsilon_{i_{s1}}|\\
                q_{t,i_{s2}} = \epsilon_{i_{s2}}\\
                q_{t,i} = \epsilon_i - \epsilon_{i-1}, \quad \forall i \in [i_{s2} + 1, \dots, 9]
            \end{cases}
        \end{align}

        \begin{align}
            \label{eq:tso_orders_risk_direction_pos_neg}
            \forall t \in T_{RR}, & \mkern9mu if \mkern6mu(bn_{t}  > 0) \mkern9mu \& \mkern9mu (bn_{t} + \epsilon_1 < 0),\notag\\
            & \begin{cases}
                \sigma_{t,i} = 1 \quad \forall i \in \{1, \dots, i_{s1}\}\\
                \sigma_{t,i} = -1 \quad \forall i \in \{i_{s2}, \dots, 9\}
            \end{cases}
        \end{align}
    \end{itemize}

    \item Computing the price of each slice, taking into account both the price of the alternative and the probability of occurrence of this slice. If we still look at upward balancing needs $bn_{t}$ for the sake of the explanation, the following reasoning is applied for a given stack of need $i$ of volume $\sum\limits_{1 \leq j \leq i} q_{t,j}$, associated with the probability $\alpha_i$:
    \begin{itemize}
        \item The cost associated with compensating this stack of needs is given by $C^{alt, up}(\sum\limits_{1 \leq j \leq i} q_{t,j})$.
        \item In addition, there is a $\alpha_i$ chance that the actual imbalance volume is lower than $\sum\limits_{1 \leq j \leq i} q_{t,j}$, meaning that the TSO will have to compensate in the downward direction with its alternative if the market activates it. To be precise, the formula giving the expected cost of downward compensation would be: $\int_{0}^{i} C^{alt, dn}(\sum\limits_{1 \leq j \leq i} q_{t,i} - \epsilon_{j}) dj$. For computation time reasons, an approximation is used here and the associated costs are estimated at $C^{alt, dn}(\epsilon_{k^{d}} - \epsilon_{i})$, where $k^{d} = \alpha_{i} - \frac{1}{2} * (\alpha_{i} - \alpha_{min})$. In that case, $\epsilon_{k^{d}}$ is an approximation of the probability expectation of the downward need to be compensated for. 
        \item On the other hand, there is a $(1 - \alpha_i)$ chance that the actual imbalance volume is higher than $\sum\limits_{1 \leq j \leq i} q_{t,j}$, meaning that the TSO would have to ask for additional upward power with its alternative. Again, associated costs are estimated at  $C^{alt, up}(\epsilon_{k^{u}} - \epsilon_{i})$, with $k^{u} = \alpha_{i} + \frac{1}{2} * (\alpha_{max} - \alpha_{i})$.
    \end{itemize}
    
    More formally, the price of slice $i$ can be written in Equations \ref{eq:2.4_risk_prices_up} and \ref{eq:2.4_risk_prices_down} which detail respectively positive and negative balancing needs directions:
    \begin{align}
        \label{eq:2.4_risk_prices_up}
        \forall t \in T_{RR}, \mkern9mu \forall i \in [1,\dots,9], \mkern9mu & if \mkern6mu bn_{t} > 0, \notag\\
        p_{t,i} = \frac{1}{\sum\limits_{1 \leq j \leq i} q_{t,j}} * \biggr[ C^{alt, up}(\sum\limits_{1 \leq j \leq i} q_{t,j}) + \alpha_i * C^{alt, dn} (\epsilon_{k^{d}} & - \epsilon_{i}) + (1-\alpha_i) * C^{alt, up}(\epsilon_{k^{u}} - \epsilon_{i}) \biggr]
    \end{align}

    \begin{align}
        \label{eq:2.4_risk_prices_down}
        \forall t \in T_{RR}, \mkern9mu \forall i \in [1,\dots,9], \mkern9mu & if \mkern6mu bn_{t} < 0, \notag\\
        p_{t,i} = \frac{1}{\sum\limits_{i \leq j \leq 9} q_{t,j}} * \biggr[ C^{alt, dn}(\sum\limits_{i \leq j \leq 9} q_{t,j}) + \alpha_i * C^{alt, dn} (\epsilon_{k^{d}} & - \epsilon_{i}) + (1-\alpha_i) * C^{alt, up}(\epsilon_{k^{u}} - \epsilon_{i}) \biggr]
    \end{align}
\end{enumerate}

\chapter{Balancing Mechanism} 
\label{ch:4. - BalancingMechanism}

\section{Overview of the actual Balancing Mechanism process}
\label{sec:4.2 - BM Overview}
\begin{figure}[H]
    \centering
    \includegraphics[width = 0.9 \textwidth]{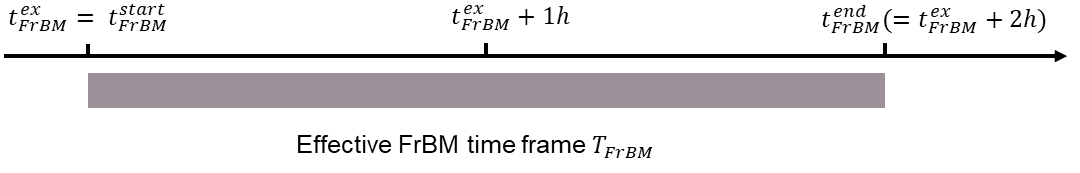}
    \caption{Historical time frame of the French Balancing Mechanism}
    \label{fig:4.2_historical_frbm_timeframe}
\end{figure}

The Balancing Mechanism module in ATLAS aims at emulating the eponymous historical balancing process used by the French TSO RTE, which we will call FrBM (for French Balancing Mechanism). The FrBM has been operational since the beginning of the 21$^{th}$ century, and is now partially being replaced by common European balancing markets. Indeed, its role is to activate the so-called "tertiary reserves"--which mainly correspond to the combination of mFRR and RR reserve types in the new European nomenclature-- to correct observed or forecasted supply-demand imbalances in the French control area. Currently, it is still being run after (i.e. closer to real-time than) the European RR market: it serves as a backup option for the French TSO, and is used to resolve any remaining imbalance still existing after the RR market. In its historical configuration, the time frame $T_{FrBM}$ of the Balancing Mechanism is illustrated in Figure \ref{fig:4.2_historical_frbm_timeframe}. 

\begin{figure}[H]
    \centering
    \includegraphics[width = 0.9 \textwidth]{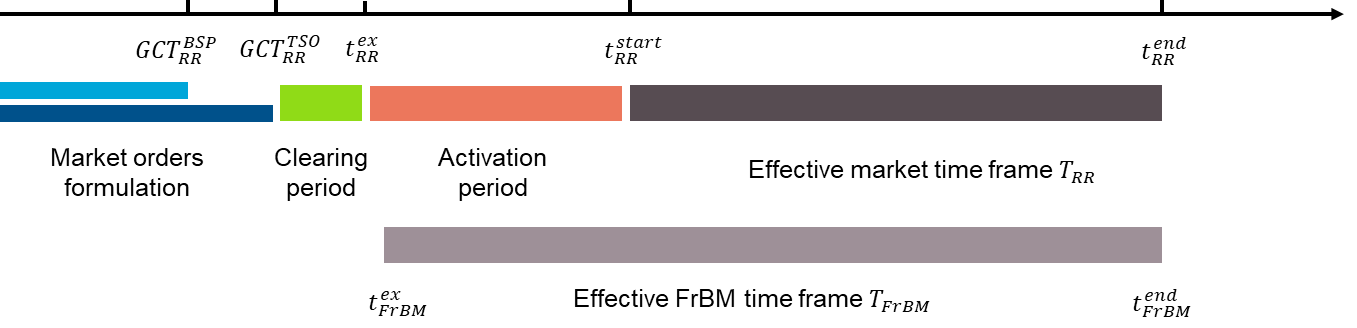}
    \caption{Current time frame of the French Balancing Mechanism, following the RR market}
    \label{fig:4.2_rr__frbm_timeframe}
\end{figure}

The execution date $t_{FrBM}^{ex}$ also corresponds to the beginning $t_{FrBM}^{start}$ of the effective period $T_{FrBM}$ (called "neutralization leadtime") that lasts until $t_{FrBM}^{end}$, during which no unit in the French area is allowed to change its planned generation or consumption program for any time until $t_{FrBM}^{end}$. Note that said units are still able to make offers on markets whose effective time frame is after $t_{FrBM}^{end}$. For instance, if we take $t_{FrBM}^{ex} = 10h$ and $t_{FrBM}^{end} = 12h$, a unit at $t = 11h$ is not allowed to plan a change to its power output for any time until 12h, but can plan a change for time steps after 12h. Historically, the FrBM was run at an hourly frequency, with a maximum neutralization leadtime of 2 hours. The new sequence, in which the FrBM is sequentially executed after the RR market, is represented in Figure \ref{fig:4.2_rr__frbm_timeframe}. The operational time frame of the FrBM begins after the clearing stage of the last RR market, at $t_{RR}^{ex}$. Both sequences and time frames are possible in ATLAS, depending on the type of market simulated.

Each BSP connected to the French transport network (the part of the power grid operated by the French TSO) is required to submit its entire available capacity in both directions (upward and downward) to the FrBM through reserve orders\footnote{Currently, in ATLAS the distinction between a connection to the transport or the distribution network is not implemented. By default, every unit is required to submit its entire capacity for the Balancing Mechanism of its TSO. We are looking to refine this assumption in future work. This approximation is nonetheless justified for thermal and hydraulic units.}. These orders are described as "specific" and "implicit", as opposed to the "standard" and "explicit" market orders that can be found on RR or mFRR markets for instance (see Sections \ref{sec:Introduction - RR market} and \ref{sec:Introduction - mFRR market}). Indeed, on the FrBM, BSP orders indicate the price at which they should be activated (set by BSPs themselves) but do not specify any maximum or minimum quantity, nor any activation duration. BSPs instead send to the TSO the generation or consumption plans of each of their units, as well as all their technical and operational constraints, leaving to the TSO the role of computing feasible power ranges. In ATLAS, BSPs are currently assumed to have no strategic behavior on the FrBM, and the price of their orders corresponds exactly to their production cost (i.e. their variable cost, plus eventual startup costs if the associated unit is offline).

The resulting problem from the TSO point of view consists in two steps: computing its balancing need, and activating BSP orders to resolve it at minimal cost. The implementation of both steps in ATLAS is described respectively in Sections \ref{sec:4.3 - FrBM TSO balancing needs} and \ref{sec:4.4 - FrBM OP problem}. Finally, the remuneration scheme of the FrBM is "pay-as-bid", meaning that all activated BSPs are then remunerated at the price of their order multiplied by the activated quantity. 

The rest of this document presents the implementation and key methods of the Balancing Mechanism developed in ATLAS. Any information hereafter does not describe the real FrBM process, but rather the model that emulates it, along with is necessary simplifications and approximations.

\section{Nomenclature and inputs}
\label{sec:4.1 - BM Overview}
This section describes all notations used in the chapter. Some elements correspond to a specific type, which is indicated in bold and italics:
\begin{itemize}
    \item \textbf{\textit{Parameter}} refers to a parameter, which is indicated by the user before the execution of the module.
    \item \textbf{\textit{Input data}} refers to an element that is extracted from the input dataset.
    \item \textbf{\textit{Variable}} refers to an optimization variable.
\end{itemize}

Remark: For sets, the notation $A_{b}$ refers to the subset of $A$ linked with variable $b$. For instance, $Z_{ca}$ indicates the subset of market areas belonging to the control area $ca$.

\renewcommand{\arraystretch}{1.3} 
\begin{longtable}[H]{!{\color{Grey}\vrule}>{\centering\arraybackslash}p{3cm} !{\color{Grey}\vrule} p{13cm}!{\color{Grey}\vrule} }
    \arrayrulecolor{Grey} \hline
    \rowcolor{Grey} \multicolumn{2}{|c|}{\large{\textbf{Sets and global market notations}}} \label{tab:bm_nomenclature_sets}\\ 
    \hline
    \rowcolor{Grey} 
    \textbf{Notation} & \textbf{Description}\\
    \hline
    $CA$ & Set of all control areas (or control blocks)\\
    \hline
    $Z$ & Set of all market areas\\
    \hline
    $U$ & Set of all units\\
    \hline 
    $U^{unit\_type}$ & Set of all units of type $unit\_type \in [g, l, th, h, st, w, pv]$. ($g$ = generation, $l$ = flexible load, $th$ = thermal, $h$ = hydraulic, $st$ = storage, $w$ = wind, $pv$ = photovoltaic)\\
    \hline
\end{longtable}

\begin{longtable}[H]{!{\color{Grey}\vrule}>{\centering\arraybackslash}p{3cm} !{\color{Grey}\vrule} p{10.5cm}!{\color{Grey}\vrule} >{\centering\arraybackslash} p{2cm}!{\color{Grey}\vrule}}
    \arrayrulecolor{Grey} \hline
    \rowcolor{Grey} \multicolumn{3}{|c|}{\large{\textbf{Temporal variables}}} \label{tab:bm_nomenclature_temporal}\\
    \hline
    \rowcolor{Grey} 
    \textbf{Notation} & \textbf{Description} & \textbf{Units}\\
    \hline
    $t_{BM}^{ex}$ & Execution date of the Balancing Mechanism. \textbf{\textit{Parameter}} & -\\
    \hline
    $t_{BM}^{start}$ & Start date of the effective period of the Balancing Mechanism. \textbf{\textit{Parameter}} & -\\
    \hline
    $t_{BM}^{end}$ & End date of the effective period of the Balancing Mechanism. \textbf{\textit{Parameter}} & -\\
    \hline
    $\Delta t_{BM}$ & Time step of the Balancing Mechanism. \textbf{\textit{Parameter}} & min\\
    \hline
    $T_{BM}$ & Effective time frame of the Balancing Mechanism, i.e. $T_{BM} = [t^{start}_{BM}, t^{start}_{BM} + \Delta t_{BM},\, \dots \,, t^{end}_{BM} - \Delta t_{BM}]$ & -\\
    \hline
\end{longtable}

\begin{longtable}[H]{!{\color{Grey}\vrule}>{\centering\arraybackslash}p{3cm} !{\color{Grey}\vrule} p{10.5cm}!{\color{Grey}\vrule} >{\centering\arraybackslash} p{2cm}!{\color{Grey}\vrule}}
    \arrayrulecolor{Grey} \hline
    \rowcolor{Grey} \multicolumn{3}{|c|}{\large{\textbf{Zonal- and TSO-specific characteristics}}} \label{tab:bm_nomenclature_zonal_tso_charac}\\
    \hline
    \rowcolor{Grey} 
    \textbf{Notation} & \textbf{Description} & \textbf{Units}\\
    \hline
    $bn_{BM,ca,t,t_{BM}^{ex}}$ \newline (short: $\mkern3mu bn_{t}$) & Balancing needs for in control area $ca \in CA$ for time $t \in T_{BM}$, seen from $t_{BM}^{ex}$. For readability, it is shortened as $bn_{t}$ & MW\\
    \hline
    $(\Delta q)^{bal}_{z,t}$ & Commercial (power) balance of area $z \in Z$ at time $t$, equal to the sum of all power exports minus the sum of all power imports & MW \\
    \hline
\end{longtable}

\begin{longtable}[H]{!{\color{Grey}\vrule}>{\centering\arraybackslash}p{3cm} !{\color{Grey}\vrule} p{10.5cm}!{\color{Grey}\vrule} >{\centering\arraybackslash} p{2cm}!{\color{Grey}\vrule}}
    \arrayrulecolor{Grey} \hline
    \rowcolor{Grey} \multicolumn{3}{|c|}{\large{\textbf{Global unit characteristics}}} \label{tab:bm_nomenclature_unit_charac}\\
    \hline
    \rowcolor{Grey} 
    \textbf{Notation} & \textbf{Description} & \textbf{Units}\\
    \hline
    $P_{u,t,t_{m}^{ex}}^{plan}$ & Power output of unit $u \in U$ at time $t \in T_{m}$, seen from time $t_{m}^{ex}$. \textbf{\textit{Input data}} & MW\\
    \hline
    $P_{u,t}^{act}$ & Power activated by the Balancing Mechanism on unit $u \in U$ at time $t \in T_{m}$. \textbf{\textit{Variable}} & MW\\
    \hline
    $P_{u,t, t_{m}^{ex}}^{final}$ & Final power output of the unit $u$ at time $t \in T_{m}$, seen from $t_{m}^{ex}$. It corresponds to the sum of $P_{u,t,t_{m}^{ex}}^{plan}$ and $P_{u,t}^{act}$ & MW\\
    \hline
    $P_{u,t}^{max}$ & Maximum power output of unit $u \in U$ at time $t \in T_{m}$. \textbf{\textit{Input data}} & MW\\
    \hline
    $P_{u,t}^{min}$ & Minimum power output of unit $u \in U$ at time $t \in T_{m}$. \textbf{\textit{Input data}} & MW\\
    \hline
    $p_{u,t}$ & Activation price of unit $u$ at time $t$. \textbf{\textit{Input data}} & \euro\\
    \hline
    $R_{ut,t_{m}^{ex}}^{m^{R}, \sigma^{R}}$ & Procured reserves of type $m^{R} \in [FCR, aFRR, mFRR, RR]$ in direction $\sigma^{R} \in [up, down]$  on unit $u$ at time $t$, seen from $t_{m}^{ex}$. \textbf{\textit{Input data}} & -\\
    \hline
\end{longtable}

\begin{longtable}[H]{!{\color{Grey}\vrule}>{\centering\arraybackslash}p{3cm} !{\color{Grey}\vrule} p{10.5cm}!{\color{Grey}\vrule} >{\centering\arraybackslash} p{2cm}!{\color{Grey}\vrule}}
    \arrayrulecolor{Grey} \hline
    \rowcolor{Grey} \multicolumn{3}{|c|}{\large{\textbf{Hydraulic- and Storage-specific unit characteristics}}} \label{tab:bm_nomenclature_hydro_storage}\\
    \hline
    \rowcolor{Grey} 
    \textbf{Notation} & \textbf{Description} & \textbf{Units}\\
    \hline
    $E_{u,t,t_{m}^{ex}}^{stored}$ & Stored energy in the reservoir of unit $u \in U^{h} \, \cup \, U^{st}$ at time $t \in T_{m}$, seen from time $t_{m}^{ex}$ & MWh\\
    \hline
    $E_{u,t}^{max}$ & Maximum storage level in the reservoir of unit $u \in U^{h} \, \cup \, U^{st}$ at time $t \in T_{m}$ & MWh\\
    \hline
    $E_{u,t}^{min}$ & Minimum storage level in the reservoir of unit $u \in U^{h} \, \cup \, U^{st}$ at time $t \in T_{m}$ & MWh\\
    \hline
    $d_{u}^{tran}$ & Transition duration between pumping and turbining of unit $u$ of type Pumped Hydraulic Storage (PHS), that are modeled as storage units in ATLAS (i.e. $u \in U^{st}$) & min\\
    \hline
    $WV_{u,t,E_{u,t,t_{m}^{ex}}^{stored}}$ & Marginal storage value of unit $u  \in U^{h}$ at time $t$, for stored energy level $E_{u,t,t_{m}^{ex}}^{stored}$ & \euro/MWh\\
    \hline
    $\eta^c_u$ & Charge efficiency of unit $u \in U^{st}$ & -\\
    \hline
    $\eta^d_u$ & Discharge efficiency of unit $u \in U^{st}$ & -\\
    \hline
\end{longtable}

\begin{longtable}[H]{!{\color{Grey}\vrule}>{\centering\arraybackslash}p{3cm} !{\color{Grey}\vrule} p{10.5cm}!{\color{Grey}\vrule} >{\centering\arraybackslash} p{2cm}!{\color{Grey}\vrule}}
    \arrayrulecolor{Grey} \hline
    \rowcolor{Grey} \multicolumn{3}{|c|}{\large{\textbf{Load-, Photovoltaic- and Wind-specific unit characteristics}}} \label{tab:bm_nomenclature_pv_wind}\\
    \hline
    \rowcolor{Grey} 
    \textbf{Notation} & \textbf{Description} & \textbf{Units}\\
    \hline
    $P_{u,t,t_{m}^{ex}}^{for}$ & Forecast of the maximum power output of unit $u \in \{U^{w}, U^{pv}, U^{l}\}$ at time $t \in T_{m}$. \textbf{\textit{Input data}} & MW\\
    \hline
\end{longtable}

\begin{longtable}[H]{!{\color{Grey}\vrule}>{\centering\arraybackslash}p{3cm} !{\color{Grey}\vrule} p{10.5cm}!{\color{Grey}\vrule} >{\centering\arraybackslash} p{2cm}!{\color{Grey}\vrule}}
    \arrayrulecolor{Grey} \hline
    \rowcolor{Grey} \multicolumn{3}{|c|}{\large{\textbf{Other notations}}} \label{tab:bm_nomenclature_others}\\
    \hline
    \rowcolor{Grey} 
    \textbf{Notation} & \textbf{Description} & \textbf{Units}\\
    \hline
    $VoLL$ & Value of loss load.\textbf{\textit{Parameter}} & \euro/MWh\\
    \hline
    $p^{spill}$ & Spillage penalty. \textbf{\textit{Parameter}} & \euro/MWh\\
    \hline
    $p^{redispatch}$ & Redispatch penalty. \textbf{\textit{Parameter}} & \euro/MWh\\
    \hline
    $E^{VoLL}_{ca,t}$ & Loss load energy in control area $ca$ at time $t$. \textbf{\textit{Variable}} & MWh\\
    \hline
    $E_{ca,t}^{spill}$ & Spilled energy in control area $ca$ at time $t$. \textbf{\textit{Variable}} & MW\\
    \hline
\end{longtable}

\renewcommand{\arraystretch}{1.1} 

\section{TSO balancing needs computation}
\label{sec:4.3 - FrBM TSO balancing needs}
From now on, we place ourselves at the scale of a control area $ca$ associated with a given TSO, and we look at a Balancing Mechanism process executed at $t_{BM}^{ex}$, on the operating time frame $T_{BM} = [t_{BM}^{start}; t_{BM}^{end}]$.

The balancing needs computation for this TSO is directly given by Equation \ref{eq:4.3_tso_needs}, and corresponds to the difference between the sum of consumption plans of all load units in $ca$ (taken in absolute value by convention, see Section \ref{sec:TSO needs - imbalance}) and the sum of all power output from generation units in $ca$, with an additional term indicating the commercial power balance of the control area:
\begin{align}
    \forall ca \in \mkern3mu CA,  &\mkern9mu \forall t \in T_{BM},\notag\\
    bn_{t} = \sum_{u \in U^{l}} | P_{u,t, t_{BM}^{ex}}^{plan} |  - & \sum_{u \in U^{g}}  P_{u,t, t_{BM}^{ex}}^{plan} + \sum_{z \in Z_{ca}} (\Delta q)^{bal}_{z,t}
    \label{eq:4.3_tso_needs}
\end{align}

As a reminder, the commercial power balance $(\Delta q)^{bal}_{z,t}$ corresponds to the sum of all exports minus the sum of all imports of market area $z$.

\section{Reserve activation problem formulation}
\label{sec:4.4 - FrBM OP problem}
The reserve activation process takes as inputs the following information, for a given TSO associated with its control area $ca$, and for an operating time frame $T_{BM}$: 
\begin{itemize}
    \item The volume of balancing needs $bn_{t}, \mkern9mu \forall t \in T_{BM}$.
    \item The activation price $p_{u,t}$ for each unit in its control area, in both upward and downward directions (usually symmetrical, except when startup costs are taken into account).
    \item Generation / consumption plans $P_{u,t,t_{BM}^{ex}}^{plan}$, and operational constraints of each unit $u$ in its control area.
\end{itemize}

Because of the specific features of reserve orders on the BM previously mentioned, the reserve activation problem that is solved by the TSO is effectively a unit commitment problem made at the scale of the control area. In that sense, it is extremely close to the optimization problem used in both Day-Ahead Orders and Portfolio Optimization modules, described in \cite{little_atlas_nodate}, especially regarding the application of operational constraints that bound the power output of units. A similar optimization problem to the optimal dispatch presented in \cite{little_atlas_nodate} is consequently used in the BM, and this section will directly refer to this article when a section of the problem is identical.

\subsection{Redefinition of global objective function and of main control variables}
\label{sec:4.4.1 - FrBM power output redefinition}
The main optimization variable used in the original problem used in the day-ahead market is the power output of each unit $u$, noted $P_{u,t}$ at time $t$. For the BM, a generation or consumption plan already exists for each unit, and the quantity to optimize is then the evolution of power output compared to this planning. This quantity will be noted $P_{u,t}^{act}$, standing for activated power. The final power output $P_{u,t,t_{BM}^{ex}}^{final}$ resulting from the BM process eventually corresponds to the sum of the planned generation or consumption and the activated power:
\begin{equation} \label{eq:4.4.1_frbm_power_variable}
    \forall t \in T_{BM}, \mkern9mu \forall u \in U, \quad P_{u,t,t_{BM}^{ex}}^{final} = P_{u,t,t_{BM}^{ex}}^{plan} + P_{u,t}^{act}
\end{equation}

$P_{u,t}^{act}$ is not directly bounded and can be both positive or negative. Every physical bound consecutive of operating constraints are applied and verified on $P_{u,t,t_{BM}^{ex}}^{final}$, as this represents the actual power output. $P_{u,t}^{act}$ is however the control variable included in the objective function $\pi_{BM}$ that performs a cost minimization, described in Equation \ref{eq:4.4.1_frbm_objective_function}.

\begin{align} \label{eq:4.4.1_frbm_objective_function}
    \pi_{BM}(P_{u,t}^{act}, E^{VoLL}_{ca,t}, E^{spill}_{ca,t}) = \min & \biggr[ \sum_{t \in T_{BM}} \sum_{u \in U_{ca}} P_{u,t}^{act} * p_{u,t} * \frac{\Delta t_{BM}}{60} \notag\\ 
    & + \sum_{t \in T_{BM}} E^{VoLL}_{ca,t} * VoLL + \sum_{t \in T_{BM}} E^{spill}_{ca,t} * 26,000 \notag \\ 
    &+ \sum_{t \in T_{BM}} | P_{u,t}^{act} | * p^{redispatch} \biggr]
\end{align}

Where:
\begin{itemize}
    \item $E^{VoLL}_{ca,t}$ is the unsupplied energy and corresponds to the part of the unflexible load that cannot be provided for a given time $t \in T_{BM}$ in the control area $ca$. Its associated cost is the $VoLL$, standing for Value of Loss Load. This value is a parameter that can be chosen by the user, and is by default set at 26,000 \euro/MWh.
    \item $E^{spill}_{ca,t}$ represents the spilled energy, which is the surplus of energy that cannot be consumed in the control area $ca$ at time $t$. Notably, this variable does not include curtailment of renewable energy sources: it represents the situation where generation units in $ca$ are technically not able to decrease their generation anymore. A spillage penalty is associated with it, and is set at 26,000 \euro/MWh. Note that this was decided for optimization purposes (cf. the following paragraph), and it does not represent the actual spillage cost  $p^{spill}$, which is a parameter set by the user.
    \item Finally, the last term of the objective function is added to prevent the Balancing Mechanism from performing economic redispatch. Indeed, its sole purpose is to resolve the balancing needs, and not to reschedule generation plans if it finds an optimal solution to its unit commitments that differs from the input situation. This concretely means that it should avoid as much as possible counter-activations, i.e. activations of power that are not in the direction that contributes to solving balancing needs. Counter-activations cannot completely be prohibited, as this could lead to solver unfeasibility. To penalize counter-activations, the absolute value of activated power is added to the objective function, and penalized by the coefficient $p^{redispatch}$ whose value is set by the user\footnote{This value has to be chosen carefully because of its interaction with $p^{spill}$ and $VoLL$. For the redispatch penalty to be effective, it should outweigh any order price $p_{u,t}$ submitted by BSPs. However, if it is chosen too high (higher than $p^{spill}$ and $VoLL$), the optimization will actually prioritize unsupplied energy or spilled energy to meet the balancing needs, even if it is able to activate power instead. It is recommended to choose a value in the range [5,000; 10,000]}. 
\end{itemize}

\subsection{Main supply-demand balance constraint}
\label{sec:4.4.2 - FrBM notice delay}
The main constraint that enforces the balance between supply and demand is described in Equation \ref{eq:4.4.2_frbm_supply_demand}:

\begin{equation} \label{eq:4.4.2_frbm_supply_demand}
    \forall t \in T_{BM}, \quad \sum_{u \in U_{ca}}  P_{u,t, t_{BM}^{ex}}^{act} + E_{ca,t}^{VoLL} * \frac{60}{\Delta t_{BM}} - E_{ca,t}^{spill} * \frac{60}{\Delta t_{BM}} = bn_{t}
\end{equation}

Conversions from energy to power are required for unsupplied energy and spilled energy, explaining the coefficient $\frac{60}{\Delta t_{BM}}$.

\subsection{Bounds of unsupplied energy and spilled energy}
\label{sec:4.4.3 - FrBM unsupplied spilled bounds}
Both $E_{ca,t}^{VoLL}$ and $E_{ca,t}^{spill}$ are defined as positive, but are not bounded by an upper limit:

\begin{equation}\label{eq:4.4.3_frbm_unsupplied_bound}
    \forall t \in T_{BM}, \quad E_{ca,t}^{VoLL} \geq 0
\end{equation}

\begin{equation}\label{eq:4.4.3_frbm_spill_bound}
    \forall t \in T_{BM}, \quad E_{ca,t}^{spill} \geq 0
\end{equation}

\subsection{Notice delay constraint}
\label{sec:4.4.4 - FrBM notice delay}
The notice delay constraint is associated with the unit characteristic $d_{u}^{notice}$, and represents the preparation period required for unit $u$ to respond to any request of power output modification, i.e. the delay between the reception of this request and the beginning of the ramping state. This constraint is not included in the core optimization problem that was constructed for the day-ahead market, as it does not appear to be relevant for the time frame of this market. However, it is definitely impactful when looking at balancing processes, and is added to the optimization problem with the following constraint:

\begin{equation} \label{eq:4.4.4_frbm_notice_delay}
    \forall t \in T_{BM}, \mkern9mu \forall u \in U, \mkern9mu if \mkern6mu  t - t_{BM}^{ex} < d_{u}^{notice}, \quad P_{u,t}^{act} = 0
\end{equation}

\subsection{Applying operating constraints to generation and consumption units}
\label{sec:4.4.5 - FrBM operational constraints}
The Optimal Dispatch function described in Chapter 3 of \cite{little_atlas_nodate} is also the basis upon which most generation and consumption units are modeled in the Balancing Mechanism:

\begin{itemize}
    \item For thermal units, the entire modeling method is applied to $P^{final}_{u,t}$ (Section 3.2), notably all constraints from 3.2.2 to 3.2.39.
    \item For wind and photovoltaic units, constraints 3.3.1 and 3.3.2 of Section 3.3 are applied to $P^{final}_{u,t}$.
    \item For non dispatchable generation and load, constraints 3.4.1 of Section 3.4 are applied to $P^{final}_{u,t}$.
    \item For flexible load units, constraints 3.5.1 and 3.5.2 of Section 3.5 are applied to $P^{final}_{u,t}$.
    \item For hydraulic units, constraints 3.6.1 and 3.6.2 of Section 3.6 are applied to $P^{final}_{u,t}$.
\end{itemize}

The constraints applied to storage units differ from the description of the Optimal dispatch in \cite{little_atlas_nodate}: 

\begin{itemize}
    \item The optimization done to identify the time steps when it is preferable to buy or sell is not performed. The activated power is separated into positive and negative components (Equation \ref{eq:bm_storage_pos_neg}), and both are subject to the following constraints:
    \begin{gather}
        P^{act}_{u,t,t^{ex}_{BM}} = \sigma^{sell}_{u,t} * P^{sell}_{u,t} - (1 - \sigma^{sell}_{u,t}) * P^{buy}_{u,t}\label{eq:bm_storage_pos_neg} \\
        0 \leq P^{sell}_{u,t} \leq \sigma^{sell}_{u,t} * (P^{max}_{u,t} - P^{for}_{u,t,t^{ex}_{BM}})\label{eq:bm_storage_sell_lim} \\
        (1 - \sigma^{sell}_{u,t}) * (P^{min}_{u,t} - P^{for}_{u,t,t^{ex}_{BM}}) \leq P^{buy}_{u,t} \leq 0\label{eq:bm_storage_buy_lim} \\
        P^{for}_{u,t,t^{ex}_{BM}} + P^{sell}_{u,t} \leq  P^{max}_{u,t} * \frac{\Delta t}{60}\label{eq:bm_storage_max_power} \\
        P^{for}_{u,t,t^{ex}_{BM}} - P^{buy}_{u,t}\geq  P^{min}_{u,t} * \frac{\Delta t}{60} 
        \label{eq:bm_storage_min_power} \\
        E^{stored}_{u,t} \leq E^{max}_{u,t} \label{eq:bm_storage_max_energy}\\
        E^{stored}_{u,t} \geq E^{min}_{u,t} \label{eq:bm_storage_min_energy}
    \end{gather}
    The storage level is tracked by Equation \ref{eq:bm_storage_tracking}:
    \begin{equation}
        \forall t \in T_{BM},  \quad E^{stored}_{u,t} = E^{stored}_{u,t-1} +P^{buy}_{u,t} * \eta^c_u - \frac{P^{sell}_{u,t}}{\eta^d_u} - \delta_{u}^{EV} * E^{disp}_{u, t^{start}-1,t} \label{eq:bm_storage_tracking}
    \end{equation}
    \item An additional constraint is applied to Pumped Hydraulic Storage (PHS) units $u_{PHS}$ in the Balancing Mechanism. It is related to the transition duration between pumping and turbining $d^{tran}$, which forces the unit to stay at a null power output for at least $d^{tran}$ before switching from a positive to negative power output (or conversely, at it is assumed to be symmetrical in ATLAS). However, this constraint is currently approximated in the Balancing Mechanism, given that the $\Delta t_{BM}$ used in the model is usually not below 15 minutes. In practice, it is sufficient to prevent a PHS unit with a $d^{tran} \geq \Delta t_{BM}$ from switching mode during a BM time frame, which is implemented in Equations \ref{eq:bm_phs_transition_duration_pos} (restriction to turbining mode) and \ref{eq:bm_phs_transition_duration_neg} (restriction to pumping mode). Equation \ref{eq:bm_phs_transition_duration_infeasible} describes the case where a transition is already planned within the BM time frame. In that situation, the power output is supposed to be unchangeable to avoid any unfeasibility.
\begin{align}
    \forall u \in U^{PHS}, \mkern9mu if \mkern9mu d^{tran}_{u} \geq \Delta t_{BM}, & \notag\\
    \label{eq:bm_phs_transition_duration_infeasible}
    if \mkern9mu (\min\limits_{t \in T_{BM}} P_{u,t, t^{ex}_{BM}}^{plan} < 0) \mkern9mu \& \mkern9mu (\max\limits_{t \in T_{BM}} P_{u,t, t^{ex}_{BM}}^{plan} > 0), & \quad P_{u,t}^{acc} = 0 \quad \forall t \in T_{BM}\\
    \label{eq:bm_phs_transition_duration_pos}
    if \mkern9mu (\min\limits_{t \in T_{BM}} P_{u,t, t^{ex}_{BM}}^{plan} \geq 0) \mkern9mu \& \mkern9mu (\max\limits_{t \in T_{BM}} P_{u,t, t^{ex}_{BM}}^{plan} \geq 0), & \quad P_{u,t}^{acc} \geq 0 \quad \forall t \in T_{BM}\\
    \label{eq:bm_phs_transition_duration_neg}
    if \mkern9mu (\min\limits_{t \in T_{BM}} P_{u,t, t^{ex}_{BM}}^{plan} \leq 0) \mkern9mu \& \mkern9mu (\max\limits_{t \in T_{BM}} P_{u,t, t^{ex}_{BM}}^{plan} \leq 0), & \quad P_{u,t}^{acc} \leq 0 \quad \forall t \in T_{BM}
\end{align}
\end{itemize}

\subsection{Previously procured reserves}
\label{sec:4.4.6 - FrBM procured reserves}
All manual reserves (RR or mFRR) previously procured are considered available on the Balancing Mechanism. On the other hand, previously procured automatic reserves (FCR and aFRR) cannot be activated in this process. This is implemented in the optimization by constraining the power output by Equations \ref{eq:bm_reserves_constraints_up} and \ref{eq:bm_reserves_constraints_down}:

\begin{align}
    \forall u \in U & , \mkern9mu \forall t \in T_{BM}, \notag\\
    \label{eq:bm_reserves_constraints_up}
    P_{u,t, t^{ex}_{BM}}^{plan} + P_{u,t}^{acc} \leq & \mkern6mu P_{u,t}^{max} - (R^{aFRR, up}_{u,t} + R^{FCR, up}_{u,t})\\
    \label{eq:bm_reserves_constraints_down}
    P_{u,t, t^{ex}_{BM}}^{plan} + P_{u,t}^{acc} \geq & \mkern6mu P_{u,t}^{min} + (R^{aFRR, dn}_{u,t} + R^{FCR, dn}_{u,t})
\end{align}

\section{Balancing Mechanism outputs}
\label{sec:4.5 - FrBM output}
Following variables and quantities are extracted for all time $t \in T_{BM}$ at the end of the Balancing Mechanism:
\begin{itemize}
    \item The overall and final power output of each unit $P_{u,t,t_{BM}^{ex}}^{final}$, for each unit $u \in U$. 
    \item Both the unsupplied energy $E_{ca,t}^{VoLL}$ and the spilled energy $E_{ca,t}^{spill}$.
    \item Total balancing costs $c_{ca,t}^{BM}$, defined in Equation \ref{eq:4.5_frbm_total_balancing_costs}, consisting of the final value of the objective function from which are subtracted two terms. First, the entire term corresponding to economic redispatch, as it was only introduced to control the behavior of the optimization and does not represent a physical quantity. Second, the quantity $E_{ca,t}^{spill} * (26,000 - p^{spill})$. $p^{spill}$ is a parameter chosen by the user and represents the actual spillage cost. The last subtraction then corrects the spillage cost from the value required for optimization reliability (equal to $26,000$) to the actual spillage value.

    \begin{equation} \label{eq:4.5_frbm_total_balancing_costs}
        \forall t \in T_{BM}, \quad c_{ca,t}^{BM} =  \pi_{BM}(P_{u,t}^{act}, E^{VoLL}_{ca,t}, E^{spill}_{ca,t}) - \sum_{t \in T_{BM}} | P_{u,t}^{act} | * p^{redispatch} - E^{spill}_{ca,t} * (26,000 - p^{spill})
    \end{equation}
\end{itemize}

\chapter{Post Clearing Aggregation} 
\label{ch:Post Clearing Aggregation}

This module is made to be run directly following the Clearing module of a balancing market process. Its role is to gather order quantities accepted by the aforementioned Clearing, and to accordingly update relevant properties of the studied dataset. This task is performed by the Portfolio Optimization module (see \cite{little_atlas_nodate}) for day-ahead and intraday markets. However, in ATLAS, balancing market processes are assumed to be performed at a unit-based level, which means that the Portfolio Optimization can be replaced by a much lighter module. 

The properties updated by the Post Clearing Aggregation are:
\begin{itemize}
    \item The power output of all units. For a given unit $u$, the new power output corresponds to the sum of the last plan and the quantity activated by the Clearing for all orders $o \in O_{u}$ formulated with unit $u$ on balancing market $m$ (Equation \ref{eq:post_clearing_aggreg_power}):

    \begin{equation}
        \label{eq:post_clearing_aggreg_power}
        \forall u \in U, \mkern9mu \forall t \in T_{m}, \quad P^{plan}_{u,t,t^{ex}_{m}} \gets P^{plan}_{u,t,t^{ex}_{m}} + \sum\limits_{o \in O_u \mkern6mu | \mkern6mu t^{start}_{o} = t} -\sigma_{o} * q^{acc}_{o}
    \end{equation}

    \item The aggregated power output in all Portfolios, which is simply the sum of the updated power output of all its units (Equation \ref{eq:post_clearing_aggreg_portfolio}):

    \begin{equation}
        \label{eq:post_clearing_aggreg_portfolio}
        \forall pf \in PF, \mkern9mu \forall t \in T_{m}, \quad P^{plan}_{pf,t,t^{ex}_{m}} \gets \sum\limits_{u \in pf} P^{plan}_{u,t,t^{ex}_{m}}
    \end{equation}

    \item Finally, for units $u$ functioning with a reservoir (i.e. Hydraulic and Storage units), the stored energy $E^{stored}_{u,t,t^{ex}_{m}}$ at time $t$ is updated according to the previously planned stored energy at time $t$ and to all activated market orders before $t$ (Equation \ref{eq:post_clearing_aggreg_energy}):

    \begin{align}
        \label{eq:post_clearing_aggreg_energy}
        \forall u \in U^{h} \cup U^{st}, \mkern9mu
        \forall t \in T_{m}, \quad E^{stored}_{u,t,t^{ex}_{m}} \gets E^{stored}_{u,t,t^{ex}_{m}} + \frac{\Delta t_{m}}{60} * \sum\limits_{t' \in T^{m} \mkern6mu | \mkern6mu t' \leq t} \biggr[ \sum\limits_{o \in O_u \mkern6mu | \mkern6mu t^{start}_{o} = t'} \sigma_{o} * q^{acc}_{o} \biggr]
    \end{align}
    
\end{itemize}

\bibliography{bibliography} 

\end{document}